\documentclass[twocolumn]{aastex6}

\usepackage{amssymb}
\usepackage{amsfonts}
\usepackage{amsmath}
\usepackage[varg]{txfonts}
\usepackage{natbib}
\usepackage{color}
\usepackage{graphicx}
\usepackage{hyperref}

\newcommand{\trm}[1]{\textrm{#1}}

\shorttitle{$^{44}$Ti and $^{56}$Ni in a Cassiopeia~A like 3D Supernova Model}

\shortauthors{A.~Wongwathanarat et al.}

\begin{document}

\title{Production and Distribution of $^{44}$Ti and $^{56}$Ni in a
Three-dimensional Supernova Model Resembling Cassiopeia~A}

\author{Annop Wongwathanarat\altaffilmark{1}$^{,}$\altaffilmark{2},
        Hans-Thomas Janka\altaffilmark{2},
        Ewald M{\"{u}ller}\altaffilmark{2},
        Else Pllumbi\altaffilmark{2,3},
        and
        Shinya Wanajo\altaffilmark{4,5}}
\altaffiltext{1}{RIKEN, Astrophysical Big Bang Laboratory, 2-1 Hirosawa,
  Wako, Saitama~351-0198, Japan; {e-mail: annop.wongwathanarat@riken.jp}}
\altaffiltext{2}{Max-Planck-Institut f\"ur Astrophysik,
       Karl-Schwarzschild-Str. 1, 85748 Garching, Germany}
\altaffiltext{3}{Physik Department, Technische Universit\"at M\"unchen, 
James-Franck-Str.~1, 85748 Garching, Germany}
\altaffiltext{4}{Department of Engineering and Applied Sciences, Sophia 
University, Chiyoda-ku, Tokyo 102-8554, Japan}
\altaffiltext{5}{RIKEN, iTHES Research Group, 2-1 Hirosawa,
  Wako, Saitama~351-0198, Japan}

\begin{abstract}
The spatial and velocity distributions of nuclear species synthesized
in the innermost regions of core-collapse supernovae (SNe) can yield
important clues about explosion asymmetries and the operation of the
still disputed explosion mechanism. Recent observations of radioactive
$^{44}$Ti with high-energy satellite telescopes (NuSTAR, INTEGRAL)
have measured gamma-ray line details, which provide direct evidence 
of large-scale explosion asymmetries in Supernova~1987A, and in
Cassiopeia~A (Cas~A) even by mapping of the spatial brightness
distribution (NuSTAR). 
Here, we discuss a three-dimensional (3D) simulation of a
neutrino-driven explosion, using a parametrized neutrino engine,
whose $^{44}$Ti distribution is mostly concentrated in
one hemisphere pointing opposite to the neutron-star (NS) kick velocity.
Both exhibit intriguing resemblance to the observed morphology
of the Cas~A remnant, although neither progenitor nor explosion
were fine-tuned for a perfect match. 
Our results demonstrate that the asymmetries
observed in this remnant can, in principle, be accounted for by a
neutrino-driven explosion, and that the high $^{44}$Ti abundance 
in Cas~A may be explained without invoking rapid rotation or a 
jet-driven explosion, because neutrino-driven explosions genericly
eject large amounts of high-entropy matter. The recoil acceleration
of the NS is connected to mass-ejection asymmetries and is opposite
to the direction of the stronger explosion, fully compatible with
the gravitational tug-boat mechanism. Our results also imply that Cas~A
and SN~1987A could possess similarly ``one-sided''
Ti and Fe asymmetries, with the difference that Cas~A is viewed
from a direction with large inclination angle to the NS motion, whereas
the NS in SN~1987A should have a dominant velocity component pointing
toward us.
\end{abstract}

\keywords{supernovae: general --- supernovae: special: Cas~A --- 
hydrodynamics --- instabilities --- 
nuclear reactions, nucleosynthesis, abundances --- neutrinos}

\section{Introduction}

Radioactive nuclei such as $^{56,57}$Ni, $^{55,60}$Co, and $^{44}$Ti,
which are freshly synthesized during the explosion, power the
electromagnetic emission of supernovae (SNe) over months and years
by their decay to stable $^{56,57}$Fe, $^{55}$Mn, $^{60}$Ni, and 
$^{44}$Ca, respectively.
Moreover, their production in the innermost regions of the exploding
star qualifies such nuclei as sensitive probes of the conditions 
near the blast-wave origin and possibly even of the explosion
mechanism and the shock-wave dynamics during the earliest phases of
SN outbursts. These aspects are strong motivation for observational
efforts and theoretical 
studies of the nucleosynthesis of radioactive species with the 
goal to connect explosion models to observations and to deduce
important constraints for the underlying processes. 

Because of its diagnostic potential for the explosion dynamics
and its long half life of about 60 years\footnote{Nuclear decay
information is available from the National Nuclear Data Center at
{\tt http://www.nndc.bnl.gov/}.}, $^{44}$Ti
is of particular interest and can potentially be detected directly
or indirectly also in SNe and 
in young SN remnants (SNRs). This has indeed been achieved for
SN~1987A and Cassiopeia~A (Cas~A)
\citep[for reviews, see][]{Vink2005,Vink12,Diehl13}. 
Besides spectroscopic analysis and light-curve fitting of SNe, 
e.g.\ in the case of SN~1987A (see, for example, the recent works
by \citealt{Jerkstrandetal11} and \citealt{Seitenzahletal14}),
X-ray and gamma-ray observations of SN~1987A 
\citep[e.g.][]{Grebenevetal12,Boggsetal15} and Cas~A 
\citep[e.g.][]{Iyudinetal1994,Vinketal2001,Renaudetal06,Grefenstetteetal14,Siegertetal15} 
allow to obtain estimates of the $^{44}$Ti yield and to
deduce information on the velocity and spatial distributions of this
nucleus.

Particularly interesting in this respect are recent, long-exposure
measurements of the $^{44}$Ti emission from SN~1987A and Cas~A 
with the space-based INTErnational Gamma-Ray Astrophysics 
Laboratory \citep[INTEGRAL; e.g.][]{Renaudetal06,Siegertetal15,WangLi16}
and the Nuclear Spectroscopic Telescope
Array (NuSTAR) focusing high-energy X-ray telescope 
\citep{Boggsetal15,Grefenstetteetal14}. These
measurements provide clear evidence of the presence of such
inner, radioactive ejecta, and, moreover, reveal
large asymmetries of the $^{44}$Ti ejection 
closely connected to the inner ``engine'' of the explosion.
The imaging of the $^{44}$Ti distribution of Cas~A is an exciting
additional piece in a growing wealth of observational data that
reveal the three-dimensional morphology of this young SNR
in fascinating detail, including nebular emission lines of
N, O, Ne, Si, S, Ar, Ca, and constraints on the neutron-star (NS)
kick \citep[e.g.][]{Hughesetal2000,Fesen2001,Fesenetal2001,Gotthelfetal2001,HwangLaming2003,LamingHwang03,Hwangetal04,Fesenetal2006b,Smithetal2009,Delaneyetal10,Isenseeetal2010,Fesenetal2011,HwangLaming12,Isenseeetal2012,MilisavljevicFesen2013,MilisavljevicFesen15}.
The NuSTAR map of the Cas~A $^{44}$Ti distribution \citep{Grefenstetteetal14}
in combination with the determined direction of the 
NS motion \citep{Thorstensenetal2001,Fesenetal06} provides extremely valuable 
hints to the explosion dynamics that manifests itself in the recoil
acceleration of the compact remnant and the anisotropic
expulsion of nucleosynthetic products originating from the 
innermost SN ejecta. The inferred $^{44}$Ti yield of $\sim$$(1.0-1.7)
\times10^{-4}\,M_\odot$ by NuSTAR \citep{Grefenstetteetal14,Grefenstetteetal17}
and INTEGRAL \citep{Siegertetal15,WangLi16}
also provides a clue of the nucleosynthesis-relevant
conditions, which allow for a $^{44}$Ti production that 
is about three times greater than recent theoretical 
estimates invoking spherically symmetric explosions 
\citep[e.g.,][]{Peregoetal15}. The need of asymmetric explosions
to explain the large yields of $^{44}$Ti in SN~1987A and Cas~A was
pointed out first by \citet{Nagatakietal97,Nagatakietal98}.

Here we present results from three-dimensional (3D) hydrodynamical
SN simulations that exhibit morphological properties with
intriguing similarity to those of the Cas~A SNR.
Our results demonstrate that the neutrino-driven mechanism
with associated hydrodynamic instabilities is able
to explain the NS kick and the spatial asymmetries of
the $^{44}$Ti knots observed in a core-collapse SN like
Cas~A. We also post-process our 3D hydrodynamic calculations
for the production of $^{44}$Ti and $^{56}$Ni using tracer
particles and a large reaction network of 6300 nuclear species.
The discussed SN simulations are based on one
model drawn from a larger pool of 3D simulations of 
neutrino-powered explosions published in a series
of papers by \citet{Wongwathanaratetal10b,Wongwathanaratetal13,
Wongwathanaratetal15}.
The considered model is case
W15-2, whose onset of the explosion and first seconds
were computed by \citet{Wongwathanaratetal10b,Wongwathanaratetal13}, 
and whose continued evolution to shock breakout from the stellar surface
was subsequently followed as model W15-2-cw by \citet{Wongwathanaratetal15}. 
The original progenitor of
this explosion model was a red supergiant star (RSG) of
15\,$M_\odot$ with a $\sim$4.4\,$M_\odot$ helium core of
slightly more than $8\times 10^{10}$\,cm in radius, whereas
Cas~A goes back to a Type-IIb SN and therefore a star
that had stripped most of its hydrogen \citep{Krauseetal08,Restetal11}.
In order to avoid the dynamical consequences of the extended 
hydrogen envelope we repeated our previous simulations for the
same stellar model but after having removed the H-envelope
except its innermost $\sim$0.3\,$M_\odot$.

For our demonstration of the possibility in principle
to reproduce the observed explosion asymmetries,
it is not necessary that the employed model of the progenitor
optimally matches the estimated properties of the Cas~A 
progenitor \citep[as discussed, e.g., in][]{Orlandoetal16}.
It is also not mandatory to strive for a perfect
agreement of the fundamental parameters of the explosion
model with those determined for the Cas~A SN. Hence,
instead of performing a tedious and
computing-intense fine tuning of explosion parameters, we
can focus on the existing W15-2 model with its peculiar explosion
geometry, even though this simulation falls somewhat short of 
producing the explosion energy and iron yield inferred from
detailed analyses of the Cas~A SNR. In view of the
stochastic nature of the hydrodynamic instabilities, i.e.\
convective overturn and standing accretion shock instability
\citep[SASI;][]{Blondinetal03}, which are responsible
for the growth of initial seed perturbations to the
observable ejecta asymmetries, we can expect that cases
with morphologies similar to that of W15-2 will also be
found in sufficiently large sets of models with slightly
different explosion energies and iron yields.

Our paper is structured as follows. In Sect.~\ref{sec:numerics}
we will describe our numerical methods, initial models,
input physics, and post-processing for the nucleosynthesis,
in Sect.~\ref{sec:results} our results for the $^{44}$Ti
and $^{56}$Ni nucleosynthesis and the spatial and velocity
distributions of these nuclei in comparison to the NuSTAR
observations of Cas~A, and in Sect.~\ref{sec:conclusions}
we will finish with a summary and conclusions, also with
an eye on possible aspects in common with SN~1987A.

\section{Numerical setup}
\label{sec:numerics}

\subsection{Numerical method}
\label{sec:method}

Our simulation of a SN~IIb-like case, starting at $\sim$1400\,s
after core bounce, is performed with the finite-volume Eulerian
multifluid hydrodynamics code {\sc Prometheus} \citep{Fryxelletal91,
  Muelleretal91,Muelleretal91b}. The multidimensional Euler equations
are integrated using the dimensional splitting technique of
\citet{Strang68}. The code utilizes the piecewise parabolic
method \citep[PPM;][]{ColellaWoodward84}, and employs a Riemann solver
for real gases \citep{ColellaGlaz85}. To prevent numerical artifacts
created by the odd-even decoupling \citep{Quirk94} the AUSM+ Riemann
solver of \citet{Liou96} is applied inside grid cells with strong
grid-aligned shocks.

For spatial discretization we utilize the Yin-Yang overlapping grid
technique in spherical geometry \citep{KageyamaSato04} implemented in
the {\sc Prometheus} code to alleviate the restrictive
Courant-Friedrich-Lewy (CFL) time-step condition in polar regions
\citep{Wongwathanaratetal10}. In our simulation an 
angular resolution of 2$^\circ$ is used. The radial grid is
logarithmically spaced, spanning from $\sim$$3.1\times10^5$\,km to
$10^{9}$\,km with a relative grid resolution $\Delta r/r$ of
$\sim$0.5\% and $\sim$1\% inside and outside of the progenitor star,
respectively. The total number of (static) Eulerian grid zones is
$1198\times47\times137\times2$ initially. The innermost radial grid
zones are successively discarded as the SN shock propagates to larger
radii to relax the CFL condition. The radius of the inner grid
boundary, which is treated with an outflow condition,
is thus time-dependent, and is placed at 
approximately $2\%$ of the minimum SN shock radius. At the end of the
simulation the number of radial grid zones has decreased to 466 cells.

\subsection{Initial model}
\label{sec:inimod}

Our SN simulation is based on a 15\,$M_\odot$ red supergiant
progenitor, model s15s7b2 of \citet{WoosleyWeaver95}, which
was denoted as pre-collapse model W15 in \citet{Wongwathanaratetal15}.
The first seconds of the neutrino-driven explosion of this model
were simulated in 3D by \citet{Wongwathanaratetal10b,Wongwathanaratetal13},
using the {\sc Prometheus-Hotb} code with the Yin-Yang grid.
The code contains an approximate, gray treatment of neutrino transport
\citep{Schecketal06} connected with a boundary condition for
the luminosities of all neutrino species imposed at a contracting
Lagrangian radius placed at 1.1\,$M_\odot$ considerably inside the
neutrinosphere. By prescribing (time-dependent) neutrino luminosities
at this inner grid boundary it is possible to regulate the strength
of the neutrino heating and thus to tune the energy of the
neutrino-driven explosion to a desired value.

The particular model considered here is W15-2, which explodes with
an energy of about $1.5\times 10^{51}\,\mathrm{erg} = 1.5\,\mathrm{bethe}
= 1.5$\,B \citep[1.13\,B after 1.4\,s and 1.47\,B 
finally;][]{Wongwathanaratetal10b,Wongwathanaratetal13,Wongwathanaratetal15}. 
The evolution until shock breakout from the red supergiant was
tracked as model W15-2-cw by \citet{Wongwathanaratetal15}.

We stress that the asymmetries of the 3D explosion model developed
stochastically mainly by convective overturn in the neutrino-heating
layer. Convection, which is the dominant hydrodynamic instability
in the postshock region under the conditions of this model, is 
seeded by initial random perturbations with an amplitude of 0.1\% 
of the radial velocity, imposed as cell-by-cell random pattern
at the beginning of the simulation on a spherically symmetric (1D)
post-collapse model 
about 10\,ms after bounce. The large-scale asymmetries that 
characterize the final distribution of iron-group elements, silicon,
and oxygen \citep[see][]{Wongwathanaratetal15} were neither imposed
by hand nor did they grow from pre-existing large-scale perturbations,
e.g.\ in the oxygen-burning shell of the progenitor star \citep[for such
possible asymmetries, which are not considered here, 
see][]{ArnettMeakin11,Couchetal15,Muelleretal16}.

\begin{figure}
\centering
\resizebox{\hsize}{!}{\includegraphics{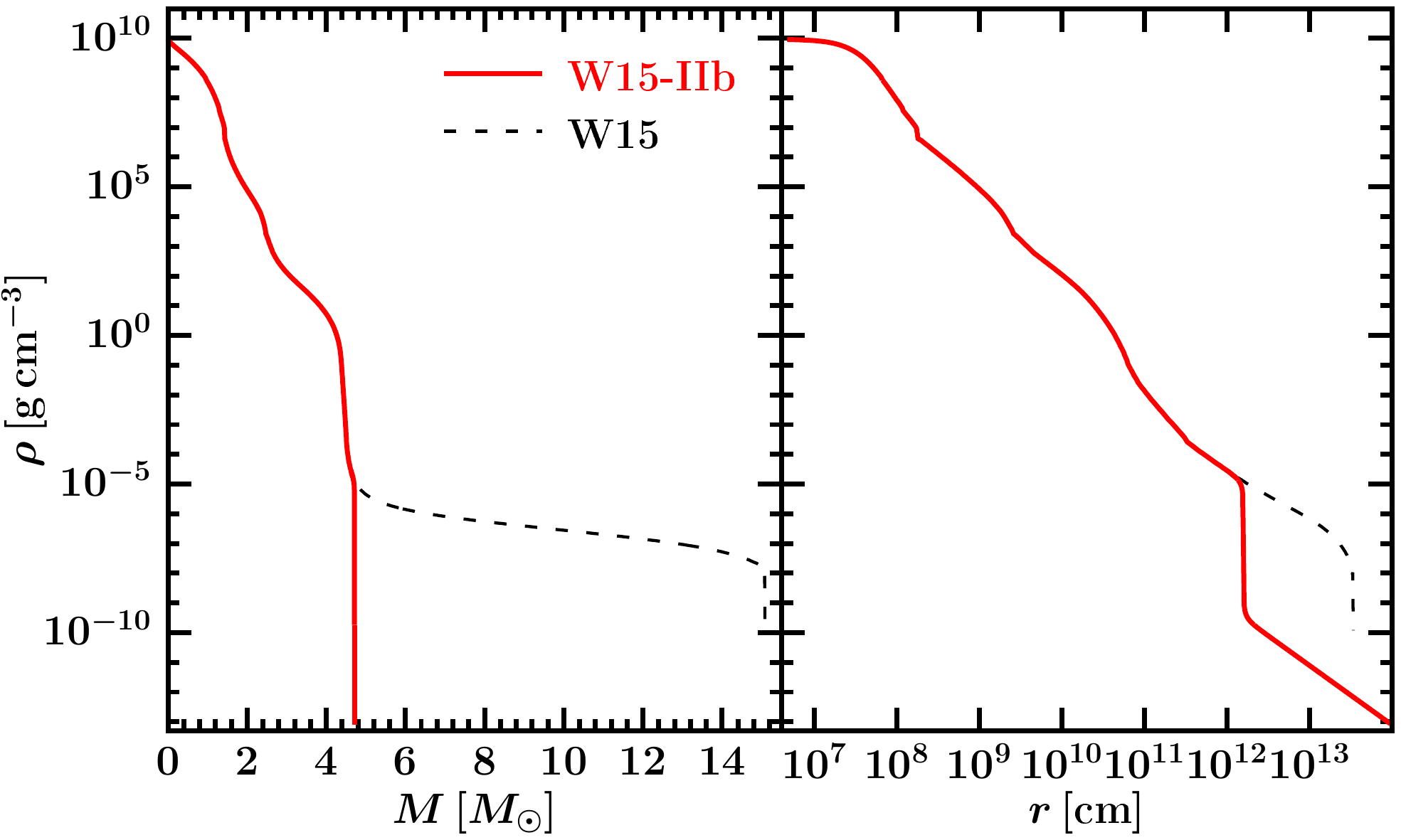}}
\caption{Density profiles versus enclosed mass (left panel) 
and radius (right panel) of progenitor model W15 (dashed black 
lines) and W15-IIb (solid red lines).}
\label{fig:rhovs}
\end{figure}

Since Cas~A was diagnosed to be a Type~IIb SN 
\citep{Krauseetal08,Restetal11}, 
we repeat the 3D long-time simulation in the present work with the
W15 progenitor model after having removed the hydrogen envelope down
to a rest of $\sim$0.3\,$M_\odot$, see model W15-IIb in 
Figure~\ref{fig:rhovs}, which shows the density profile versus
enclosed mass and radius of the modified progenitor model in
comparison to the original W15 model. The stellar radius of 
W15-IIb becomes $R_*\approx1.5\times10^{7}$\,km. Outside of this
radius we adopt the treatment by \citet{Wongwathanaratetal15} and
assume the presence of circumstellar matter with density 
and temperature profiles following an $r^{-2}$ decline. 
Since the onset and early phases of the explosion are not 
affected by our modification of the hydrogen envelope, we 
map initial data from model W15-2-cw at a post-bounce time of
$t_\mathrm{pb}= 1431$\,s when the SN shock has nearly 
reached $R_*$.
We then follow the subsequent evolution of the SN ejecta until
approximately 1.25 days after core bounce. This new explosion
simulation is termed W15-2-cw-IIb.  

By artificially removing most of the hydrogen we destroy the 
self-consistent, hydrostatic structure of the near-surface 
layers and the remaining hydrogen envelope becomes unrealistic 
in its radial profile. This is likely to corrupt the shock 
propagation through the hydrogen shell and the shock breakout
from the stellar surface. However, this manipulation has no
effect on the large-scale asymmetries imposed by the explosion
mechanism on the innermost SN ejecta such as the $^{56}$Ni and 
$^{44}$Ti yields. Moreover, our modified model is also good enough
to permit the demonstration of consequences associated with
the absence of the massive hydrogen envelope,
in particular for the maximum 
expansion velocities that can be retained by the mentioned
nucleosynthesis products. As shown by \citet{Wongwathanaratetal15},
the detailed composition-shell structure of the SN
progenitor, especially the density profile in the He- and
H-layers, has a strong influence on the propagation of the 
SN shock, on the development of secondary mixing instabilities
and reverse shocks at the composition-shell interfaces, and on 
the interaction of these instabilities and reverse shocks
with the initial ejecta asymmetries imposed 
by the central explosion mechanism itself. In the case of the 
lack of a massive hydrogen envelope, for example, the reverse
shock forming at the He/H interface after the passage of the
SN shock is absent and the corresponding deceleration of 
fast-moving inner ejecta, especially also of nickel and 
titanium, does not take place. We will discuss these effects
in Sect.~\ref{sec:veldistribution}.

Starting with a neutrino-driven explosion with highly aspherical 
mass ejection from the beginning, we are unable to quantitatively
assess a possible reduction of the strength of Rayleigh-Taylor (RT) 
instabilities
in material of intermediate mass numbers $Z$ due to the lack of
a strong reverse shock from the He/H interface. Such a reduction
is suggested by the works of 
\citet{Ellingeretal12,Ellingeretal13}, where a strong reverse
shock had a substantial influence on the RT growth at the C-O/He
composition interface. In contrast to our explosion modeling,
\citet{Ellingeretal12,Ellingeretal13} launched
their explosions in 1D and followed the propagation of a spherical 
blast wave from the C-O core into the He layer.
However, the initial asymmetries created by
the neutrino-driven mechanism in collaboration with hydrodynamic
instabilities during the very first second of the explosion
affect the iron and titanium nucleosynthesis and imprint 
large-scale asphericities on the innermost ejecta, the outgoing
SN shock, and the SN explosion
as a whole. They are therefore crucial in seeding
the secondary mixing instabilities that grow at the C-O/He
and He/H shell interfaces of the progenitor, and the presence
of pronounced initial explosion asymmetries causes the radial 
mixing to be much more efficient and to penetrate deeper (well
down into the region of Fe formation) than in the case of
explosions initiated in spherical symmetry \citep[see,
e.g.,][]{Kifonidisetal03,Kifonidisetal06,Wongwathanaratetal15}. 
Since the RT growth factor at the C-O/He-core interface in our 
model is huge after the passage of the SN shock 
\citep[see figure~5 in][]{Wongwathanaratetal15}, the development
of RT instability at this interface depends primarily
on the initial perturbations in the metal-rich
ejecta (including the layers containing $^{56}$Ni and $^{44}$Ti)
and not on the lack or presence of the reverse shock from the
He/H interface. As a consequence, we observe the same highly
deformed ejecta geometry on the largest scales for model
W15-2-cw-IIb and for the RSG explosion W15-2-cw, before in the 
latter case the reverse shock
from the He/H interface decelerates the core of intermediate-mass
elements and RT instability at the base of the hydrogen shell
triggers the fragmentation of the inner ejecta to
smaller-scale structures
\citep[compare figure~7, top row, of][with the results reported
in Sect.~\ref{sec:spacedistribution} below]{Wongwathanaratetal15}.

\subsection{Input physics}
\label{sec:inputphysics}

In our simulation of model W15-2-cw-IIb, 
gravity is taken into account in the hydrodynamics
equations. We use the spherical harmonics expansion technique of
\citet{MuellerSteinmetz95} to solve the Poisson equation with source
terms for the Newtonian self-gravity of the SN ejecta and the
gravitational contribution of a point mass that is placed at the 
coordinate origin and accounts for the central compact remnant and 
additional material that starts as ejecta gas but gets absorbed 
through the inner grid
boundary in course of the evolution by fallback and the successive
outward relocation of the inner grid boundary. 

The tabulated equation of state (EoS) of \citet{TimmesSwesty00} is
employed for describing the stellar plasma as a mix of 
arbitrarily degenerate and relativistic electrons and positrons, 
blackbody radiation, and ideal Boltzmann gases of a defined set of 
fully ionized nuclei, taking into account corrections for Coulomb 
effects. We consider 11 nuclear species, namely protons, nine
$\alpha$-nuclei from $^4$He to $^{56}$Ni (excluding $^{32}$S,
$^{36}$Ar, $^{48}$Cr, and $^{52}$Fe from the $\alpha$ chain),
and a ``tracer nucleus'' X that (schematically) represents 
iron-group species synthesized in neutrino-heated ejecta under
conditions of neutron excess. The advection of nuclear species with
plasma motions is treated with the consistent multifluid advection
scheme of \citet{PlewaMueller99}.

In the discussed simulation of model W15-2-cw-IIb nuclear burning 
plays no role, because
nuclear reactions become unimportant in the expanding SN ejecta
during the late stages and on the time scales considered. The 
simulation is therefore
devoted to studying the long-time evolution of the spatial and
velocity distribution of the nucleosynthesis products obtained
in the earlier phases, i.e., in models W15-2
and W15-2-cw, from which we adopt the initial data for the 3D
evolution of model W15-2-cw-IIb. Nuclear fusion processes were taken
into account in the computations of models W15-2 and W15-2-cw 
through a small $\alpha$-reactions network, connecting the 11
nuclear species mentioned above, details of which
were given by \citet{Wongwathanaratetal13,Wongwathanaratetal15}. 
The small network
is able to provide rough information about the assembling of
heavy nuclei in the SN outflows, but it is unable to give accurate 
results for the yields of individual nuclear species. The
production of $^{44}$Ti, for example, is massively overestimated
by the small network \citep[see Table~3 of][]{Wongwathanaratetal13}.
For a more accurate assessment of the nucleosythesis of $^{56}$Ni and 
$^{44}$Ti we therefore perform a post-processing analysis with a 
large network, employing tracer-particle information that we extract
from our 3D simulations of models W15-2, W15-2-cw, and W15-2-cw-IIb
as described in the next section.

\subsection{Post-processing of the nucleosynthesis}
\label{sec:nucleosynthesis}

The post-processing for nucleosynthetic yields is performed with
a nuclear reaction network of 6300 nuclear species 
between the proton-drip line and the neutron-drip line, reaching
up to the $Z$ = 110 isotopes, as described in \citet{Wanajo06}. 
The latest reaction library of REACLIB V2.0 \citep{Cyburtetal10} 
is employed, in which experimentally evaluated values are adopted 
whenever available. We also test a possible uncertainty of the most 
influential reaction for $^{44}$Ti production,
$^{44}$Ti($\alpha$, $p$)$^{47}$V, by 
dividing the forward and inverse rates in REACLIB V2.0 by a factor 
of two, according to the new experimental evaluation in 
\citet{Margerinetal14}. Using the hydrodynamical trajectories 
described below, nucleosynthesis calculations are started when the 
temperature decreases to $10^{10}$\,K, assuming initially free protons 
and neutrons with mass fractions $Y_e$ and $(1-Y_e)$, respectively. 
Nuclear statistical equilibrium is immediately established at such 
high temperature, and the detailed initial composition is 
unimportant. For the tracer-particle
trajectories with maximum temperatures below $10^{10}$\,K, the 
calculations are started from the beginning of the hydrodynamic
simulation (i.e., shortly after core bounce) with
the composition of the pre-SN model.

\begin{figure}
\centering
\resizebox{\hsize}{!}{\includegraphics{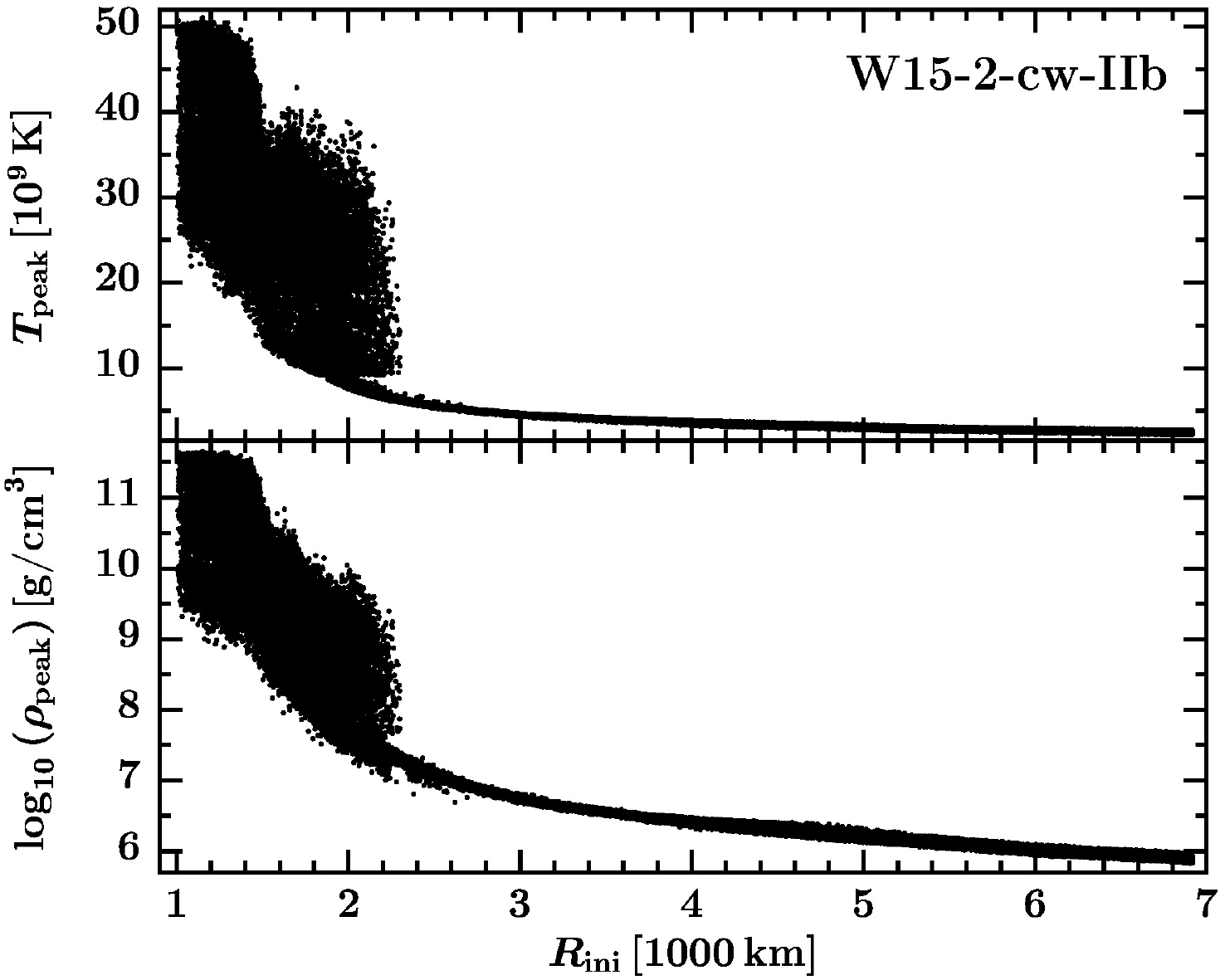}}
\caption{Peak temperature and peak density versus initial
(i.e., post-bounce) radius for all tracer particles that
contribute to the $^{44}$Ti and $^{56}$Ni production.}
\label{fig:trhoprofiles}
\end{figure}

A total number of 131,072 tracer particles is initially (i.e.,
some 15 milliseconds after core bounce) placed between $\sim$1000\,km
and $\sim$6900\,km, covering the entire Si-layer and the inner
shells of the O-layer, corresponding to a mass resolution of 
$3.49\times10^{-6}$\,M$_\odot$. This setting is suitable
to probe the whole volume where $^{56}$Ni (iron-group material)
and $^{44}$Ti are nucleosynthesized during the explosion
(the corresponding peak temperatures
and peak densities are shown in Fig.~\ref{fig:trhoprofiles}).
It is important to note that we do not co-evolve the tracer
particles with the hydrodynamic simulation but use the data 
files stored from the 3D hydrodynamics run to construct the 
tracer histories only after the simulation. This is achieved by a
4th order Runge-Kutta time integrator for the particle trajectories.
Linear interpolation of the velocity field is used to obtain
velocities at the spatial positions of the particles. The data 
of our simulation are stored every 100 hydrodynamics time steps.
These time intervals between hydrodynamics outputs are further 
divided into 10 substeps, and the velocity
field at the intermediate steps is
obtained by linear interpolation in time.

Despite the higher-order time integration, the resolution
limitations associated with the discrete time sampling by the
data outputs (which is much coarser than the small time steps
applied in the simulation) can lead to integration errors of
the particle evolution. This problem should be kept in mind 
and will be a matter of discussion in Sect.~\ref{sec:veldistribution}.

The majority of our initial particles, namely 107,910, are ejected, 
while 23,162 particles fall through the inner grid boundary and are
removed from further analysis. We discriminate between two types
of nucleosynthesis-relevant ejecta and the corresponding particles.
On the one hand, there are shock-heated ejecta, which are 
immediately accelerated outward by the expanding SN shock.
On the other hand, there are neutrino-processed ejecta, which
fall inward below $\sim$250\,km to be affected by electron
capture and/or neutrino interactions, which can change the
electron fraction, $Y_e$, and the entropy of the plasma.
While the shock-heated matter retains the $Y_e$ from the
pre-collapse conditions, the electron fraction of the 
neutrino-processed material is reset by the neutrino reactions
computed with our gray transport approximation. This
implies considerable uncertainties not only because our neutrino 
transport treatment is approximative, but also, more generally,
because the final neutron-to-proton ratio is very sensitive to 
many details of the neutrino physics, some of which \citep[such as 
collective neutrino-flavor oscillations and the lepton-emission
self-sustained asymmetry, LESA, recently found by 
\citealt{Tamborraetal14}; for a review, see 
e.g.,][]{Mirizzietal16} are not yet fully understood.

A subset of 75,862 of the expelled particles can be considered
as shock-heated ejecta, and another 27,222 particles represent
neutrino-processed ejecta. This corresponds to 
$\sim$0.265\,$M_\odot$ and $\sim$$9.51\times10^{-2}\,M_\odot$ 
for the
shock-heated and neutrino-processed ejecta, respectively.
(A small number of 4826 remaining
particles did not produce $^{44}$Ti and $^{56}$Ni at any
significant amounts and therefore is irrelevant for the 
investigations of this paper.)
Because, for the reasons mentioned above, $Y_e$ is uncertain
in the neutrino-processed ejecta, we will consider two cases 
for the nucleosynthetic post-processing of the corresponding
particles in order to test the sensitivity of the nucleosynthesis
to $Y_e$: 1) We adopt $Y_e$ as obtained in our 3D simulations
with the approximative neutrino treatment at a temperature of
$T\simeq 5\times10^{9}$\,K in outflowing material. The 3D
long-time simulation with the corresponding post-processing
results for nucleosynthesis will be denoted as 
W15-2-cw-IIb-Ye$_\mathrm{sim}$. 
2) Alternatively, we assume that all of the SN ejecta, also
the neutrino-heated ones, inherit their $Y_e$ values directly
from the progenitor, i.e., we take the $Y_e$ data from
the progenitor profile, where $Y_e\ge 0.4979$ for pre-collapse 
radii of $r\gtrsim$\,1160\,km and enclosed masses larger
than 1.287\,$M_\odot$, 
while the more neutron-rich deeper layers are accreted into
the newly formed neutron star (NS). (The final mass cut is close
to 1.35\,$M_\odot$.)

Typical entropies in the neutrino-heated ejecta at
$T = 5$\,GK range from $\sim$10\,$k_\mathrm{B}$ per nucleon to
$\sim$30\,$k_\mathrm{B}$ per nucleon, while the shock-heated
material has entropies between $\sim$7\,$k_\mathrm{B}$ and
$\sim$12\,$k_\mathrm{B}$ per nucleon during the relevant time of 
$^{44}$Ti and $^{56}$Ni nucleosynthesis.

As we will see in Sect.~\ref{sec:production}, the nucleosynthetic
yields of $^{56}$Ni and, in particular, of $^{44}$Ti differ
considerably between the two investigated cases. Because of a better 
match with the Cas~A remnant, we will mainly focus
on the second case as our preferred and more interesting model,
which we term W15-2-cw-IIb.

We have tested the convergence of our post-processing analysis
by comparing the yields for our full set of tracer particles
with yields estimated on grounds of randomly selected subsets of
these particles. We found that typically less than 10\% of all
particles are sufficient to obtain the produced $^{44}$Ti and
$^{56}$Ni masses with accuracies of better than a few percent. 
Of course,
good representations of the $^{44}$Ti and $^{56}$Ni distributions
in 3D space require the use of significantly more than a few 
1000 tracer particles.

\begin{figure*}
\centering
\resizebox{0.49\hsize}{!}{\includegraphics{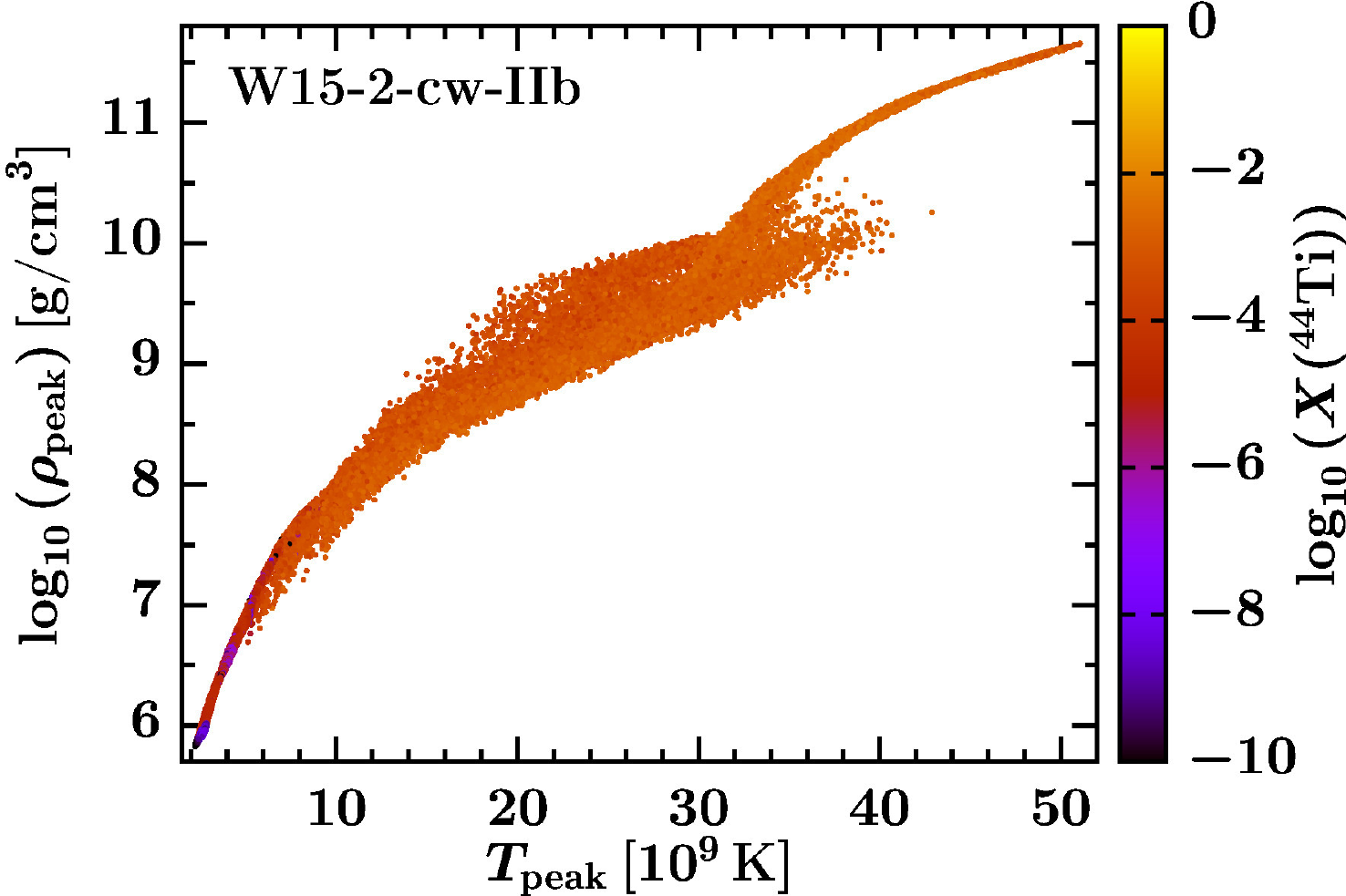}}
\resizebox{0.49\hsize}{!}{\includegraphics{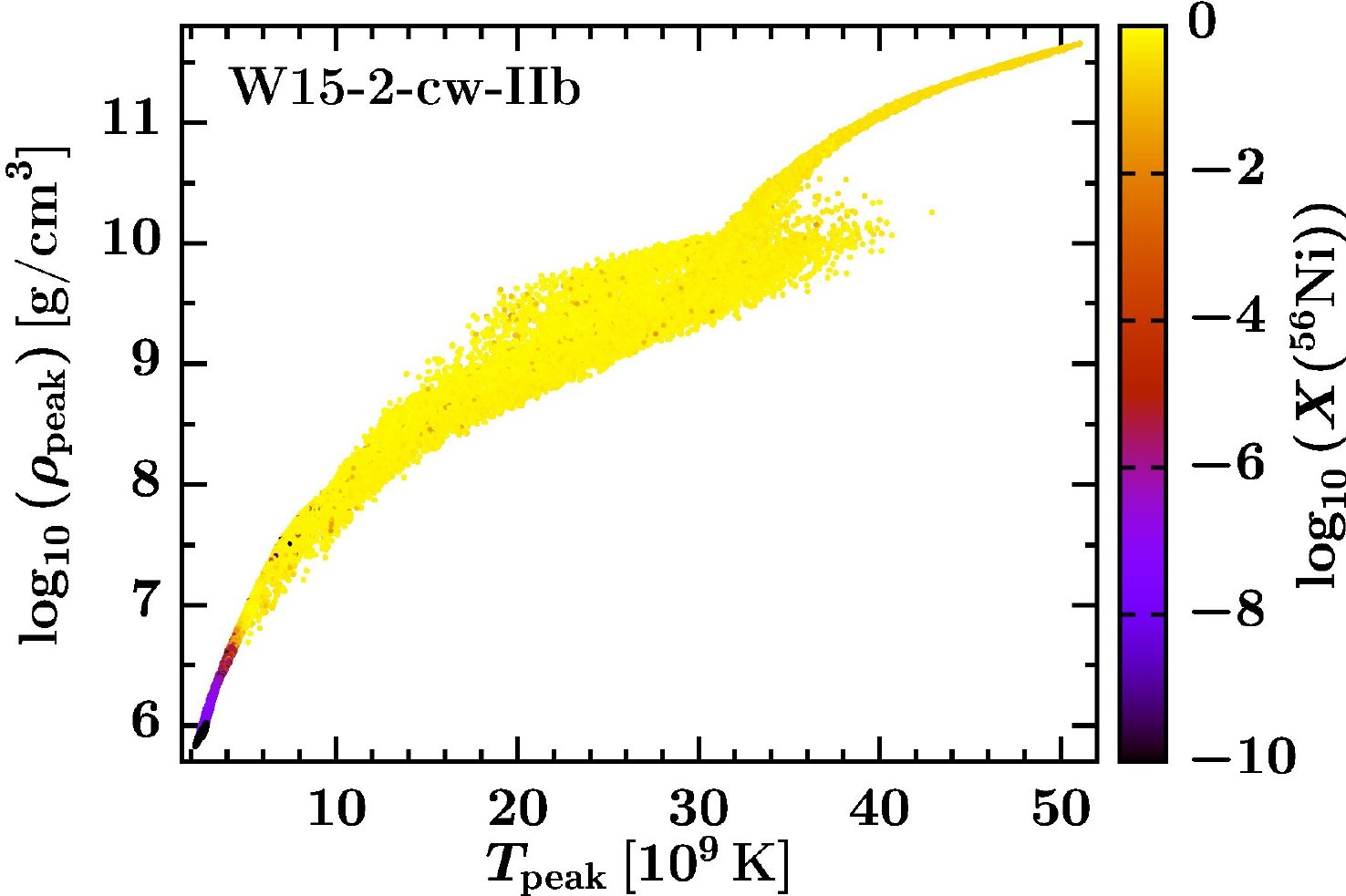}}
\caption{Mass fractions of nucleosynthesized $^{44}$Ti 
({\em left}) and $^{56}$Ni ({\em right}), color coded for all
tracer particles of model W15-2-cw-IIb in the peak
temperature-peak density plane.}
\label{fig:trhobands}
\end{figure*}


\begin{table}
\caption{Yields from nucleosynthetic post-processing of
tracer particles with $Y_e = 0.5$ (upper three lines) and 
$Y_e$ from the simulation with approximate neutrino 
transport (lower three lines), using our standard REACLIB 
V2.0 nuclear rates.}
\centering
\begin{tabular}{ccc}
\hline
\hline
Model & M($^{44}$Ti) [M$_\odot$] & M($^{56}$Ni) [M$_\odot$] \\
\hline
W15-2-cw-IIb            & $1.57\times10^{-4}$ & $9.57\times10^{-2}$ \\
W15-2-cw-IIb-shock      & $8.66\times10^{-6}$ & $4.20\times10^{-2}$ \\
W15-2-cw-IIb-$\nu$proc & $1.49\times10^{-4}$ & $5.38\times10^{-2}$ \\
\hline 
W15-2-cw-IIb-Ye$_\trm{sim}$        & $1.58\times10^{-5}$ & $4.29\times10^{-2}$ \\
W15-2-cw-IIb-shock      & $8.66\times10^{-6}$ & $4.20\times10^{-2}$ \\
W15-2-cw-IIb-$\nu$proc-Ye$_\trm{sim}$ & $7.16\times10^{-6}$ & $0.10\times10^{-2}$ \\
\hline
\end{tabular}
\label{tab:yield}
\end{table}

\section{Results}
\label{sec:results}

Radioactive nuclei such as $^{44}$Ti and $^{56}$Ni, which are 
assembled in the innermost regions
of SN explosions, are of great relevance for the
diagnostics of the explosion mechanism. Comparisons of 3D model
predictions with the NuSTAR map of the spatial $^{44}$Ti 
distribution are particularly interesting, because they
allow one to identify morphological peculiarities that 
are connected to asymmetries of the blast wave during the
earliest moments of its outward propagation.
The $^{44}$Ti distribution can therefore 
provide extremely valuable information on the working of the 
central engine that initiates and powers the SN explosion.
It is also important to investigate whether our model could, at 
least in principle, produce a $^{44}$Ti yield 
compatible with the large mass of this nucleus 
inferred from NuSTAR and INTEGRAL in the case of Cas~A.

Once again we emphasize that our 3D explosion model was
selected as one case from the model set of 
\citet{Wongwathanaratetal13,Wongwathanaratetal15} but was not
fine tuned to match the progenitor and explosion properties
of Cas~A. It is also important to repeat that the explosion 
geometry did not emerge from predefined or artificially
imposed asymmetries but developed
purely stochastically by the growth of hydrodynamic instabilities
from small, random initial perturbations in the neutrino-heating 
layer (see Sect.~\ref{sec:inimod}).

\subsection{Production of $^{44}$Ti and $^{56}$Ni}
\label{sec:production}

Figure~\ref{fig:trhoprofiles} shows the peak temperature 
and peak density versus ``initial'' (post-bounce) radius,
$R_\mathrm{ini}$, of all particles 
that contribute to the production of $^{44}$Ti and $^{56}$Ni.
Shock-heated and neutrino-processed ejecta can be well
distinguished. The shock-heated particles reach
$T_\mathrm{peak}\lesssim 9$\,GK and $\rho_\mathrm{peak}\lesssim 
10^{7.6}$\,g\,cm$^{-3}$ and align in rather narrow stripes of  
temperature and density values as functions of initial radius,
because the shock front is only moderately aspherical.
The neutrino-processed ejecta form clusters of higher 
$T_\mathrm{peak}$ and $\rho_\mathrm{peak}$ at post-bounce
radii of less than $\sim$2200\,km. While most of these particles
are neutrino-heated to temperatures up to about 40\,GK in 
the convective flows within the postshock layer (peak densities 
$\rho_\mathrm{peak}\lesssim 2\times 10^{10}$\,g\,cm$^{-3}$),
some of this matter (starting from smaller $R_\mathrm{ini}
\lesssim 1500$\,km and reaching to even more extreme peak values
of density and temperature) is accreted onto the nascent neutron
star before being blown out again in the neutrino-driven wind.

The electron fraction in the neutrino-processed ejecta,
which is set by the competition of $\nu_e$ and $\bar\nu_e$ 
absorption and emission processes, 
sensitively determines the efficiency of producing $^{44}$Ti
and $^{56}$Ni. For this reason the yields of both nuclei
differ strongly between the two considered cases, namely
model W15-2-cw-IIb, where, for a sensitivity test,
we adopt the electron fraction, $Y_e$,
from the progenitor conditions, compared to model
W15-2-cw-IIb-Ye$_\mathrm{sim}$, where we
take $Y_e$ from our 3D simulations with approximative neutrino
transport. In the latter case, $Y_e$ is shifted slightly to
the neutron-rich side (values at $T\simeq 5\times10^{9}$\,K
lie in the range between 0.456 and 0.498),
which strongly suppresses the creation of both $^{44}$Ti
and $^{56}$Ni as discussed by \citet{Magkotsiosetal10},
where the theoretical framework of the production of these
radioactive isotopes was depicted in great detail. As the
$^{44}$Ti yield becomes maximal at $Y_e\approx0.5$ with other
conditions fixed \citep{Wanajoetal13}, our two investigated
cases roughly bracket the upper and lower limits of the 
$^{44}$Ti production in our model.

Table~\ref{tab:yield} lists the total yields of models
W15-2-cw-IIb and W15-2-cw-IIb-Ye$_\mathrm{sim}$ as well as
the individual contributions from shock-heated and 
neutrino-processed material in both cases, all obtained by the
nucleosynthetic post-processing of our tracer particles. In model
W15-2-cw-IIb-Ye$_\mathrm{sim}$ the shock-heated ejecta component
(indicated by the model-name extension ``shock'')  
dominates the expelled mass of $^{56}$Ni by far, whereas both 
ejecta components contribute roughly equally to the making
of $^{44}$Ti. In model
W15-2-cw-IIb roughly 95\% of the $^{44}$Ti and about 56\% of
the $^{56}$Ni are assembled in the neutrino-processed ejecta
(model-name extension ``$\nu$proc''),
where $Y_e$ is essentially 0.5. Effectively, this means 
that model W15-2-cw-IIb produces more than twice as much $^{56}$Ni
as W15-2-cw-IIb-Ye$_\mathrm{sim}$ ($\sim$0.096\,$M_\odot$ compared
to $\sim$0.043\,$M_\odot$) and 10 times more $^{44}$Ti
($\sim$$1.6\times 10^{-4}$\,$M_\odot$ compared to 
$\sim$$1.6\times 10^{-5}$\,$M_\odot$).

We note in passing that the calculations with the
small $\alpha$-network that we employ during the hydrodynamic
simulations, massively overestimate the yield of $^{44}$Ti,
whereas they produce $^{56}$Ni in the same ballpark as the
tracer-particle based post-processing with the large nuclear 
network reported above. Table~3 of \citet{Wongwathanaratetal13}
lists nearly $3\times 10^{-3}\,M_\odot$ of $^{44}$Ti and 
0.055--0.139\,$M_\odot$ of $^{56}$Ni, depending on the 
undetermined contribution of this nucleus in the 
neutrino-processed ejecta, for model W15-2.

By far most of the $^{44}$Ti and $^{56}$Ni are assembled under
$\alpha$-rich freeze-out conditions, which are experienced
by most of the relevant shock-heated and neutrino-heated
ejecta, and where the mass fractions of these
nuclei reach their highest values. Only subdominant contributions
originate from regions of incomplete Si burning (in line with
conclusions drawn by \citet{WoosleyHoffman91} and 
\citet{Magkotsiosetal10}. This can be 
concluded from the plots of Fig.~\ref{fig:trhobands}, which show
color coded mass fractions of $^{44}$Ti (left) and $^{56}$Ni (right)
for all tracer particles of model W15-2-cw-IIb in the 
peak temperature-peak density
plane. The two images should be viewed in relation to figure~4
of \citet{Magkotsiosetal10}. All of the expelled matter
with $T_\mathrm{peak}\gtrsim 5\times 10^9$\,K crosses the 
$\alpha$-particle-rich
freeze-out region when it expands and cools during ejection,
and only a small amount of $^{44}$Ti and $^{56}$Ni is produced
by incomplete Si-burning in matter with $T_\mathrm{peak}<5\times 10^9$\,K.
The banana-shaped cluster of particles between peak densities of 
about $10^8$\,g\,cm$^{-3}$ and $2\times 10^{10}$\,g\,cm$^{-3}$ 
corresponds to neutrino-heated matter in the convective layer between
gain radius and SN shock, while the antenna-like narrow extension
toward higher densities and temperatures is matter that has
transiently been accreted into the outer layers of the proto-neutron 
star before it was blown out in the neutrino-driven wind later.
The mass ratio of $^{44}$Ca/$^{56}$Fe as stable daughter
nuclei of $^{44}$Ti and $^{56}$Ni in our model W15-2-cw-IIb
is $1.64\times 10^{-3}$ and thus compatible with the solar
ratio of these elements \citep[$=1.26\times10^{-3};$][]{Lodders03},
while our less favorable case, model 
W15-2-cw-IIb-Ye$_\mathrm{sim}$, accounts for a production
of only about 1/3 of the solar $^{44}$Ca/$^{56}$Fe value.

All tracer particles that contribute to the creation of 
significant amounts of $^{44}$Ti and $^{56}$Ni are ejected 
(i.e., accelerated outwards)
during the first second of the explosion. 
In our 3D neutrino-driven explosion model the far dominant part
of the ejected $^{44}$Ti originates from the neutrino-heated,
high-entropy
postshock layer, whose mass depends on the multi-dimensional 
processes that drive nonradial flows in this region. More plasma
is channeled through the region of neutrino-matter interactions
by such flows than typically experiencing neutrino-heating
in corresponding 1D explosion models. Moreover,
the nucleosynthesis of $^{44}$Ti is extremely sensitive 
to the neutron-to-proton ratio and entropy in the 
$\nu$-processed matter.
Spherically symmetric models of neutrino-powered explosions
are certainly not able to capture the dynamics of this layer
reliably. For example, in our attempts to perform quantitative 
comparisons of 1D and 3D simulations by varying the parameters
that determine the strength of the explosion-driving
neutrino source, we found that it is hardly
possible to reproduce the shock acceleration and the growth
of the explosion energy of our 3D model simultaneously by 
1D explosions. The yield of
$^{44}$Ti hinges critically on all characteristic properties
(i.e., mass, $Y_e$, entropy, and expansion velocity) of the
ejecta that emerge from the region where the mass cut between
compact remnant and SN outflow develops. For this reason 1D models 
are not qualified to adequately address the problem of $^{44}$Ti
production in SNe. This, in particular, also applies to 
SN calculations invoking pistons or thermal bombs for
initiating the explosions. Both methods are unable to accurately 
describe the dynamics, thermodynamics, and electron fraction of 
matter expelled from the vicinity of the mass cut.

\begin{figure}
\centering
\resizebox{\hsize}{!}{\includegraphics{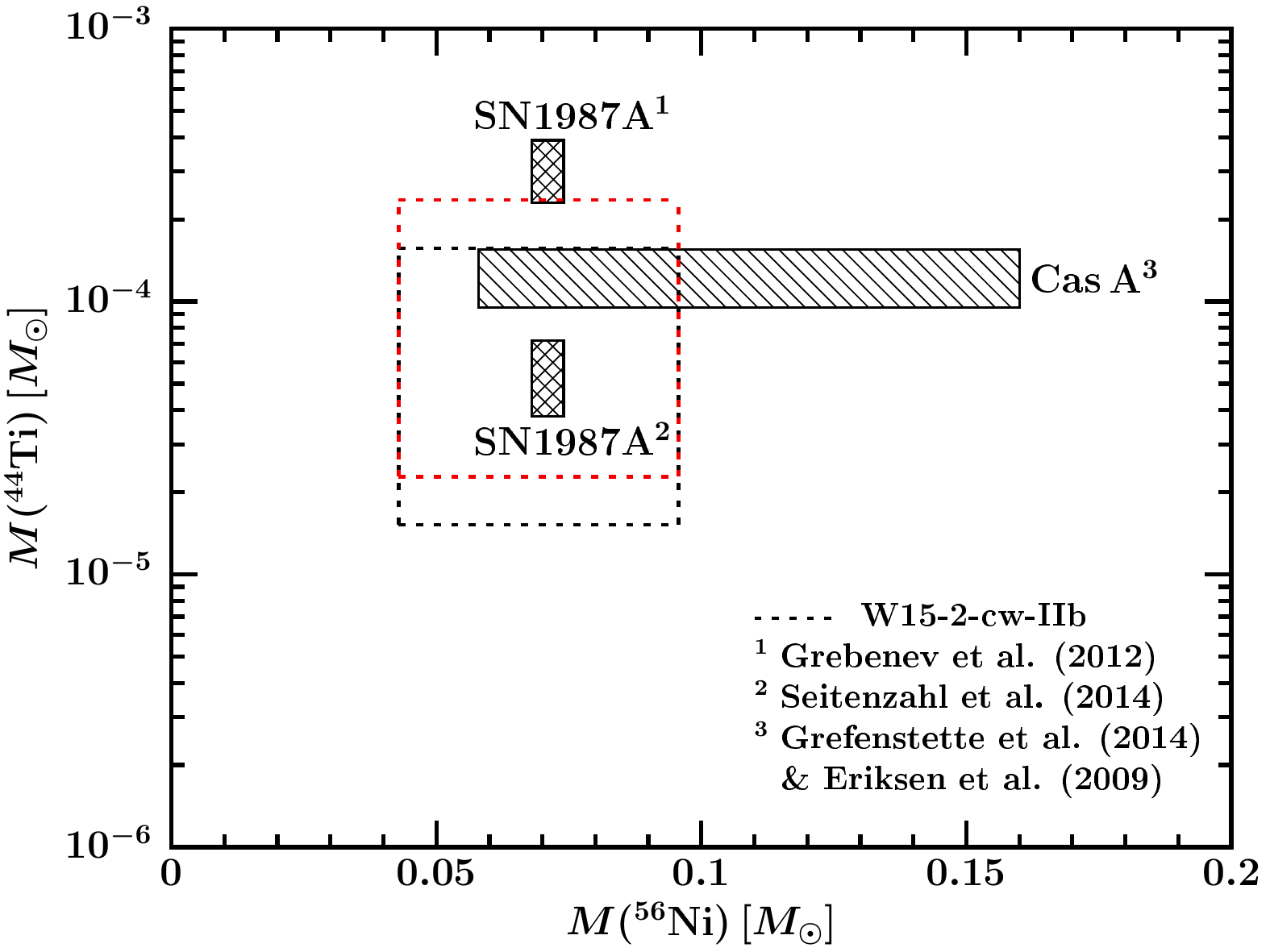}}
\caption{Comparison of observationally determined yields of 
$^{56}$Ni and $^{44}$Ti for SN~1987A and Cas~A with those obtained
by our post-processing nucleosynthesis calculations.
The hatched boxes correspond to different recent empirical approaches
\citep[data were taken from][as referenced in the Figure]
{Eriksenetal09,Grebenevetal12,Grefenstetteetal14,Seitenzahletal14}
and represent the current band width of values.
The dashed black box corresponds to the maximal masses produced
by model W15-2-cw-IIb and minimal masses produced by model
W15-2-cw-IIb-Ye$_\mathrm{sim}$ as listed in Table~\ref{tab:yield}.
The red dashed box indicates the increase of the $^{44}$Ti
production by a factor of 1.5 when the 
$^{44}$Ti$(\alpha,p)^{47}$V reaction rate is reduced by
a factor of two \citep{Margerinetal14}
instead of using our standard rate from \citet{Cyburtetal10}.}
\label{fig:mtini}
\end{figure}

It is therefore not astonishing that 3D neutrino-driven explosions
can potentially produce much more $^{44}$Ti than prediced by 
piston-initiated explosions \citep[as already surmised by][]{ 
WoosleyHoffman91},
where this isotope is only made in the shock-heated ejecta, 
which contribute to our $^{44}$Ti yield only on the 10\% level.
Our results also demonstrate that large global asymmetries,
e.g.\ by rapid rotation or even jet-driven 
(magnetohydrodynamic) explosions \citep{Nagatakietal98}, are not 
needed to explain the $^{44}$Ti masses observed in Cas~A and
SN~1987A.
The yield of our model W15-3D-cw-IIb is compatible with both
observations (see Fig.~\ref{fig:mtini}),\footnote{Results for 
observational estimates of the
ejected $^{44}$Ti masses in the same ballpark as those plotted
in Fig.~\ref{fig:mtini} were reported for
SN~1987A by \citet{Jerkstrandetal11}
($M(^{44}\mathrm{Ti}) = 1.5^{+0.5}_{-0.5}\times 10^{-4}\,M_\odot$)
and \citet{Boggsetal15}
($M(^{44}\mathrm{Ti}) = 1.5^{+0.3}_{-0.3}\times 10^{-4}\,M_\odot$)
and for Cas~A by \citet{Siegertetal15}
($M(^{44}\mathrm{Ti})= 1.37^{+0.19}_{-0.19}\times 10^{-4}\,M_\odot$),
\citet{WangLi16}
($M(^{44}\mathrm{Ti})= 1.3^{+0.4}_{-0.4}\times 10^{-4}\,M_\odot$),
and
\citet{Grefenstetteetal17}
($M(^{44}\mathrm{Ti})= 1.54^{+0.21}_{-0.21}\times 10^{-4}\,M_\odot$).}
in particular if we account for
uncertainties in the important nuclear reaction rate of the 
$^{44}$Ti$(\alpha,p)^{47}$V process. Reducing this rate by
a factor of two according to the new experimental value of
\citet{Margerinetal14} boosts the
$^{44}$Ti production by a factor of 1.5 compared to the 
values listed in Table~\ref{tab:yield}. We did not attempt to 
regulate the explosion energy of our 3D model for a precise
match of the observations. Its value of
nearly 1.5\,B 
is close to the one estimated for SN~1987A \citep[around
1.5\,B;][]{Utrobin05} and only slightly smaller
than that of Cas~A \citep[roughly 2.3\,B;]
[and references therein]{LamingHwang03,Orlandoetal16}.
Despite this energy difference both SNe seem to have produced 
$^{44}$Ti yields in the same ballpark (Fig.~\ref{fig:mtini}),
although the SN~1987A value is not determined very accurately
and there is considerable tension between the results
obtained from X-ray emission-line measurements
\citep{Grebenevetal12,Boggsetal15}, late-time spectral analysis 
\citep{Jerkstrandetal11}, and late-time light-curve fitting
\citep{Seitenzahletal14}.

Neutrino-driven explosions genericly eject large
amounts of matter with sufficiently high entropies
(between $\sim$10\,$k_\mathrm{B}$ and $\sim$30\,$k_\mathrm{B}$
per nucleon) to enable an effective production
of $^{44}$Ti, and therefore they can potentially produce 
considerable yields of this nucleus. The reason is that the explosion
energy is carried by matter that must be heated to the point
of marginal gravitational binding in the strong field of the
new-born NS before outward acceleration of the
SN shock and of the postshock plasma can set in 
\citep[for a detailed discussion, see, e.g.][]{Janka2001,Schecketal06}.
The net energy available to the explosion is then basically
provided by the recombination energy of initially free
nucleons in the neutrino-heated gas to $\alpha$-particles
and heavy nuclei. This releases at most $\sim$9\,MeV per nucleon,
of which under typical conditions 5--7\,MeV per nucleon or
$\sim$(5--7)$\times 10^{18}$\,erg\,g$^{-1}$ are 
available to power the SN blast. For the explosion energy
one therefore estimates, very roughly,
\begin{equation}
E_\mathrm{exp} \sim (1.0\,...\,1.4)\times 10^{51}\,
\left(\frac{\Delta M_{\nu-\mathrm{heat}}}{0.1\,M_\odot}\right)\,\,
\mathrm{erg}\,,
\label{eq:expleng}
\end{equation}
where $\Delta M_{\nu-\mathrm{heat}}$ is the mass of neutrino-heated
material, and the negative gravitational binding energy of the 
overlying stellar layers exterior to the SN shock as well as
positive energy contributions by explosive nuclear burning of 
shock-heated ejecta are not taken into account.
This expression means that for a net explosion energy of 
1\,$\mathrm{B} = 10^{51}$\,erg, at least $\sim$0.1\,$M_\odot$ of
neutrino-heated matter must be expelled. 

Our results for models W15-2-cw-IIb and 
W15-2-2cw-IIb-Ye$_\mathrm{sim}$, however,
demonstrate that the actual mass of
$^{44}$Ti produced in the explosion can vary strongly, depending
on the exact value of $Y_e$ in the neutrino-processed ejecta.
A reliable determination of the neutron excess
requires a more sophisticated treatment of the neutrino transport
than applied in our 3D simulations and, moreover, is sensitive
to aspects that are uncertain or hard to control 
with high accuracy. Therefore, since predictions of $Y_e$ with
a precision of a few percent are extremely difficult, we do not
want to put overly much weight on the elemental yields
but, instead, focus on the distributions of $^{44}$Ti and $^{56}$Ni
as functions of enclosed mass, radial velocity and spatial 
coordinates in the following.

\begin{figure}
\centering
\resizebox{\hsize}{!}{\includegraphics{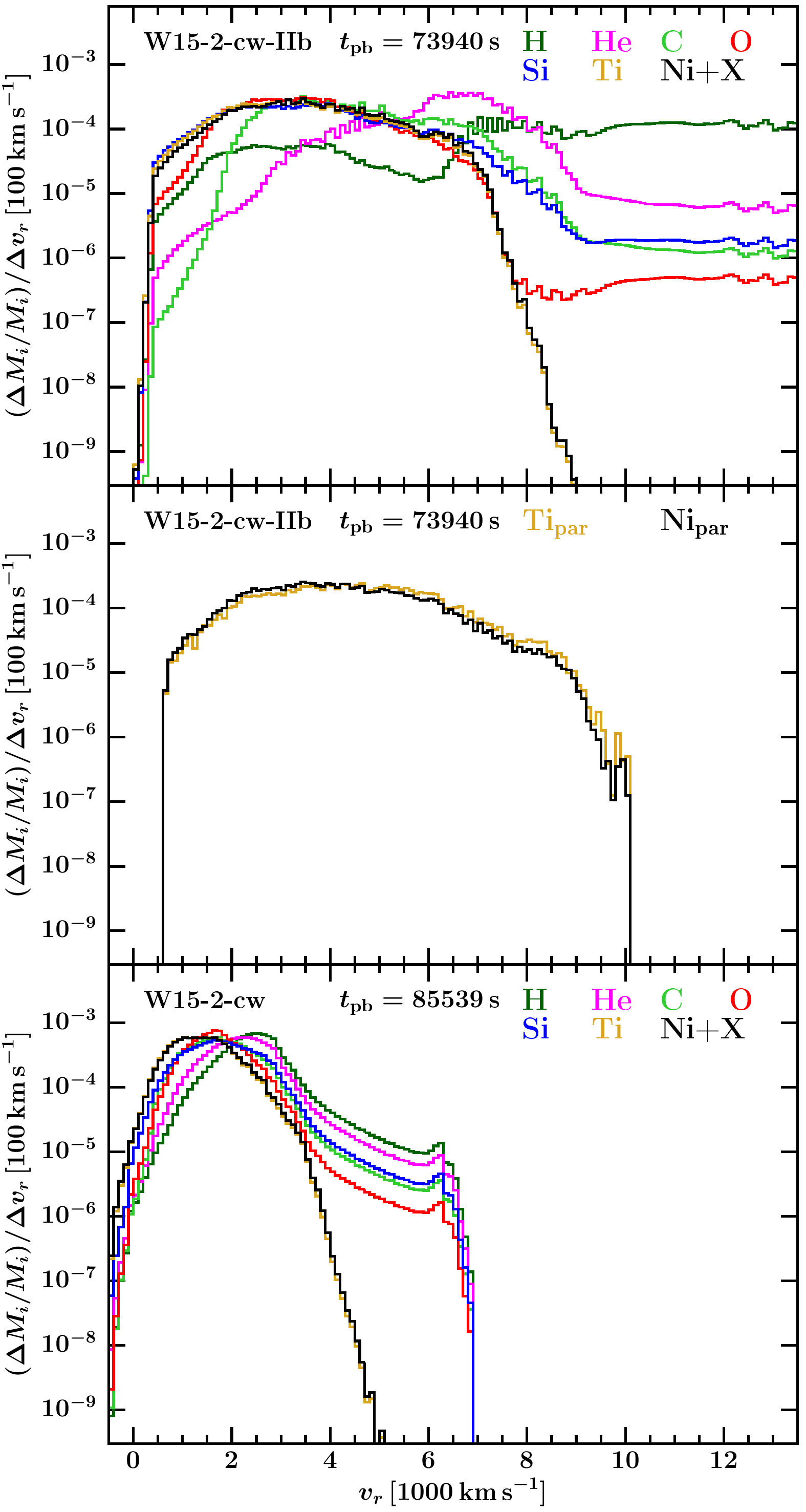}}
\caption{Mass distributions versus radial velocity for a set of
nuclear species. The distribution of each element is normalized 
by the total ejecta mass of this element.
The {\em top panel} displays results for model W15-2-cw-IIb at 
post-bounce time $t_\trm{pb}=73940$\,s as computed with a
small $\alpha$-network used during the hydrodynamic simulation.
The {\em middle panel} shows the corresponding distributions of 
$^{44}$Ti and $^{56}$Ni obtained from post-processing tracer-particle
histories with a large nuclear-burning network. For comparison,
the {\em bottom panel} presents the small-network results for the
red supergiant model W15-2-cw of \citet{Wongwathanaratetal15}
at the time of shock breakout ($t_\trm{pb}=85539$\,s). Note that
in this case the strong reverse shock from the shock passage 
through the He/H interface leads to the deceleration of the 
metal core and to fallback of roughly
$10^{-2}\,M_\odot$ of matter, which can be seen with negative 
velocities in the {\em bottom panel}.} 
\label{fig:mvsvr}
\end{figure}

\begin{figure*}
\centering
\centering
\resizebox{0.49\hsize}{!}{\includegraphics{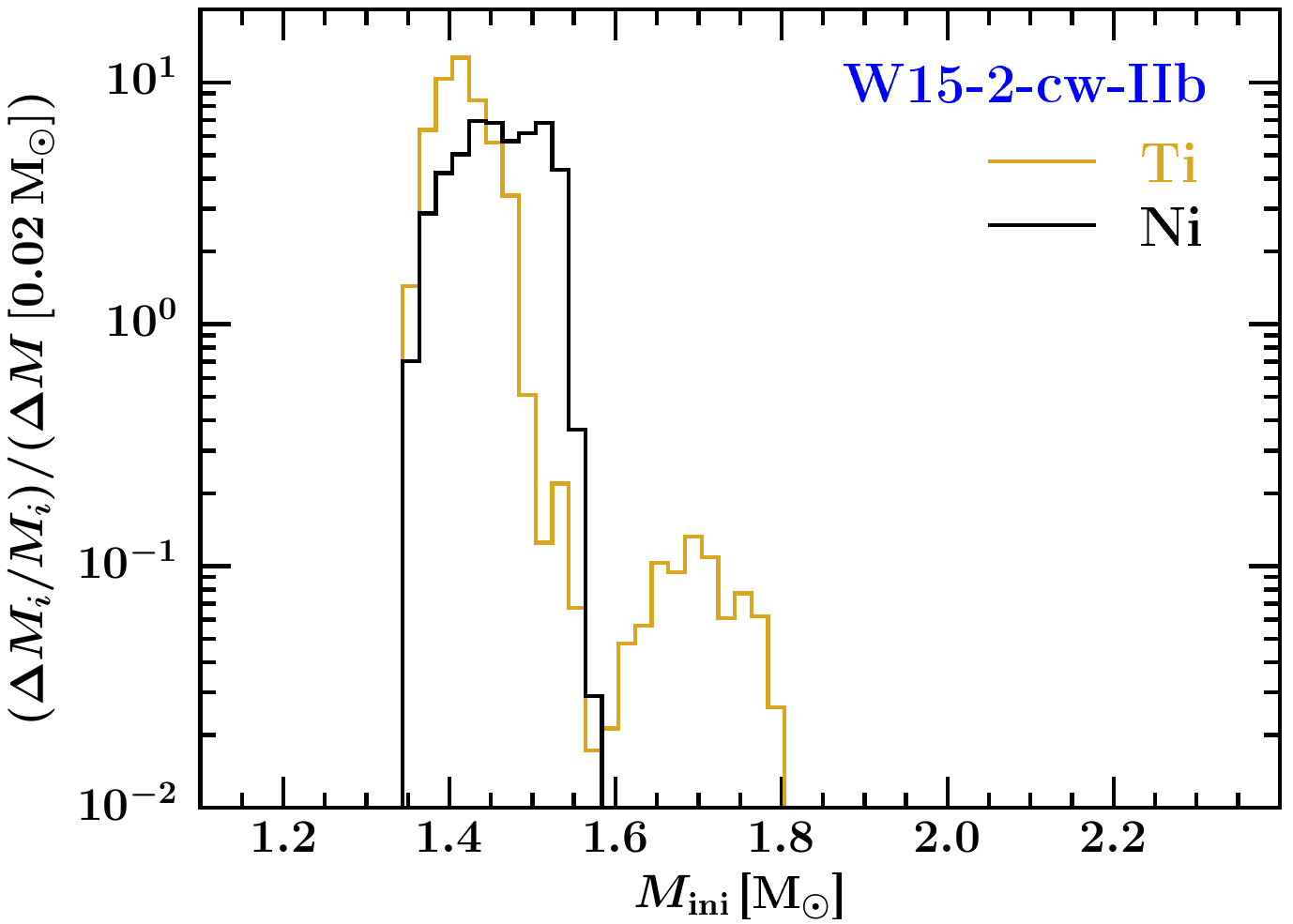}}
\resizebox{0.49\hsize}{!}{\includegraphics{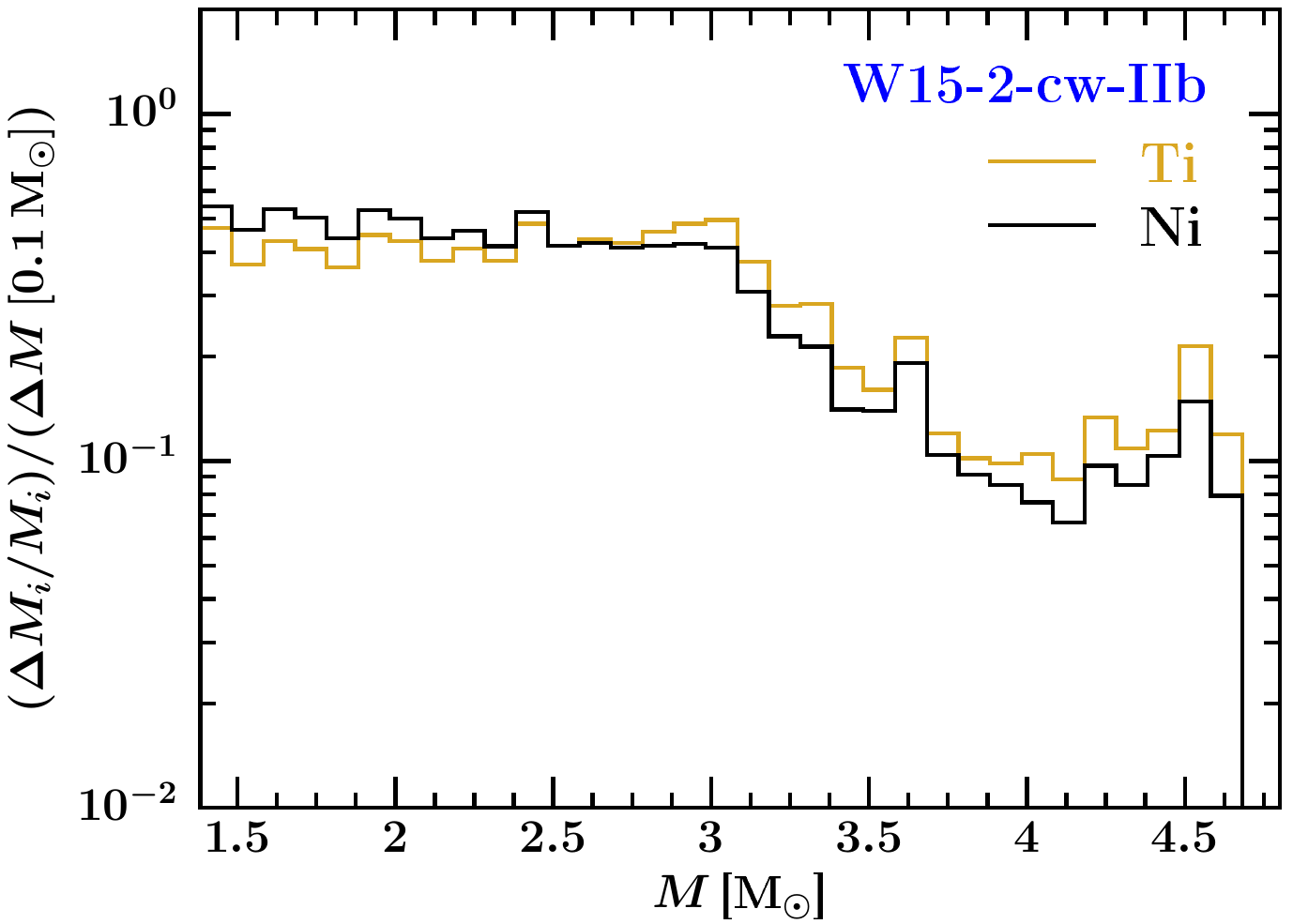}}
\caption{{\em Left:}
Mass distributions of $^{44}$Ti (gold) and
$^{56}$Ni (black) as functions of enclosed mass as
given by the progenitor structure at the beginning of our simulation.
The $x$-coordinate therefore shows the initial locations of matter
that ends up as titanium and nickel when explosive burning is over.
The distributions are obtained from our
post-processing nucleosynthesis with a large nucleosynthesis
network, and the two distributions are normalized individually
with the total yields of the radioactive species.
{\em Right:} Similar to the left panel, but for model W15-2-cw-IIb 
at $t_\trm{pb}=73940$\,s, with the mass coordinate being 
defined by the stellar mass contained in a spherical
volume of given radius.}
\label{fig:mvsm}
\end{figure*}

\begin{figure*}
\centering
\centering
\resizebox{0.49\hsize}{!}{\includegraphics{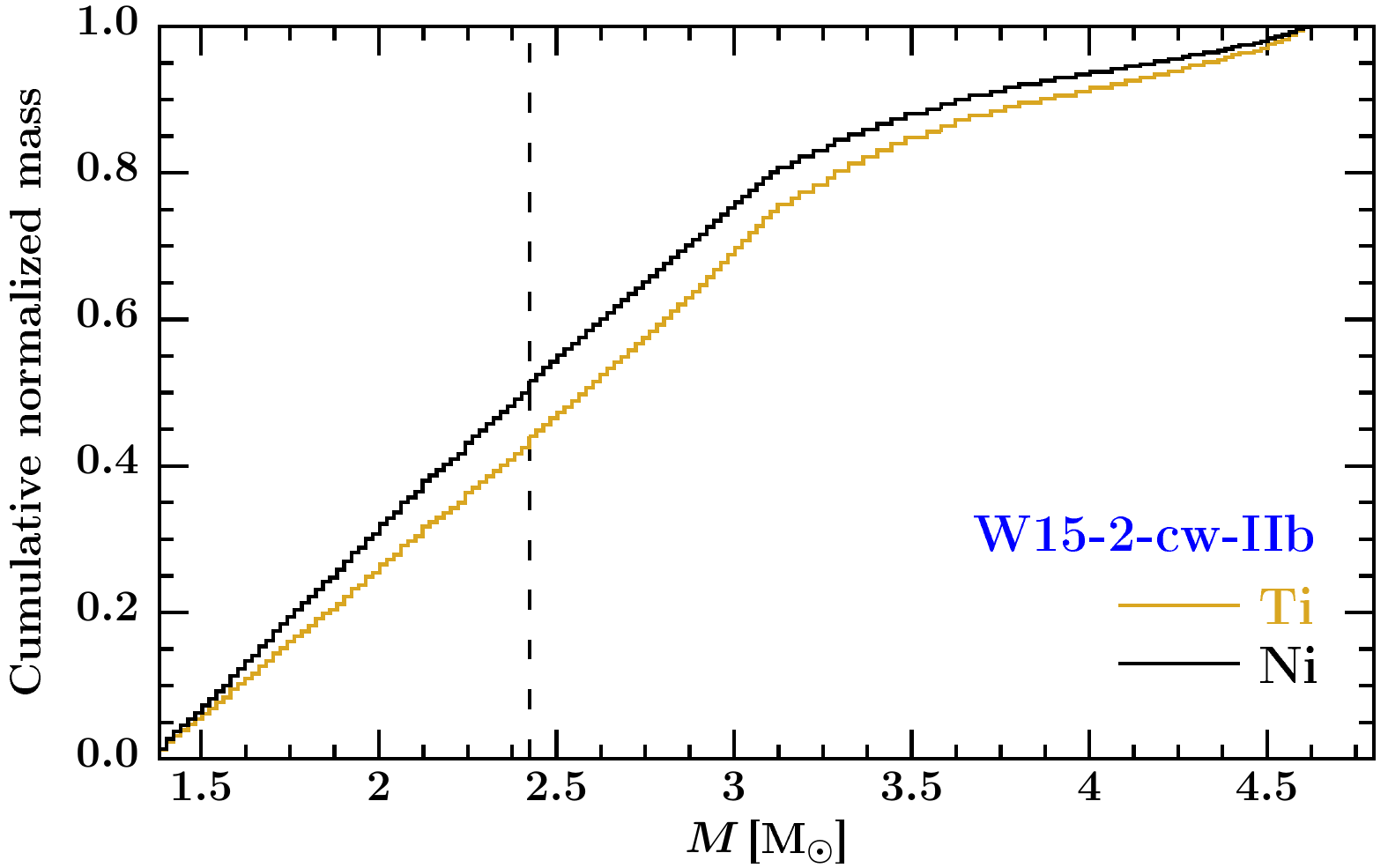}}
\resizebox{0.49\hsize}{!}{\includegraphics{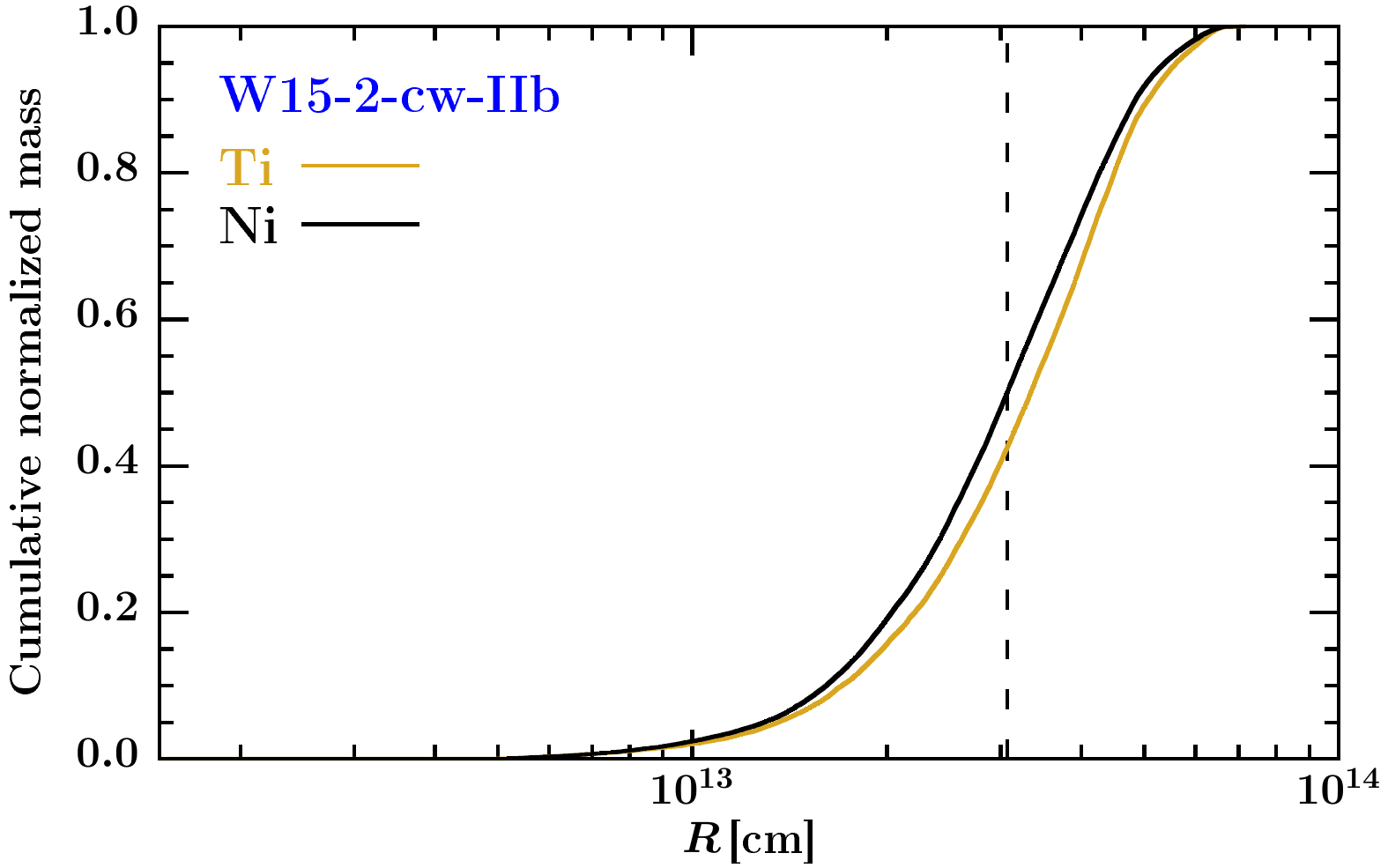}}
\caption{{\em Left:}
Cumulative mass distributions of $^{44}$Ti (gold) and
$^{56}$Ni (black) as functions of enclosed mass {\em (left)}
and radius {\em (right)} for model W15-2-cw-IIb
at $t_\trm{pb}=73940$\,s. The vertical dashed line marks
the location that encloses 50\% of the total $^{56}$Ni mass
and about 43\% of the total $^{44}$Ti mass (with radial
velocities below roughly 3500--4000\,km\,s$^{-1}$; note
that due to the nearly homologous expansion the velocity
scales essentially linearly with radius). 
It corresponds to the position where the reverse shock is
assumed in the right panel of Fig.~\ref{fig:combmap}.
About 57\% of the (post-processed) $^{44}$Ti are in the 
shell outside of this position in our model.}
\label{fig:cumul}
\end{figure*}

\newpage
\subsection{Distribution in radial velocity space}
\label{sec:veldistribution}

Hydrodynamic instabilities during the neutrino-heating
phase determine the overall asphericity of the explosion and 
the directional asymmetries of the nucleosynthetic ejecta.
The initial structures begin to freeze out at the time
when the outward acceleration of the SN shock marks the onset
of the explosion. The final morphology of the SN blast and the
velocity distributions of the chemical elements, however,
depend strongly not only on this initial explosion asymmetry 
but also on the structure of the progenitor star. The latter
has an important influence because the outgoing shock accelerates
or decelerates when travelling through the different
composition shells, depending on whether the density profile
is steeper or flatter than $r^{-3}$. Deceleration after 
the shock has passed a composition-shell interface leads to 
compression of the postshock matter into a dense shell and the
formation of a reverse shock that moves backward into the SN 
ejecta and slows down the expanding flow. 
Moreover, in regions of crossing pressure
and density gradients the growth of RT and
Richtmyer-Meshkov instabilities is seeded by the shock 
asphericity, preexisting perturbations in the progenitor,
and, most importantly, by the primary explosion asymmetries 
in the innermost SN ejecta.
The secondary instabilities thus enhance the mixing between
layers of different chemical composition and cause a 
fragmentation of the initially large, metal-rich plumes and 
clumps to smaller, more filamentary structures as discussed in
detail for 3D SN simulations of different red and
blue supergiant progenitors including W15-2 by
\citet{Wongwathanaratetal15}.

The size of such effects, in particular
the efficiency of the metal penetration into the 
hydrogen envelope and the corresponding fragmentation 
of the mixed layer, depend on the shock velocity after the shock
has crossed the C-O core, the shock deceleration in
the He-layer, and the density gradient at the He/H interface.
In red supergiants the shock accelerates strongly at a
steep density decline between the He- and H-shells and
experiences a similarly dramatic subsequent deceleration
when travelling through the extended hydrogen envelope. 
This leads to large growth rates of the RT 
instability at the shell interface and therefore efficient
metal and helium mixing into the hydrogen envelope. 
In blue supergiants the density gradient at the He/H
boundary is less pronounced, the shock acceleration at
this location is therefore far more moderate, and the shock 
deceleration in the more compact H-envelope less extreme.
Because the RT growth rates at the
He/H transition are smaller, the intrusion of metals
into the hydrogen envelope is impeded, unless the fastest
metal plumes stay in close touch with the outgoing
shock and thus can enter the hydrogen envelope before they
experience significant deceleration by interaction with
the reverse shock and with a dense ``helium wall'' that 
builds up from decelerated helium. If such a situation
holds, extended metal-enriched fingers can reach with high 
velocities deep into the surrounding hydrogen also in
the case of blue supergiants \citep[see][]{Wongwathanaratetal15}.

These results demonstrate that the presence of the
hydrogen envelope in Type-IIP supernovae has important
consequences for the final velocities and the mixing and
fragmentation of the distribution of metal-containing ejecta.
In Fig.~\ref{fig:mvsvr} we present the mass distributions
as functions of the radial velocity for a 
set of nucleosynthesis components
for model W15-2-cw at the time of shock breakout \citep[bottom
panel; data from figure~12 of][]{Wongwathanaratetal15}
compared to results of our present simulation of model
W15-2-cw-IIb, i.e., after having removed the hydrogen 
envelope except for a relic mass of $\sim$0.3\,$M_\odot$.

In model W15-2-cw the mass distributions of 
$^{56}$Ni and $^{44}$Ti reach
up to about 5000\,km\,s$^{-1}$ for a small faction
of this radioactive material, which indicates the 
extent of mixing in velocity space. The overall shapes
of the distributions of all elements (iron-group,
intermediate-mass, and light elements) resemble each
other, with the bulk ejecta velocities ranging from
$\sim$1000\,km\,s$^{-1}$ to $\sim$3000\,km\,s$^{-1}$,
and the maxima as well as the positive slopes below
the maxima exhibiting the order expected from homologous 
expansion ($v\propto r$) of spherically symmetric explosions,
where lighter elements reside at larger radii with
higher velocities. The tails of the fastest ejecta above 
the distribution maxima stretch out to roughly
7000\,km\,s$^{-1}$ and contain, besides hydrogen and
helium, also intermediate and light elements (silicon
to carbon) corresponding to the chemical composition of
the progenitor envelope. 

For a given explosion energy $E_\mathrm{exp}$,
the average velocity of the ejecta (mass $M_\mathrm{ej}$)
is approximately given
by $\bar{v}_\mathrm{ej}\sim \sqrt{2E_\mathrm{exp}/M_\mathrm{ej}}$,
and also the maximum ejecta velocity (ignoring a small amount
of mass accelerated off the stellar surface at shock 
breakout) follows roughly the scaling with the ratio
$\sqrt{E_\mathrm{exp}/M_\mathrm{ej}}$. Our red supergiant
model W15-2-cw ($E_\mathrm{exp}\approx 1.5\,$B,
$M_\mathrm{ej}\approx 13.6\,M_\odot$) yields 
$E_\mathrm{exp}/M_\mathrm{ej}\approx 0.11$\,B/$M_\odot$,
corresponding to $\bar{v}_\mathrm{ej}\sim 3300$\,km\,s$^{-1}$. 
Without the decelerating influence of the massive hydrogen 
envelope, model W15-2-cw-IIb possesses a higher value of
$E_\mathrm{exp}/M_\mathrm{ej}\approx 
1.5\,\mathrm{B}/3.3\,M_\odot \approx 0.45$\,B/$M_\odot$,
for which reason one expects about twice the ejecta velocities
of the red supergiant model.

Indeed, the main ejecta mass (helium for W15-2-cw-IIb)
exhibits velocities around 6000--8000\,km\,s$^{-1}$, and
small amounts of material expand with more than 
10,000\,km\,s$^{-1}$
(Fig.~\ref{fig:mvsvr}, top panel). The mass 
distributions of different elements are much more dissimilar
in model W15-2-cw-IIb than in model W15-2-cw. This 
suggests considerably less mixing, fully in line with the
absence of a reverse shock and RT unstable
conditions at the He/H interface. Hydrogen and helium 
naturally show the tendency of having the highest velocities
for the bulk of their mass, followed by carbon and silicon,
whereas the dominant masses of oxygen, titanium, and nickel 
are distributed in a broad maximum below 
$\sim$7000\,km\,s$^{-1}$. Only minor fractions of the
radioactive elements (of order $\lesssim$1\% of the total
$^{44}$Ti and $^{56}$Ni mass) are contained in high-velocity
tails that extend from 7000\,km\,s$^{-1}$ up to 
$\sim$9000\,km\,s$^{-1}$.

The highest velocities of significant amounts of 
$^{44}$Ti (around 7000\,km\,s$^{-1}$) in our simulations
are in good agreement with the upper bound on the 
fastest such material determined by Cas~A observations 
($5350\pm 1610$\,km\,s$^{-1}$, \citealt{Grefenstetteetal14};
$\sim$$6300\pm 1250$\,km\,s$^{-1}$, 
\citealt{Grefenstetteetal17}).\footnote{Knot 20b, the
fastest and brightest (biggest) structure in the 3D
$^{44}$Ti distribution analysed by \citet{Grefenstetteetal17},
seems to be blueshifted with a (line-of-sight) velocity of even 
$7500\pm 1600$\,km\,s$^{-1}$, but this component is not 
indisputably identified as an individual coherent feature, and
it is associated only with a single broadened $^{44}$Ti line in
the NuSTAR bandpass.}  
Besides the question how much dilute, high-velocity $^{44}$Ti
might be present above this velocity range, but hard to detect
observationally, our model results
need to be considered with some caution, too. First, we remind
the reader once again of the fact that we did not attempt to
fine tune our simulations for optimal comparison with the
observations. Detailed analysis of Cas~A suggests an explosion
energy of $\sim$2.3\,B for an ejecta mass of $\sim$4\,$M_\odot$
\citep{Vink2004,Orlandoetal16}, which implies a value of
$E_\mathrm{exp}/M_\mathrm{ej}\approx 0.575$\,B/$M_\odot$.
Therefore an optimal Cas~A model might possess bulk and
maximum velocities of $^{44}$Ti that are even 10--15\% higher
than in our model W15-2-cw-IIb. However, our simulations
currently track the evolution of the SN explosion only for
the first day. Although the ejecta are already close to
homology at this time, our models ignore important 
future effects such as the heating associated with radioactive
decays and the deceleration of the fastest $^{44}$Ti by the
reverse shock that originates from the interaction of the
ejecta with the circumstellar environment decades later. The
long-time evolution from the first day to the remnant
stage as observed at the present epoch will be the subject of
future studies, for which reason we postpone a more detailed 
quantitative comparison of the velocity information with
Cas~A observations to a later stage.

The middle panel of Fig.~\ref{fig:mvsvr} displays the 
mass distributions in velocity space for $^{44}$Ti and $^{56}$Ni 
as obtained by our post-processing of tracer particles with a
large nucleosynthesis network (as described in 
Sect.~\ref{sec:nucleosynthesis}), in contrast to the top and
bottom panels of this Figure, where
results from hydrodynamic simulations with the small
nucleosynthesis network are presented. While the distributions
for model W15-2-cw-IIb in the top and middle panels
look roughly similar for the bulk of the radioactive material,
they nevertheless exhibit an important difference: the tracer 
particle distributions possess a considerably more populated
high-velocity tail, which also reaches up to higher velocities.
While the hydrodynamic simulation of model W15-2-cw-IIb
with the small network
yields $\lesssim$1\% of the ejected $^{44}$Ti and $^{56}$Ni
with velocities of more than 7000\,km\,s$^{-1}$, the 
tracer particle analysis predicts such a small fraction 
of the radioactive nuclei to be faster than 
$\sim$8500\,km\,s$^{-1}$, a tiny amount to be 
ejected with velocities up to 9700\,km\,s$^{-1}$
instead of $\sim$9000\,km\,s$^{-1}$ for the hydrodynamic run,
and some 5 (8) percent of the $^{56}$Ni ($^{44}$Ti) to be
expelled with over 7000\,km\,s$^{-1}$.

\begin{figure*}[!]
\centering
\resizebox{0.45\hsize}{!}{\includegraphics{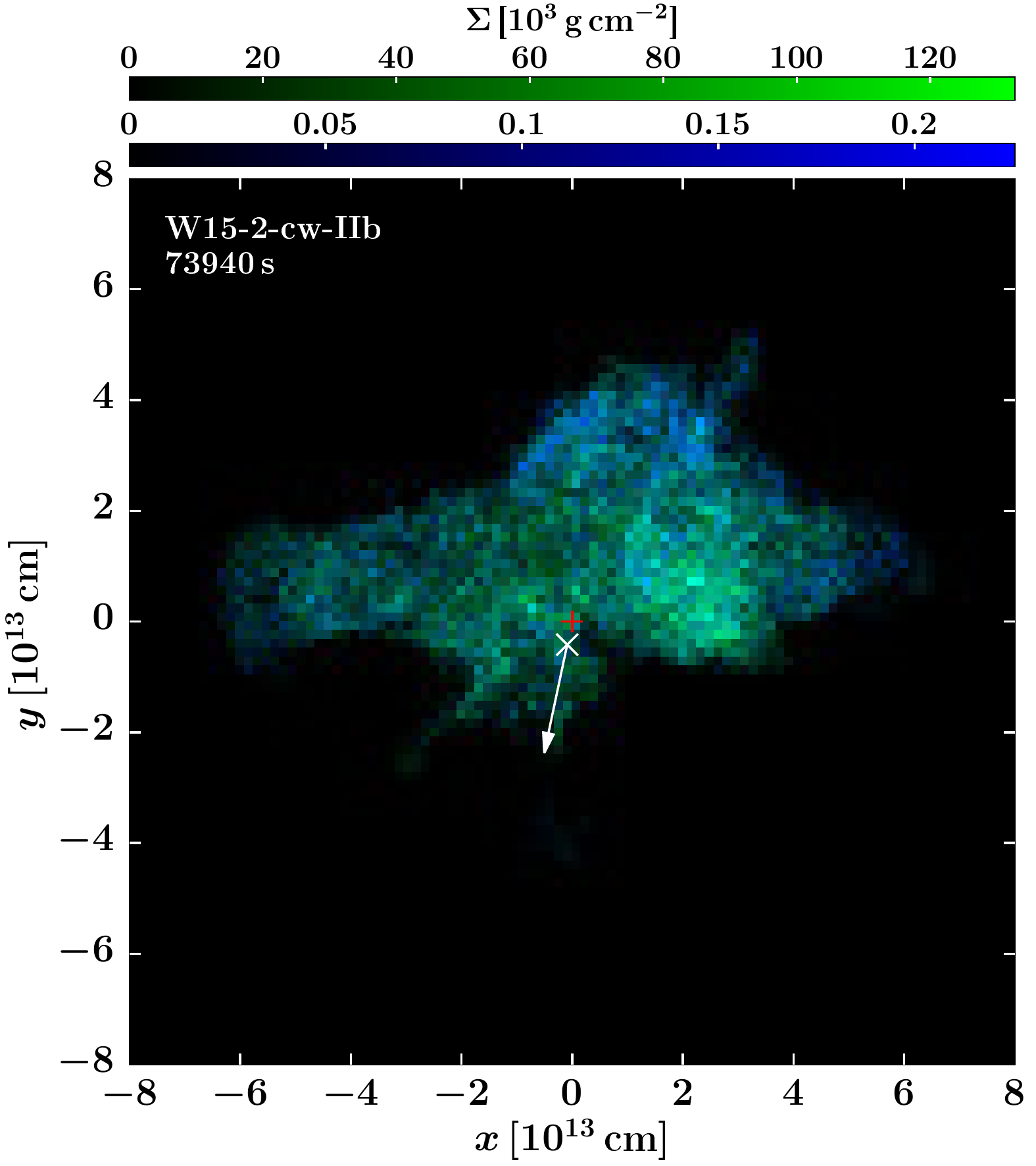}}
\resizebox{0.45\hsize}{!}{\includegraphics{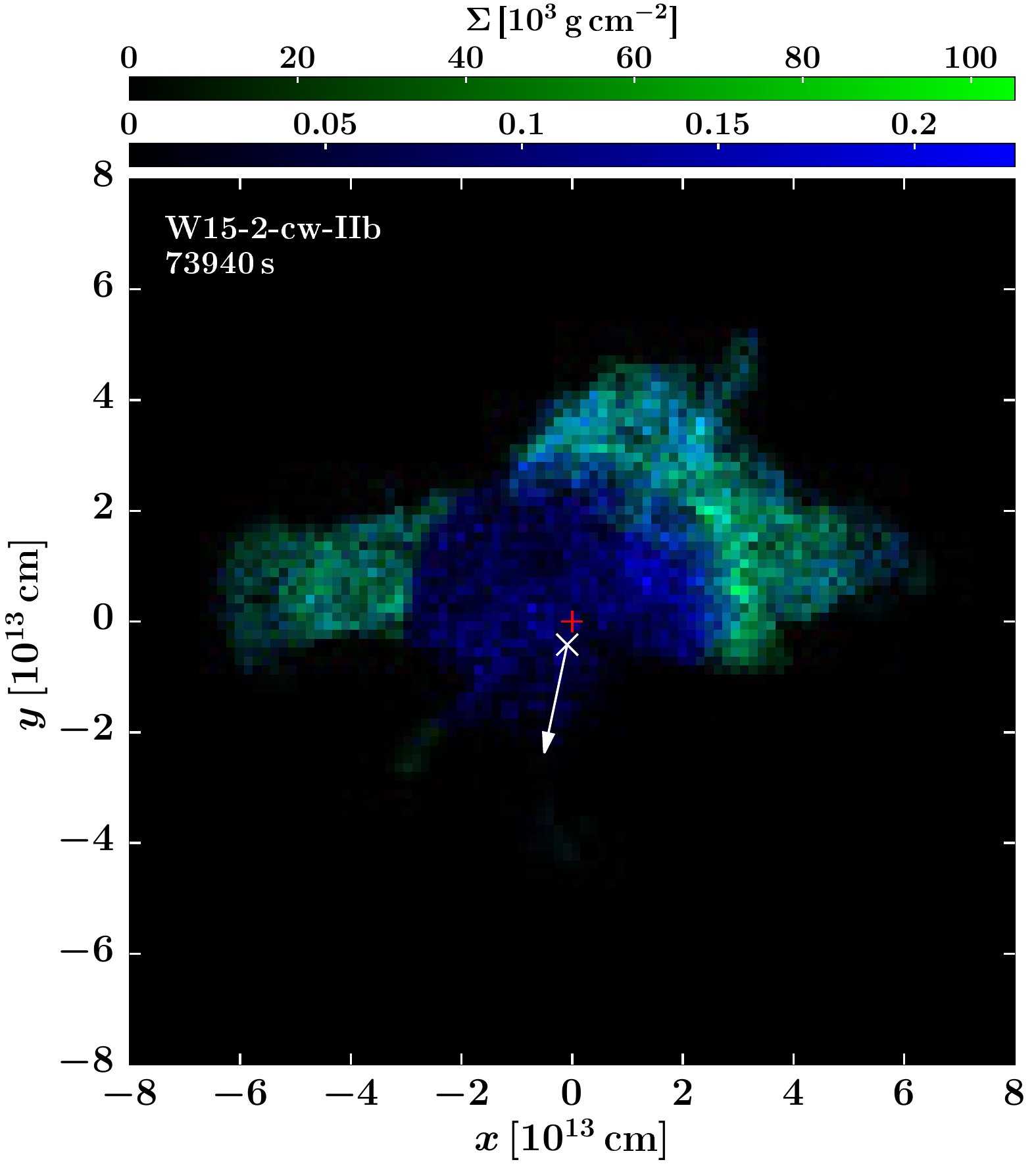}}
\caption{Column densities of $^{56}$Ni (green) and $^{44}$Ti (blue)
for model W15-2-cw-IIb at $t_\trm{pb}=73940$\,s. The integration
was performed along lines-of-sight perpendicular to a plane
containing the recoil velocity of the newly formed NS,
using the abundances of the radionuclides as computed by our
nucleosynthetic post-processing of tracer particles.
The orientation of the NS kick direction in the plane of
the image (indicated by the white arrow) was chosen to be the
same as in figures~2 and 3, and extended data figure~1 of
\citet{Grefenstetteetal14} (see Fig.~\ref{fig:nature}). The white
cross marks the current NS position, whose shift away from the
center of the explosion (red plus symbol) due to a kick velocity
of nearly 600\,km\,s$^{-1}$ \citep{Wongwathanaratetal13} is shown
for comparison with the extension of the radioactive ejecta.
While in the {\em left panel} all titanium and nickel are
displayed, the {\em right panel} assumes that after decay
to stable $^{56}$Fe, only half of the ejected nickel is visible
in the outer SN-remnant shell, because the reverse shock from
the ejecta-environment interaction has not travelled backward
to the center yet and thus has not heated the inner half
of the iron to X-ray emission temperatures.}
\label{fig:combmap}
\end{figure*}

Tests showed that this discrepancy is a consequence of numerical
inaccuracies in our calculation of the tracer-particle 
histories. As mentioned in Sect.~\ref{sec:nucleosynthesis},
we do not evolve the tracer particles online with the 
hydrodynamic model but reconstruct the particle
velocities and positions by higher-order time integration
based on a finite number of output files from the 
hydrodynamic run. Since these output files are dense in
the early phase (the first seconds of the explosion), we
do not consider this approach as problematic during the
time interval when the nucleosynthesis takes place.
Indeed, inspecting the mass distributions versus radial 
velocity at 10\,s, the results of the calculations
with small network (during the hydro run) and large
network (based on particle post-processing) exhibit
much closer similarities with respect to their shape and 
spread in 
velocity space, apart from effects that are expected
as consequences of nucleosynthesis differences between
the approximative $\alpha$-network treatment and the
more detailed calculation (for example a shift of the 
$^{44}$Ti production in the latter case
to less strongly heated, higher-velocity
material instead of a distinctly synchronous production with
$^{56}$Ni by the small $\alpha$-network). Integrating
the particle histories to late times, however, numerical
errors accumulate, in particular since data files for the
later stages are stored only with larger time intervals.

In spite of this obvious caveat, we will further analyse 
the nucleosynthesis-related aspects of our results on the
basis of the particle information, because only the 
big-network calculation provides reliable quantitative 
information on the production of $^{44}$Ti and $^{56}$Ni
for their mutual comparison (though the treatment of the 
latter is less problematic with the small network, 
see Sect.~\ref{sec:production}).
Yet, we should keep in mind that this implies an
overestimation of the abundances of these nuclei in the 
highest-velocity tails of their late-time distributions. 
However, since the problem concerns
only a small amount of matter (some percent of the total
ejected $^{44}$Ti and $^{56}$Ni masses), it is
not of great relevance for our discussion of the spatial
asymmetries in Sect.~\ref{sec:spacedistribution}.

Figure~\ref{fig:mvsm} shows the initial and final 
distributions of the bulk of $^{44}$Ti and $^{56}$Ni
in mass space. The mass coordinate of a tracer particle  
is defined by the stellar mass contained in a sphere
whose radius is given by the particle's radial position.
All of the nickel and the far dominant part of the 
titanium originate from a mass range between 1.35\,$M_\odot$
and 1.6\,$M_\odot$ in the progenitor star, which 
encompasses neutrino-processed as well as shock-heated 
ejecta (left panel of Fig.~\ref{fig:mvsm}). 
A small fraction (of order one percent) of the $^{44}$Ti is
made in matter associated with an initial mass interval from
1.6\,$M_\odot$ to 1.8\,$M_\odot$, where the outgoing 
shock raises the peak temperatures to 4--$5\times 10^9$\,K
and allows some $^{44}$Ti to be produced during incomplete 
silicon burning. The dip between the two maxima is 
connected to a depletion region called the ``chasm'' by
\citet{Magkotsiosetal10}.

\begin{figure}
\centering
\resizebox{\hsize}{!}{\includegraphics{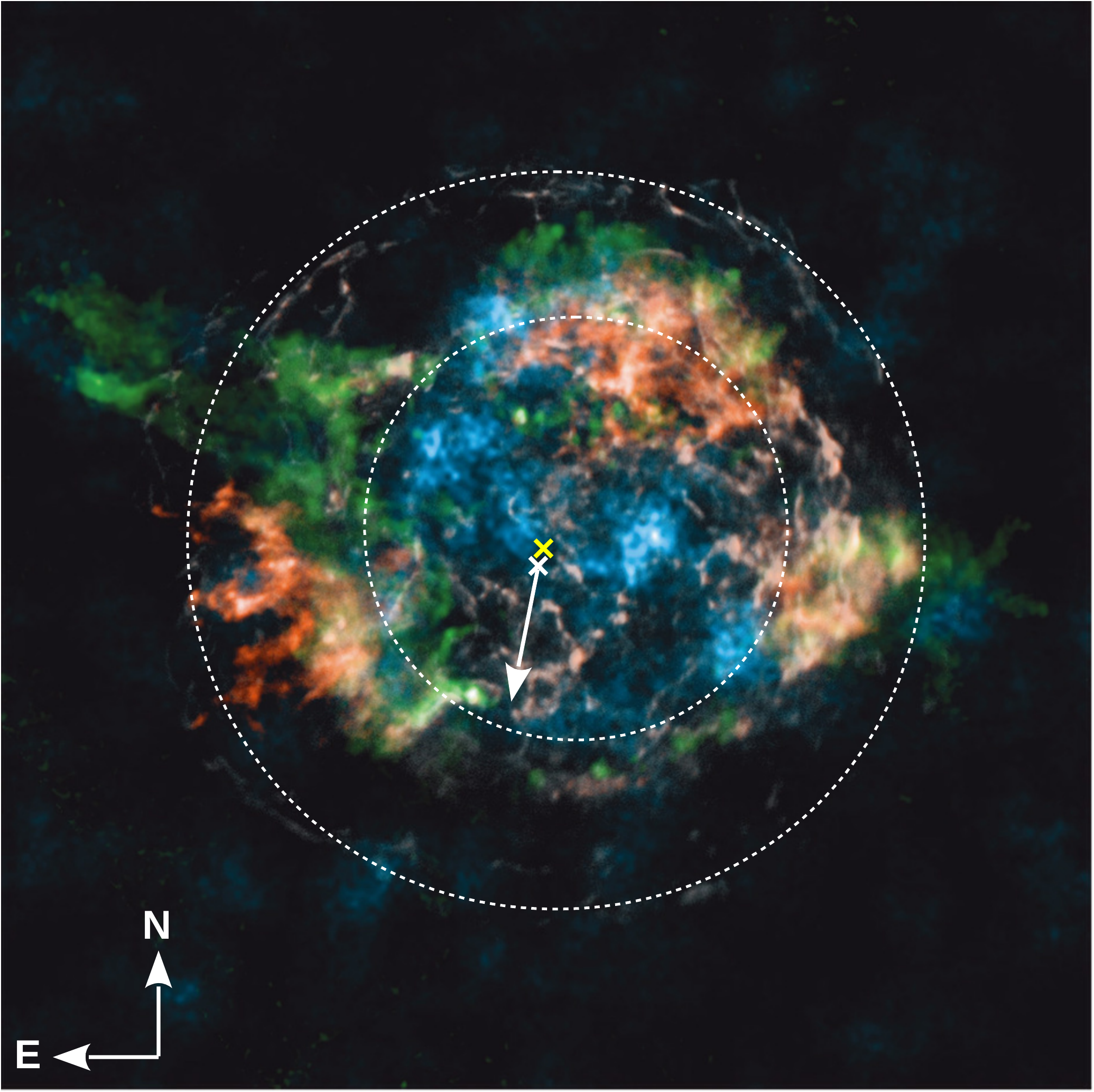}}
\caption{
Spatial distribution of $^{44}$Ti (blue) and of known Fe
K-shell emission in Cas~A. The image is adopted from figure~3
of \citet{Grefenstetteetal14} with the 4--6\,keV continuum 
emission (white) and X-ray-bright Fe (red) seen by Chandra
\citep[Fe distribution courtesy of U.~Hwang;][]{Hwangetal04,HwangLaming12}. 
The orientation in
standard astronomical coordinates is indicated by the compass
in the lower left corner. The yellow cross marks the 
geometrical center of the expansion of the explosion, 
the white cross and the arrow the current location and
the direction of motion of the central, X-ray emitting
compact object, and the outer and inner white dashed
circles the locations of the forward and reverse shocks,
respectively. These features were extracted from figure~2
and extended data figure~1 of \citet{Grefenstetteetal14},
where a detailed description of the observational data and 
corresponding references can be found.
(Reprinted by permission from Macmillan Publishers Ltd: Nature,
\citet{Grefenstetteetal14}, \copyright2014)
}
\label{fig:nature}
\end{figure}

At nearly a day the distributions of $^{44}$Ti and $^{56}$Ni 
are smeared over the whole range of ejecta masses, indicating
efficient radial transport of the innermost SN ejecta into
the carbon-oxygen and helium-layers of the exploding star
(right panel of Fig.~\ref{fig:mvsm}; corresponding cumulative
distributions versus radius and enclosed mass are displayed
in Fig.~\ref{fig:cumul}).
While the main formation of nickel and titanium takes
place in the same mass region, but with considerable 
variation of the relative production (left panel),
the close similarity of the normalized mass distributions
of both radioactive nuclei after one day
does not imply that both species are well mixed to
equal composition in all regions. As we will see from
our analysis of the spatial distributions in
Sect.~\ref{sec:spacedistribution}, there are regions 
that survive with $^{44}$Ti being enhanced relative to 
$^{56}$Ni and vice versa, compared to the average 
production ratio of the two species.

Finally, we again allude to the fact that the tracer
particles with the highest velocities are not very
reliable because of integration errors in the
particle histories. For the mass distributions in the
right panel of Fig.~\ref{fig:mvsm}, this concerns a 
subdominant mass of $^{44}$Ti and $^{56}$Ni accumulating 
in a local maximum between $\sim$4.2\,$M_\odot$ and 
$\sim$4.7\,$M_\odot$. This feature is absent in our
results from the small $\alpha$-network calculation.
The peak contains a small fraction
(a few percent) of the material that originates from the
initial mass interval of 1.35--1.55\,$M_\odot$. In the
case of nickel, both shock-heated and neutrino-heated
matter contributes, whereas in the case of titanium
the peak is supplied only by tracer particles carrying
yields from the far dominant neutrino-processed ejecta
component.

\begin{figure*}
\centering
\resizebox{0.24\hsize}{!}{\includegraphics{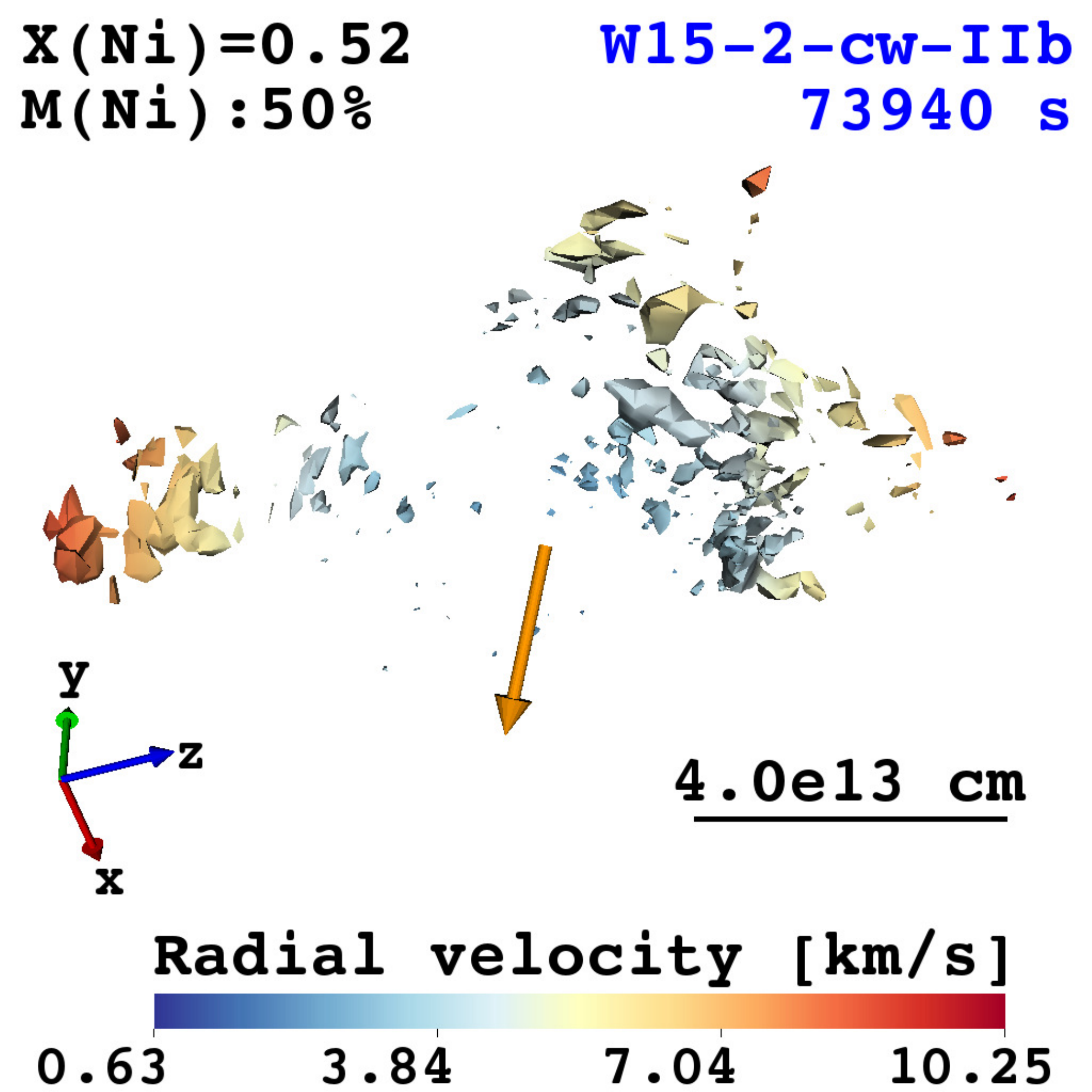}}
\resizebox{0.24\hsize}{!}{\includegraphics{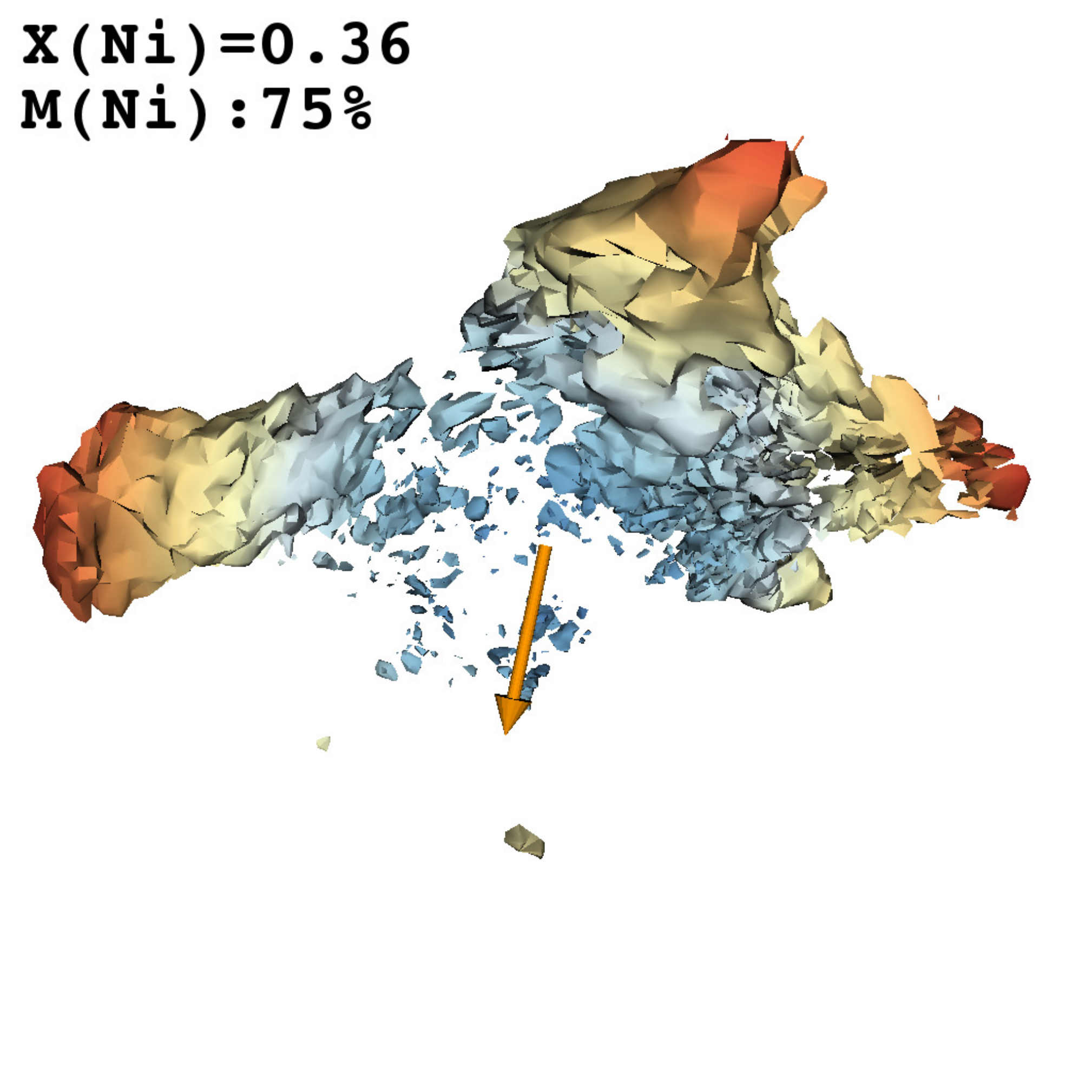}}
\resizebox{0.24\hsize}{!}{\includegraphics{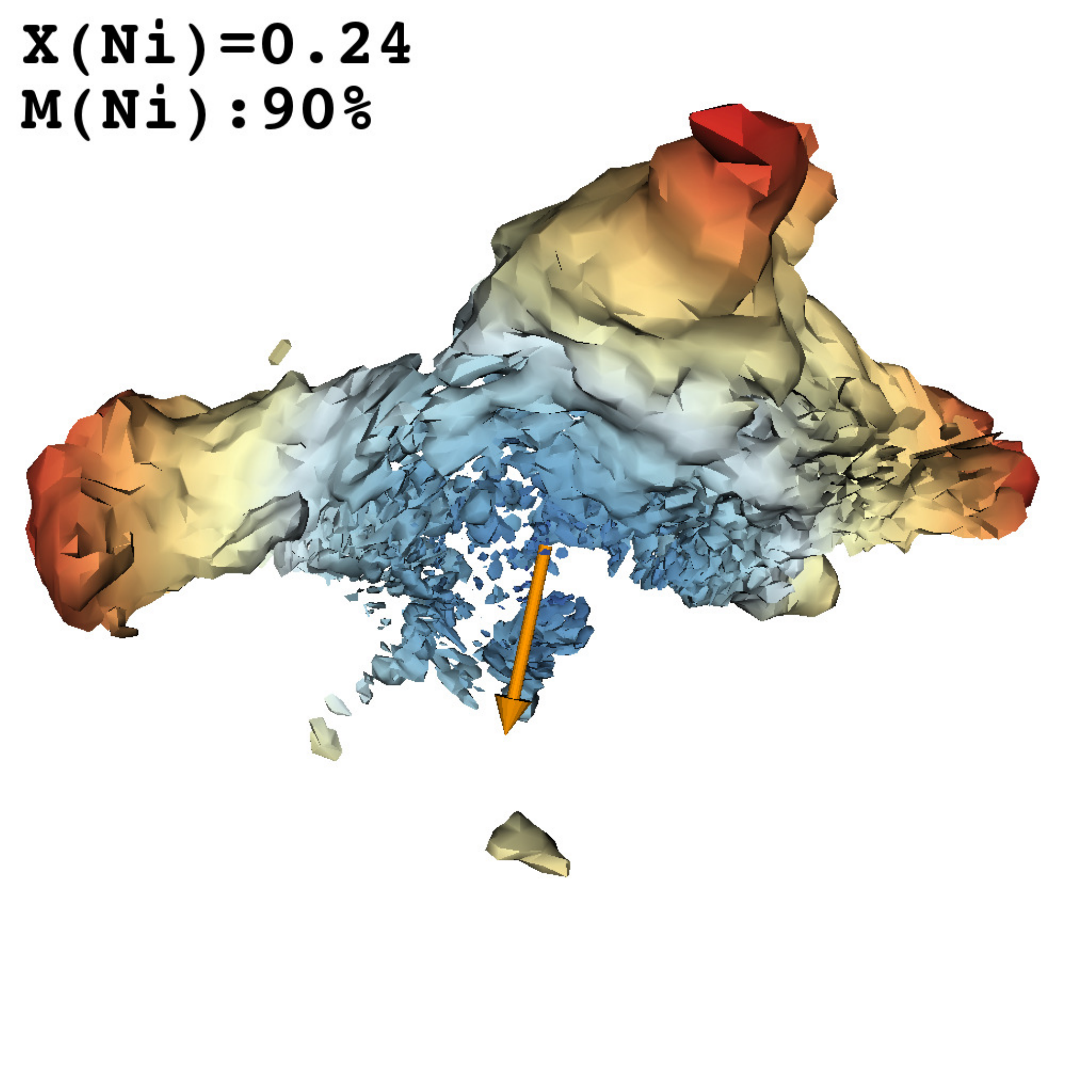}}
\resizebox{0.24\hsize}{!}{\includegraphics{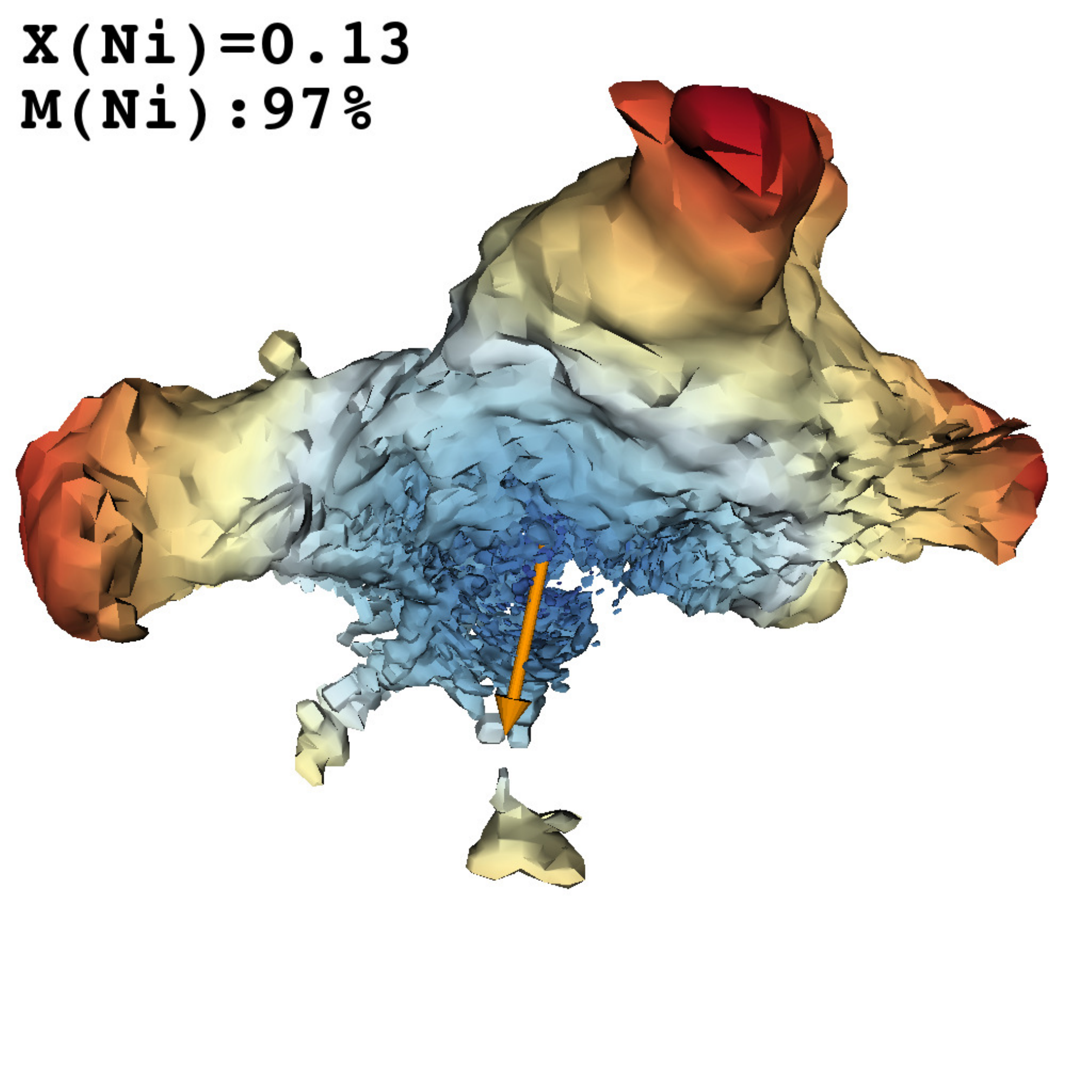}}\\
\resizebox{0.24\hsize}{!}{\includegraphics{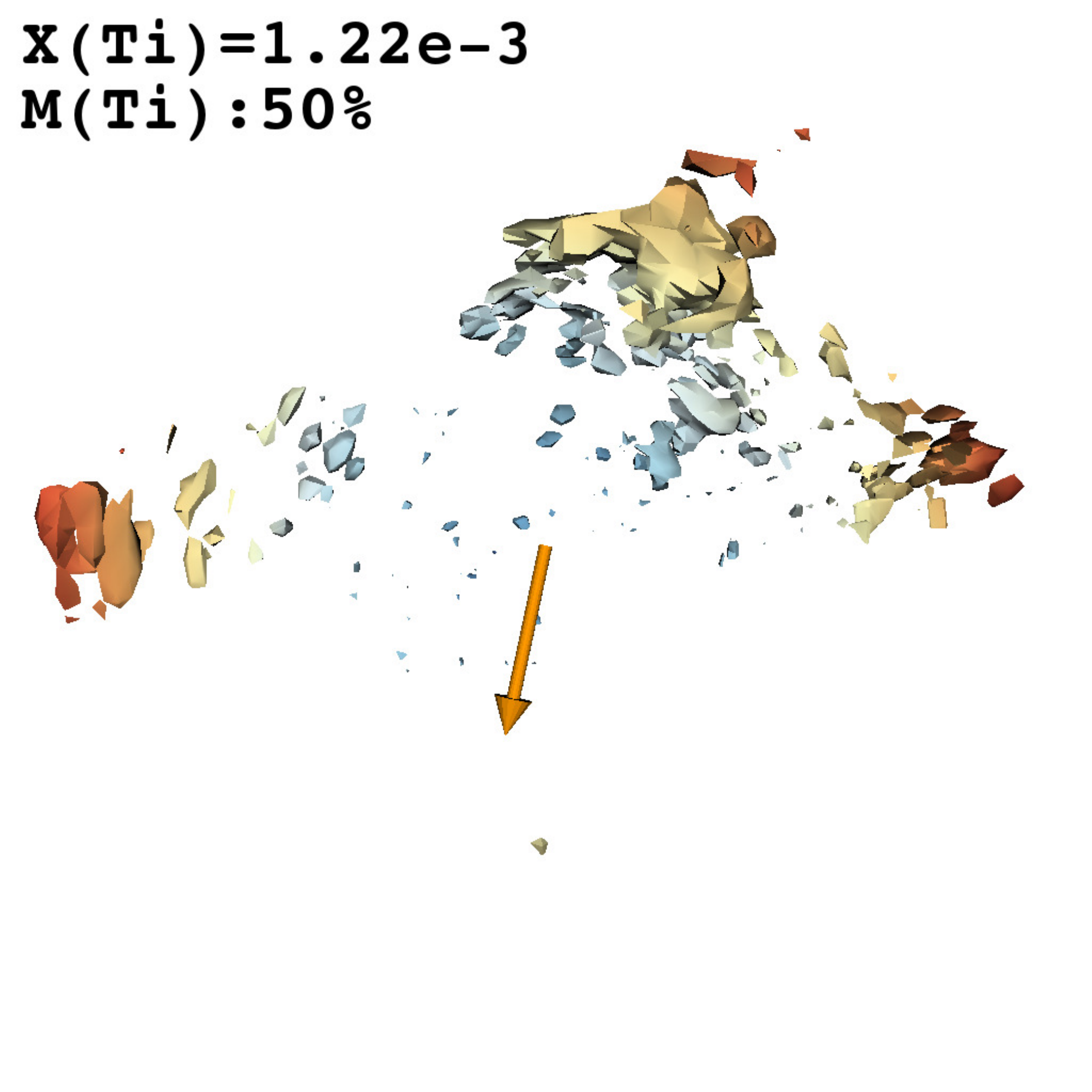}}
\resizebox{0.24\hsize}{!}{\includegraphics{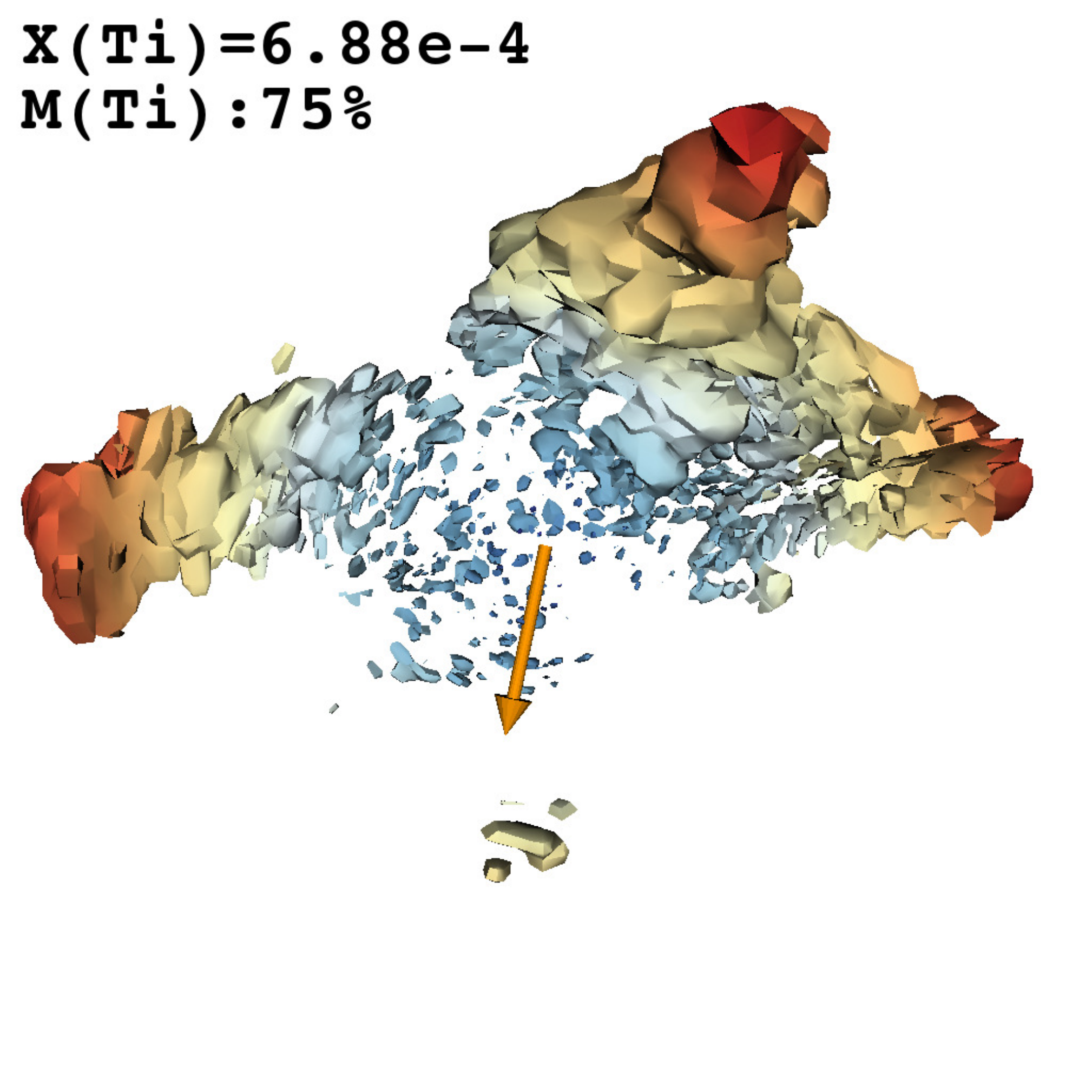}}
\resizebox{0.24\hsize}{!}{\includegraphics{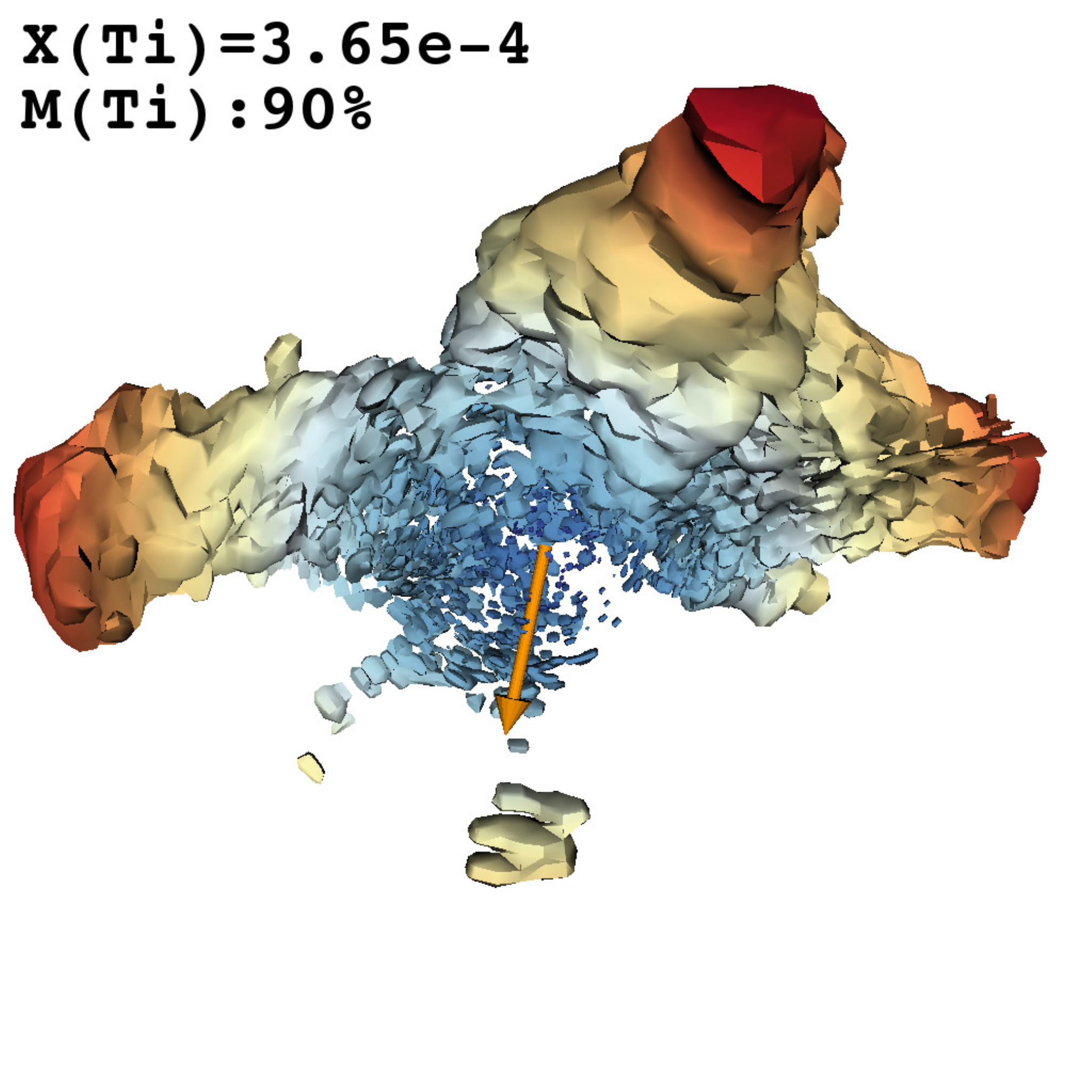}}
\resizebox{0.24\hsize}{!}{\includegraphics{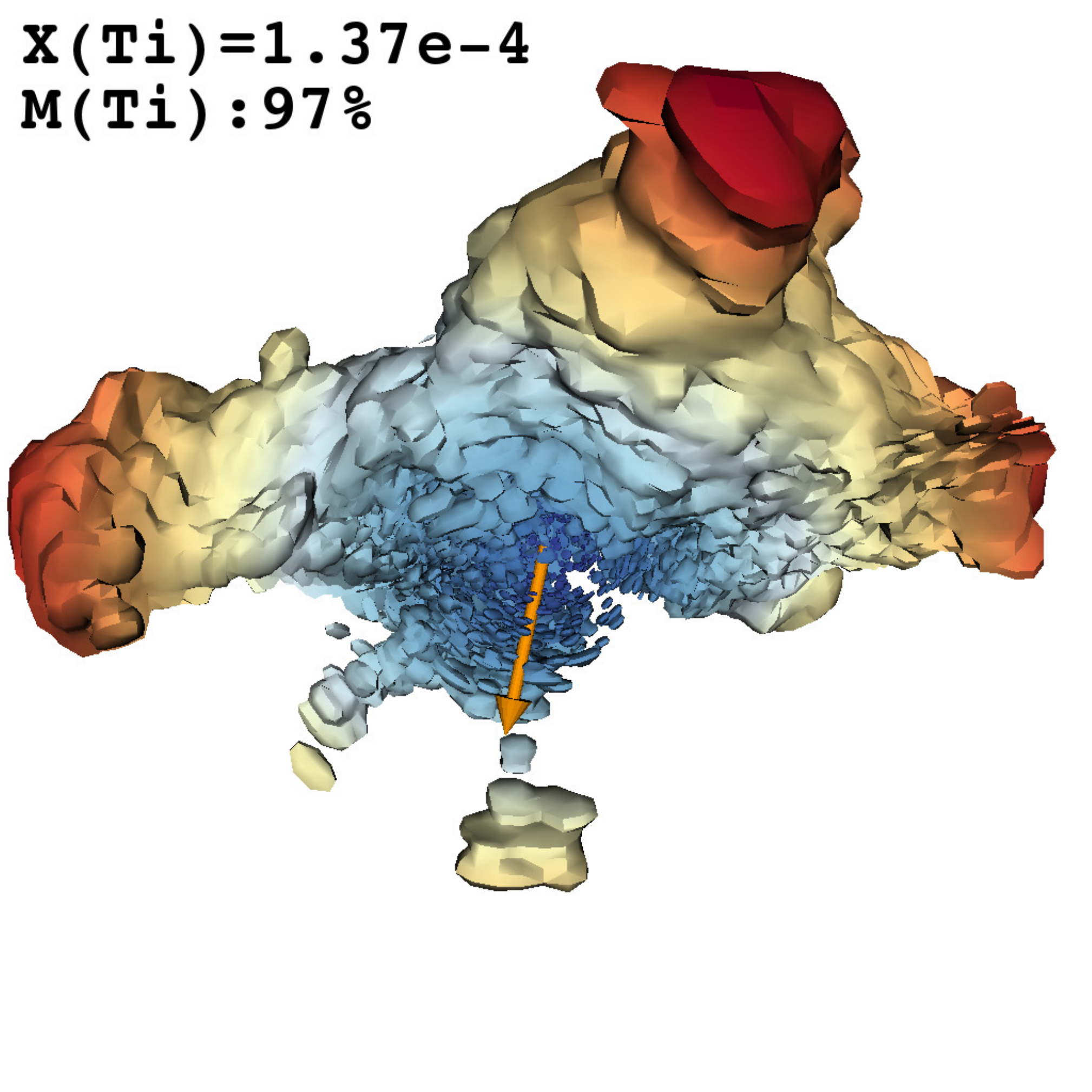}}
\caption{Isosurfaces of constant mass fractions of $^{56}$Ni 
({\em upper row}) and $^{44}$Ti ({\em lower row}) for model W15-2-cw-IIb
at $t_\trm{pb}=73940$\,s, based on the nucleosynthesis yields obtained
from our post-processing analysis of tracer particles. The different
values of the mass fractions (decreasing from left to right) result
from the requirement that the isosurfaces enclose 50\%, 75\%, 90\%,
and 97\% of the total ejecta mass of the respective element. The 
color-coding of the surfaces expresses the radial velocities 
according to the color bar placed between the two panels
on the left side. The kick direction of the NS is indicated
by the orange arrow, whose tail starts at the center of the
explosion. The viewing direction is the same as in
Fig.~\ref{fig:combmap}.}
\label{fig:niti3D}
\end{figure*}

\subsection{Spatial distributions}
\label{sec:spacedistribution}

Spatial variations of the abundance mix of $^{44}$Ti and
$^{56}$Ni can be concluded from Fig.~\ref{fig:combmap}, where
we show column densities of both nuclei as observed from a
direction perpendicular to the kick direction of the newly
formed NS, which is indicated by a white arrow. The orientation
of our image is chosen such that direct comparison with figures~2
and 3 and extended data figure~1 of \citet{Grefenstetteetal14}
(see Fig.~\ref{fig:nature}) is possible and exhibits the closest
resemblance that we can achieve.
Although nickel and titanium can be found in coexistence in the
whole mass-filled volume, the color variations reflect considerable
local differences of the abundance ratios of both nuclei,
with more intense blue colors signaling relatively higher
concentrations of $^{44}$Ti, and more intense green picturing
higher concentrations of $^{56}$Ni.

In the right panel of Fig.~\ref{fig:combmap}, we show
the nickel ejecta only at large radii but do not
display those 50\% of the expelled $^{56}$Ni mass that
are contained in an inner, spherical volume.
This is motivated by investigations of the Cas~A
SNR by \citet{MilisavljevicFesen15} and
\citet{Orlandoetal16}, who concluded that about half of
the iron produced by the explosion could reside in the
central region that has not yet been heated to X-ray emission
temperatures by the reverse shock from the ejecta-environment
interaction. After $^{56}$Ni has decayed to stable $^{56}$Fe,
ionization by the reverse shock is necessary to enable Fe 
X-ray emission from atomic transitions. In contrast, $^{44}$Ti
continues to be radioactive well into the SNR stage,
even at the age of Cas~A, and therefore radiates X-rays by
nuclear transitions. We note that somewhat less than half
of the total $^{44}$Ti mass (namely roughly 43\%) is 
located within the 50-percent-nickel sphere 
(Fig.~\ref{fig:cumul}), consistent with
the fact that the velocity distribution of $^{44}$Ti is
slightly shifted to higher velocities compared to the $^{56}$Ni
distribution (see middle panel of Fig.~\ref{fig:mvsvr}).

The overall morphological similarity between the NuSTAR image
of Cas~A (figure~2 of \citealt{Grefenstetteetal14}; $^{44}$Ti 
distribution in Fig.~\ref{fig:nature}) and our model
results (Fig.~\ref{fig:combmap}) is astounding. 
We emphasize once more that this similarity
is not by construction but by chance, resulting as a consequence of 
stochastic asymmetries that develop at the origin of the explosion.
For this reason the focus here should be on basic properties, not 
structural details. It is obvious that the NS kick vector
points away from the direction in which the highest 
concentrations of iron-group material and $^{44}$Ti are expelled
\citep[see also][]{Grefenstetteetal17}.
This is fully in line with the gravitational tug-boat mechanism
for NS acceleration in asymmetric SN explosions as discussed
by \citet{Wongwathanaratetal10b,Wongwathanaratetal13} and 
\citet{Schecketal06} \citep[see also][]{Nordhausetal10,Nordhausetal12}. 

This kick mechanism accelerates the NS over time scales of
seconds mainly by gravitational forces between the anisotropic
and clumpy innermost SN ejecta rather than by direct
hydrodynamic interaction. In the context of the 
neutrino-driven explosion mechanism, which was investigated by
3D SN simulations by \citet{Wongwathanaratetal10b,Wongwathanaratetal13}, 
hydrodynamical instabilities such as convective overturn and the
SASI naturally create large-scale asymmetries behind the SN
shock in the prelude of the explosion. Due to the aid of these
instabilities, neutrino-energy deposition revives the stalled
SN shock and initiates its outward expansion anisotropically. 
In the direction of the more powerful shock acceleration, driven
by the largest high-entropy plumes of neutrino-heated matter,
more neutrino-processed plasma is expelled and, in addition,
stronger shock-heating of the stellar shells swept up by the 
expanding blast wave allows for more efficient explosive 
nucleosynthesis.
This leads to enhanced production of elements from roughly
$^{28}$Si to the iron group (and beyond) in the direction
of the stronger explosion. The corresponding hemispheric 
differences in the total yields of these
elements can be as large as factors of 3--4 for 
highly asymmetric explosions, and even greater differences
might well be possible for more extreme cases than those obtained
in the set of 3D simulations of \citet{Wongwathanaratetal13},
or for nuclear species not included in their nucleosynthesis
treatment with a small $\alpha$-network.

In the direction where the explosion and the 
nucleosynthesis are weaker, the shock and the postshock
material accelerate outward more slowly. On this side the nascent
NS can therefore accrete for a longer period of time 
before the mass infall is quenched by the acceleration of
the SN explosion. The momentum transfer both by hydrodynamic 
accretion flows and by the gravitational attraction from more
inert, typically denser and more massive innermost SN ejecta,
pulls the NS to the side of the weaker blast wave. Thus the
NS receives a kick opposite to the direction of the stronger
shock expansion, consistent with momentum conservation during 
the explosion
\citep[for detailed discussions, see][]{Wongwathanaratetal13,Schecketal06}.

This means that the NS experiences a recoil acceleration that
points away from the hemisphere where the SN ejects more heavy
elements with atomic numbers $Z \sim 14$ and higher. The kick 
velocity of the NS depends on the stochastic explosion asymmetry,
the explosion energy, and the density around the newly formed
NS \citep[see][]{Janka2017}. The environmental density of the NS, 
in turn, depends on the compactness of the progenitor
core and determines the amount of matter that is neutrino-processed
and shock-heated in the region of explosive nucleosynthesis.

From the set of 3D simulations by \citet{Wongwathanaratetal13},
model W15-2, which we consider in the present paper,
develops a fairly large explosion asymmetry and the NS receives
a kick of 575\,km\,s$^{-1}$ until $\sim$3\,s after the onset
of the SN blast (with acceleration continuing on a low level 
for even longer time). Information on hemispheric differences
of the yields of nuclei included in the $\alpha$-network
can be found in Table~3 of \citet{Wongwathanaratetal13}, and
visualizations of the 3D distribution of the ejected nickel 
(compared to other models) can be found in figures 14 and 15
of that paper. Even for a case with rather high NS kick like
model W15-2, radionuclei as nucleosynthesis products in the
innermost SN ejecta, where the explosion asymmetry is most
extreme, are not just expelled in one hemisphere, but some
of this material can be ejected also on the side of
the kicked NS. The exact geometry, however, strongly varies
from case to case, and lower NS kicks go hand in hand with
more isotropic ejection of radioactive material 
\citep[compare the cases displayed in figures~14 and 15 
of][]{Wongwathanaratetal13}.

\begin{figure*}
\centering
\resizebox{0.24\hsize}{!}{\includegraphics{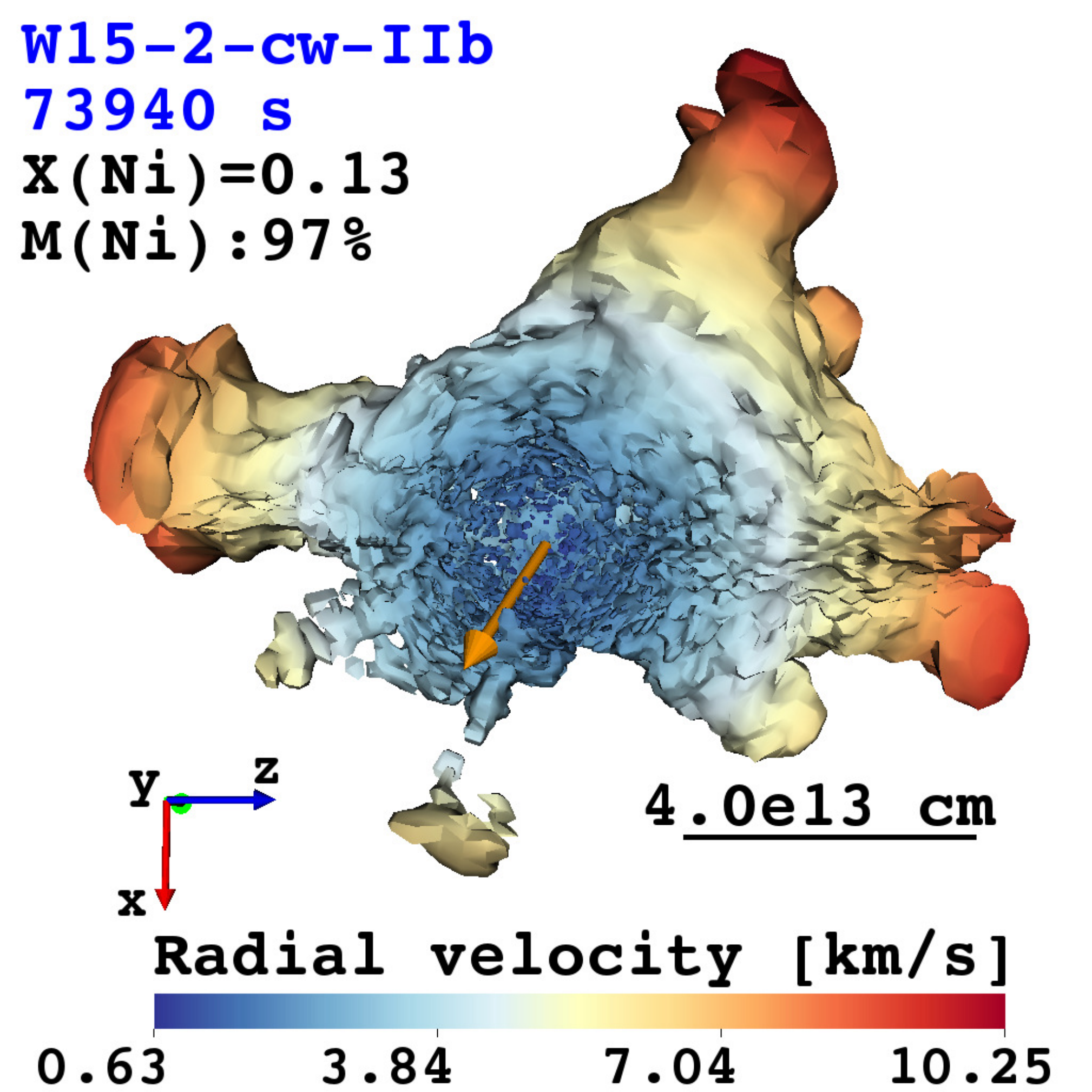}}
\resizebox{0.24\hsize}{!}{\includegraphics{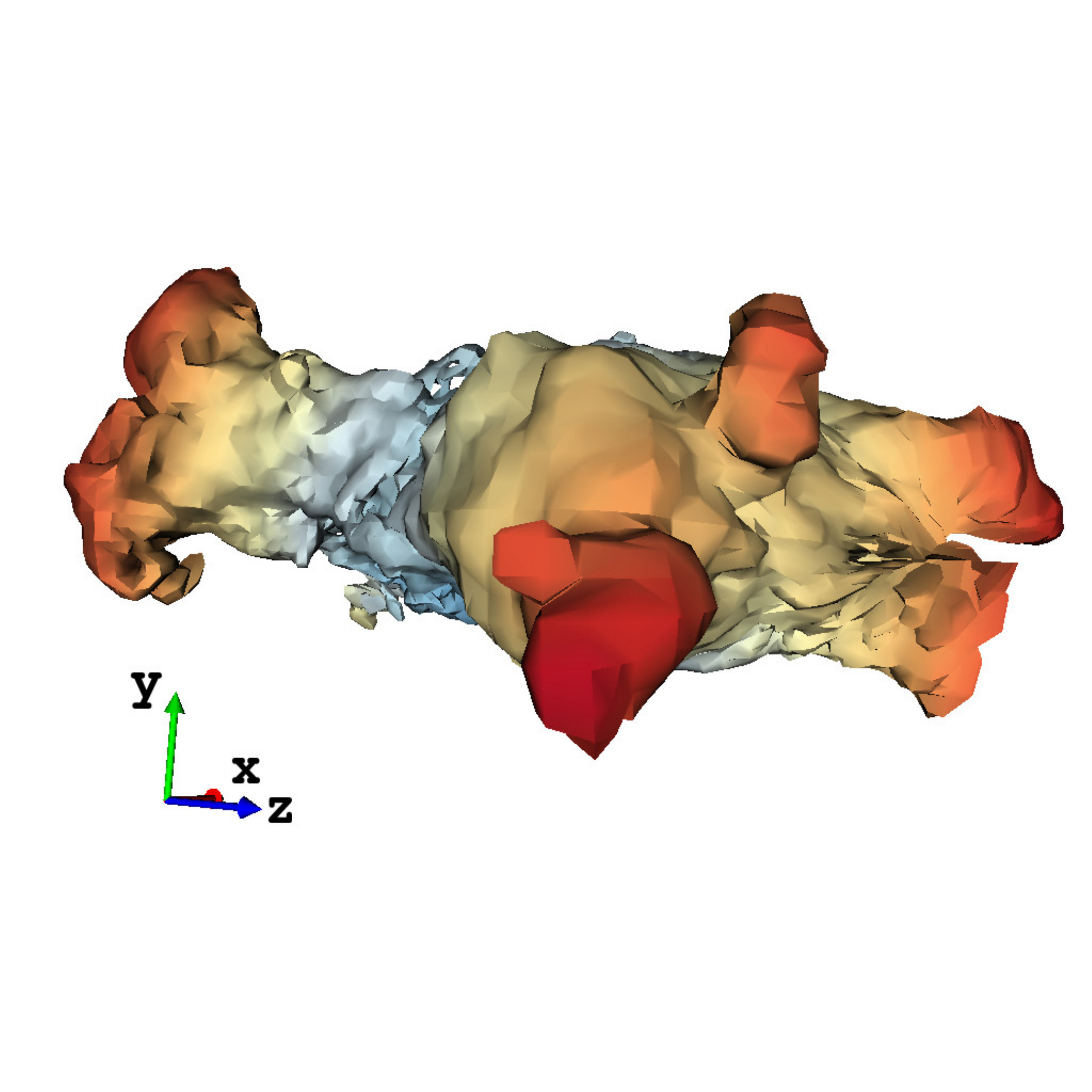}}
\resizebox{0.24\hsize}{!}{\includegraphics{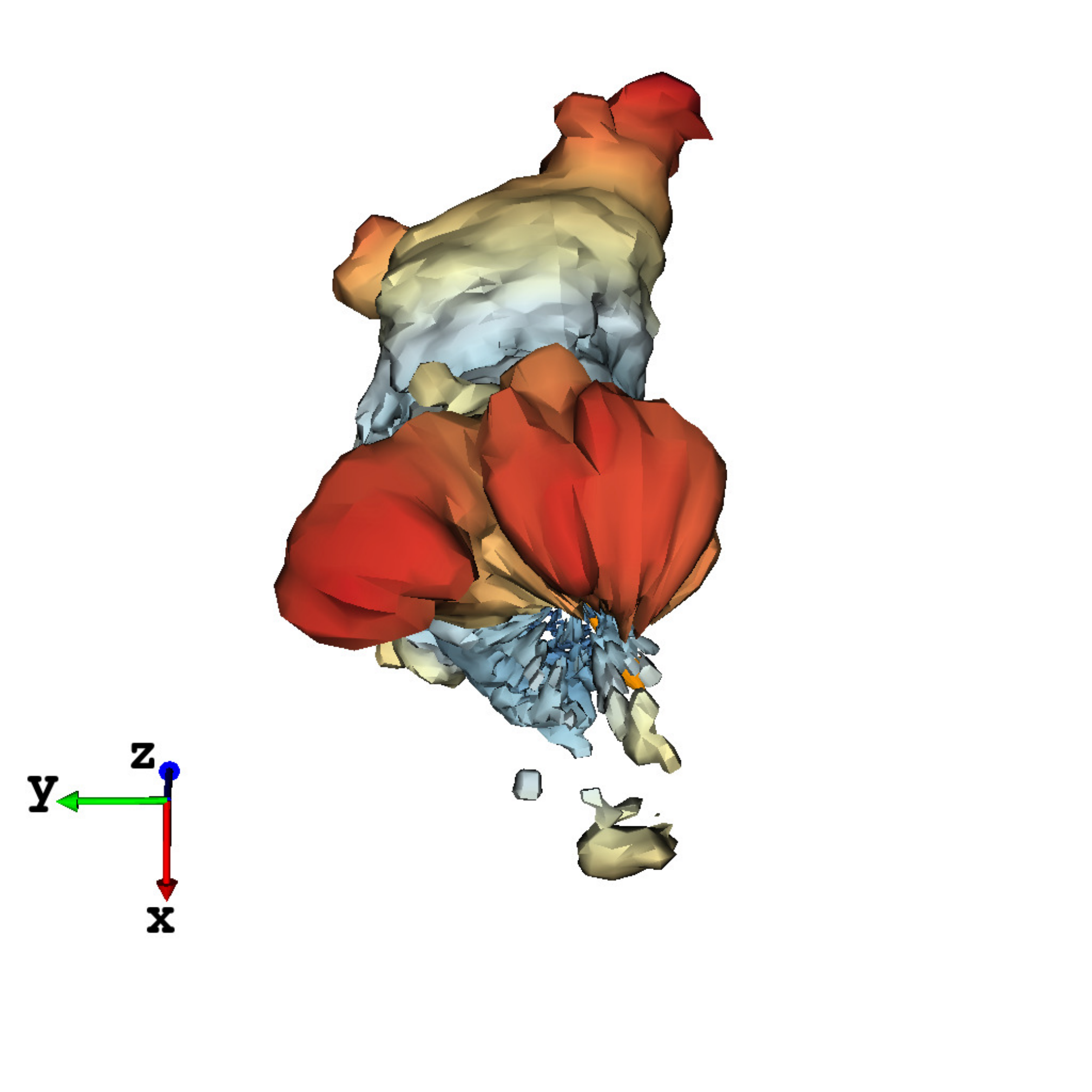}}
\resizebox{0.24\hsize}{!}{\includegraphics{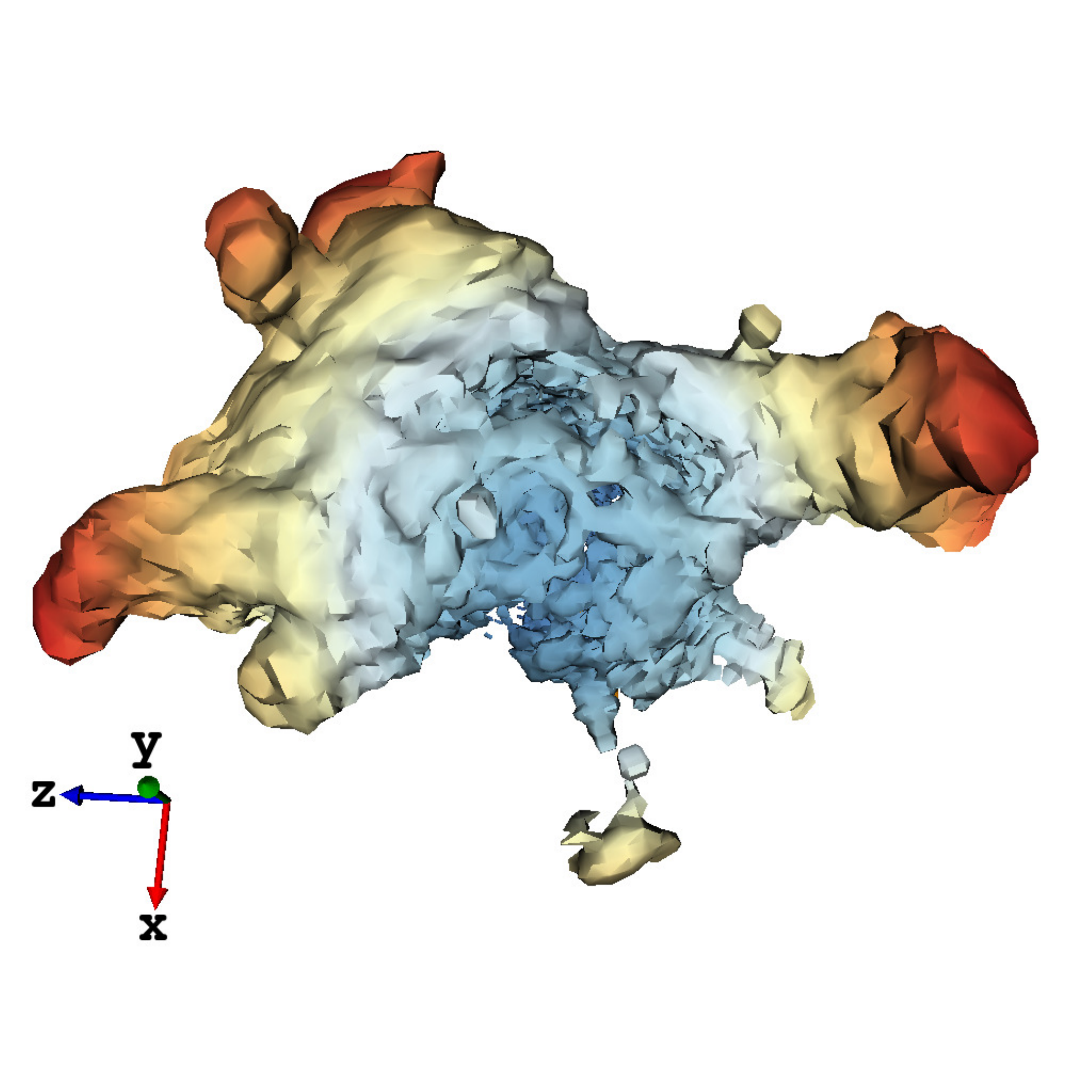}}\\
\resizebox{0.24\hsize}{!}{\includegraphics{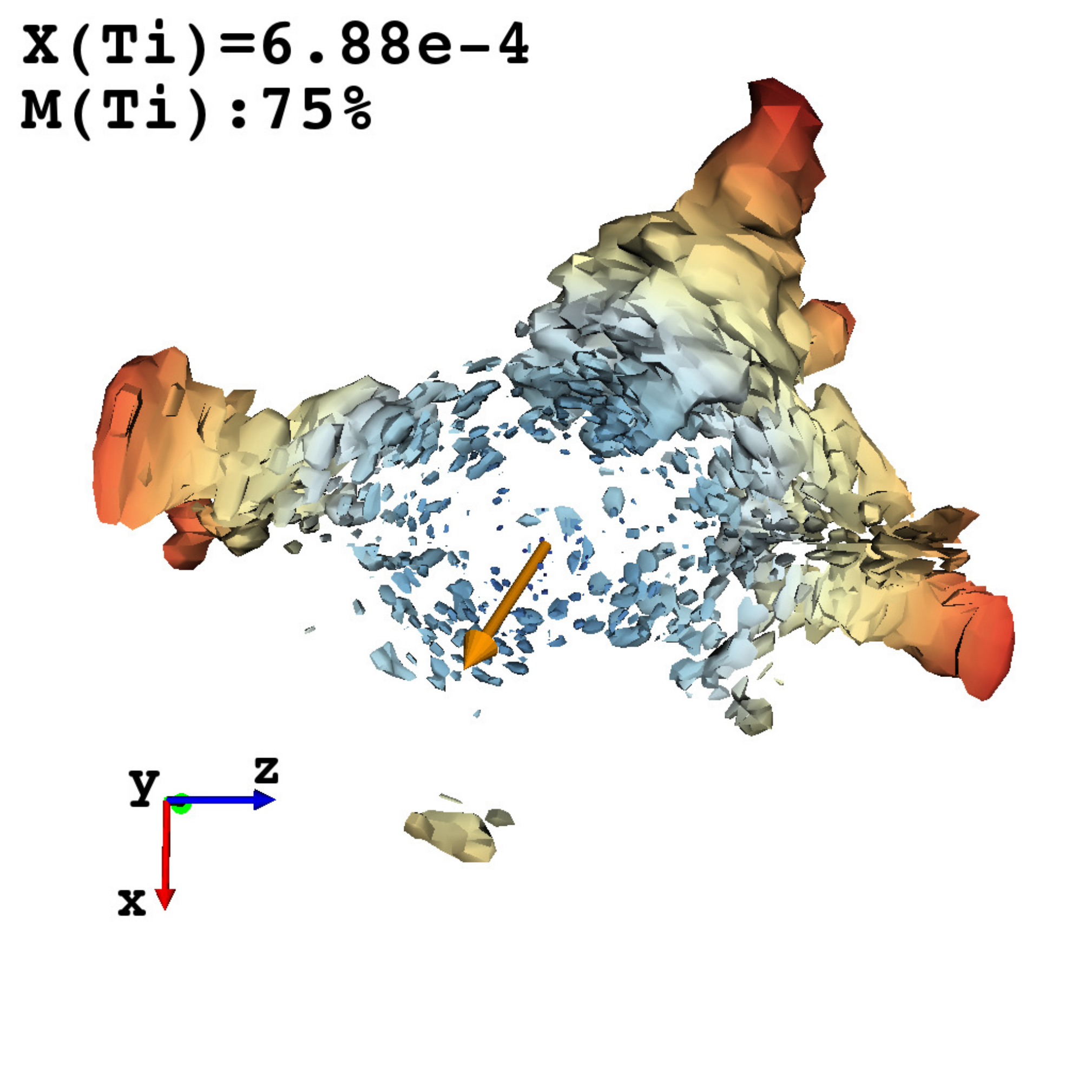}}
\resizebox{0.24\hsize}{!}{\includegraphics{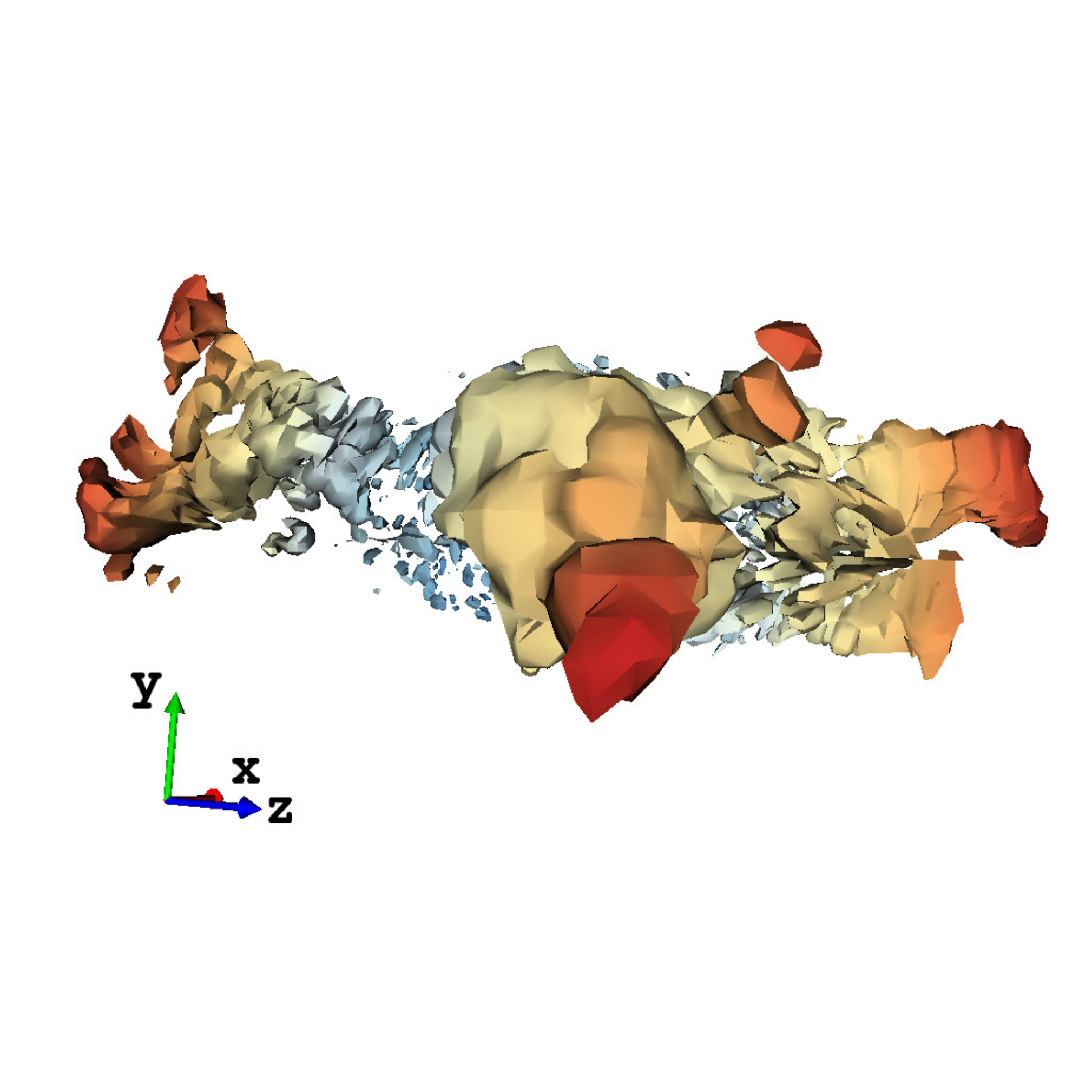}}
\resizebox{0.24\hsize}{!}{\includegraphics{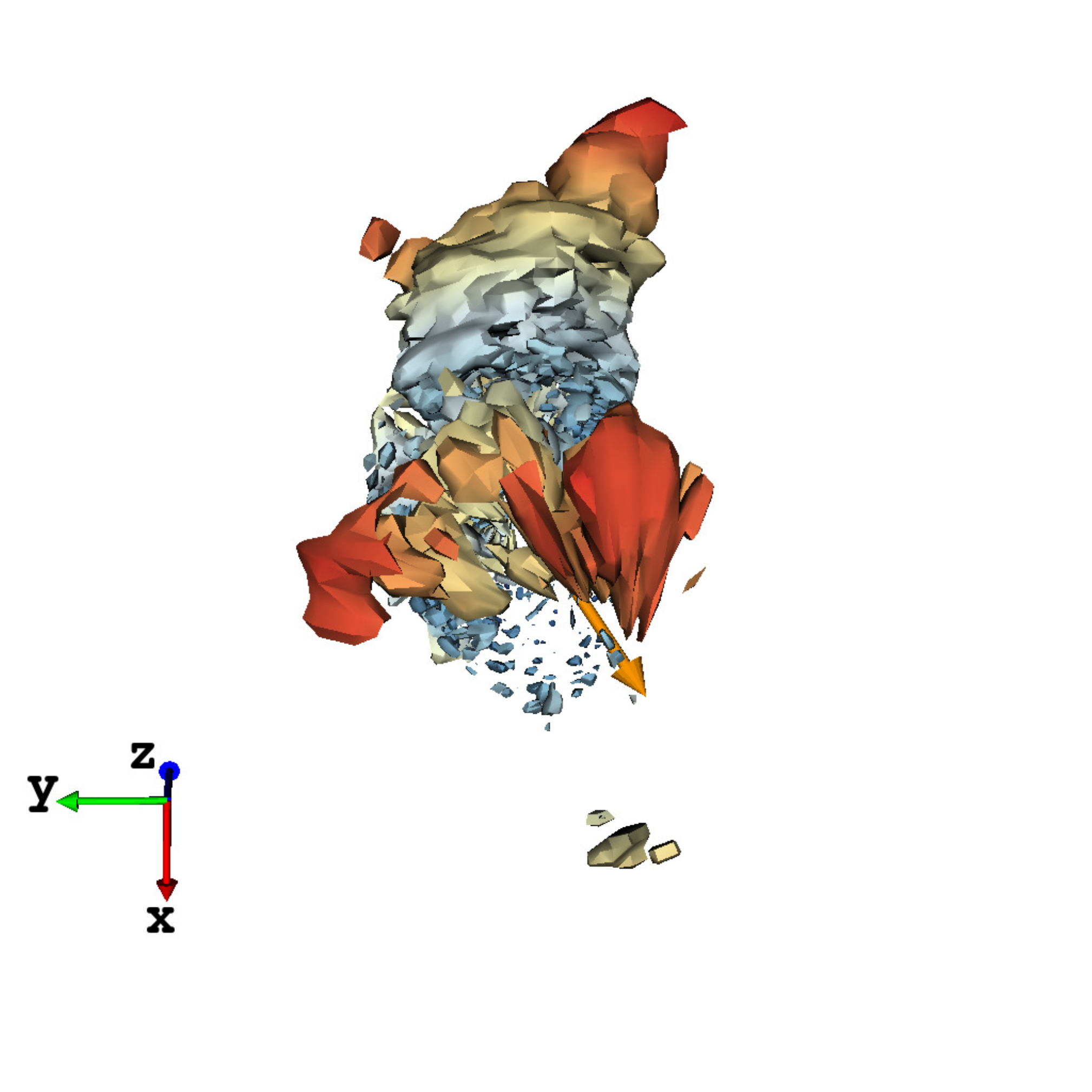}}
\resizebox{0.24\hsize}{!}{\includegraphics{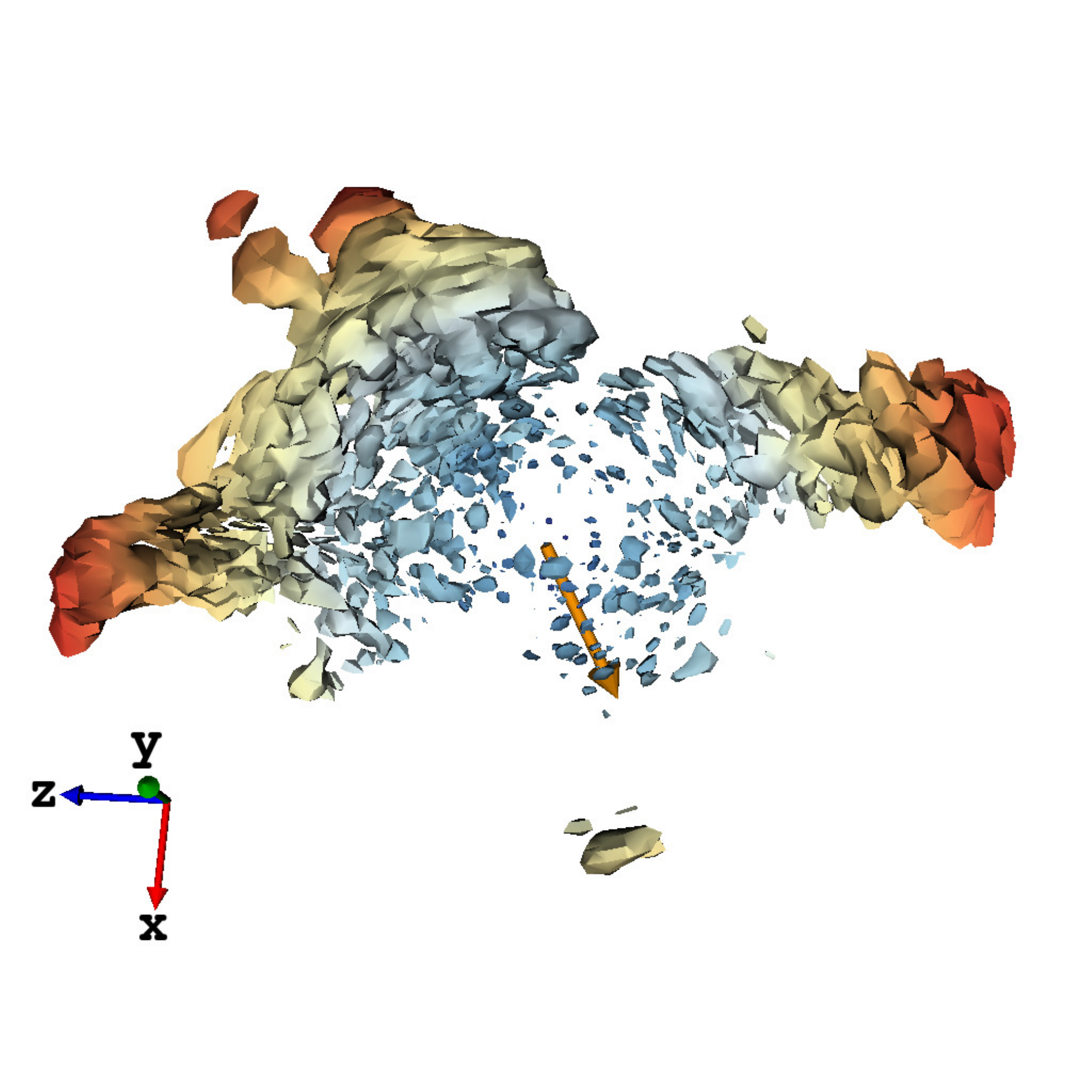}}
\caption{Isosurfaces of constant mass fractions of $^{56}$Ni
({\em upper row}) and $^{44}$Ti ({\em lower row}) from four
different viewing angles for model W15-2-cw-IIb at
$t_\trm{pb}=73940$\,s, based on the nucleosynthesis yields
obtained from our post-processing analysis of tracer particles.
In the case of nickel the isosurface enclosing 97\% of the
total ejected mass is shown, in the case of titanium we chose
the surface that encompasses 75\% of the total mass of the
nucleosynthesized material (see Fig.~\ref{fig:niti3D}).
The surfaces are color-coded by the radial velocities
according to the color bar given between the two left panels.
The NS kick direction is indicated by the orange arrow.
The images should be compared to the 3D
visualization of Cas~A (in particular the iron distribution,
see Fig.~\ref{fig:smithonian})
provided at {\tt http://3d.si.edu/explorer?modelid=45}.
Leaning on the convention used there,
the {\em left panels} display the ``front view'', which is
slightly inclined relative to the perspective shown in
Fig.~\ref{fig:niti3D} (as can be seen from the tripods on the left
side of both Figures). The {\em second panels} show the
``top view'', the {\em third panels} the ``left
view'', and the {\em right panels} provide the ``back view''.}
\label{fig:viewingangles}
\end{figure*}

In Fig.~\ref{fig:niti3D} we provide a closer comparison of the
3D distributions of $^{44}$Ti and $^{56}$Ni including volumetric 
information. The images of this Figure show, for both nuclear
species, isosurfaces corresponding to different values of the
mass fraction. These isosurfaces were determined such that they
enclose 50\%, 75\%, 90\%, and 97\% of the total mass ejected of
the considered nucleus.

The plots confirm our previous conclusions drawn on the basis 
of the mass distributions in velocity space
(Sect.~\ref{sec:veldistribution}): since
$^{44}$Ti and $^{56}$Ni are nucleosynthesized in close spatial
proximity in regions of (incomplete) silicon burning and the 
$\alpha$-particle-rich freeze-out (Sect.~\ref{sec:production}),
they are expelled in close connection and there is no process
at work that could decouple or decompose them in the ejecta.
Their distributions therefore closely resemble each other and
the two nuclei, overall, trace the same 3D geometry.

The bulk of the nickel and titanium is concentrated in relatively
small, highly enriched clumps and knots that contain half of 
the ejecta masses of these nuclei (see left panels of
Fig.~\ref{fig:niti3D}). The majority of these clumps expands
away from the center of the explosion in the hemisphere 
opposite to the direction of the NS motion. A close inspection
of the two left panels in comparison reveals that the lumps
containing 50\% of the ejected
$^{44}$Ti are considerably more extended
than those of $^{56}$Ni, meaning that titanium is clearly 
more diluted. Moreover, one can find regions of high $^{56}$Ni
concentration and little $^{44}$Ti and vice versa. Note that
all of the clumps with radial velocities of
$\lesssim$4000\,km\,s$^{-1}$ (blue and light blue colors)
in the left two columns
of Fig.~\ref{fig:niti3D} are located within
the volume of the inner sphere that is assumed
not to be reverse-shock heated in the right panel of 
Fig.~\ref{fig:combmap}. The radius of this sphere is about
half the distance from the center to the outermost tips of
the largest Ni and Ti fingers. The maximum velocities of titanium
in this inner, unshocked sphere are in the ballpark of the 
fastest material seen by NuSTAR \citep[][]{Grefenstetteetal14}.
Our model suggests that a considerable amount of $^{44}$Ti
should exist in Cas~A with higher velocities outside of the 
reverse-shock radius, which is in line with the 3D data
published recently by \citet{Grefenstetteetal17}.

The isosurfaces enclosing 90\% and 97\% of the ejecta masses for 
the two nuclei (right two columns in Fig.~\ref{fig:niti3D})
exhibit differences only in details. One example 
is a larger nickel feature at the 3:30 o'clock
position (i.e., right of the root point of the kick vector)
that contains little titanium. This feature is even
better visible in the second column of Fig.~\ref{fig:niti3D}.
Another example are the tips of the most extended, longest
finger-like structures, which are more inflated and show
higher velocities for titanium. Indications of this trend 
can already be observed in the images of the left and second
columns.

The series of plots in both rows also show significant
amounts of low-velocity radionuclei distributed more
homogeneously and in a more dilute,
volume-filling manner in the central part of the 3D 
structure formed by the $^{44}$Ti and $^{56}$Ni ejecta.
Again, most of this material expands away from the center
in the hemisphere pointing opposite to the NS kick. 
A major exception is the big feature between the six o'clock
direction and the 9:30 o'clock direction, which mostly shares 
the same hemisphere with the NS kick-velocity vector. One
might diagnose a more knotty $^{44}$Ti NuSTAR map 
instead of the more homogeneous distribution of titanium 
with radial velocities of $\lesssim$4000\,km\,s$^{-1}$.
Note, however, that our simulations do not account for the
long-time effects of radioactive decay heating of nickel
and titanium, which might have an influence on porosity
and clumping in the titanium distribution 340 years later.


\begin{figure*}
\centering
\resizebox{0.40\hsize}{!}{\includegraphics{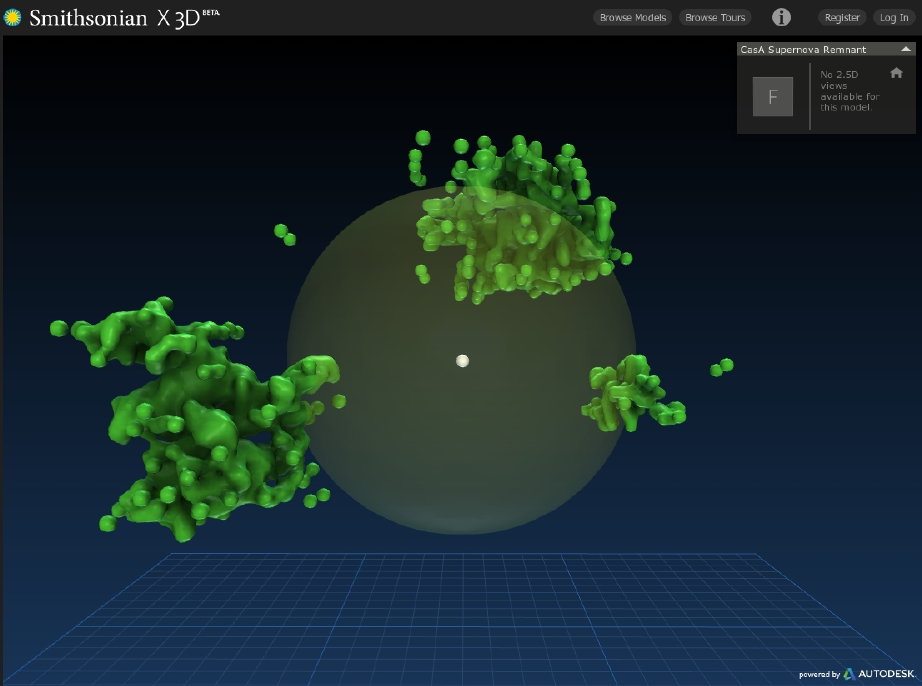}}
\hspace{20pt}
\resizebox{0.40\hsize}{!}{\includegraphics{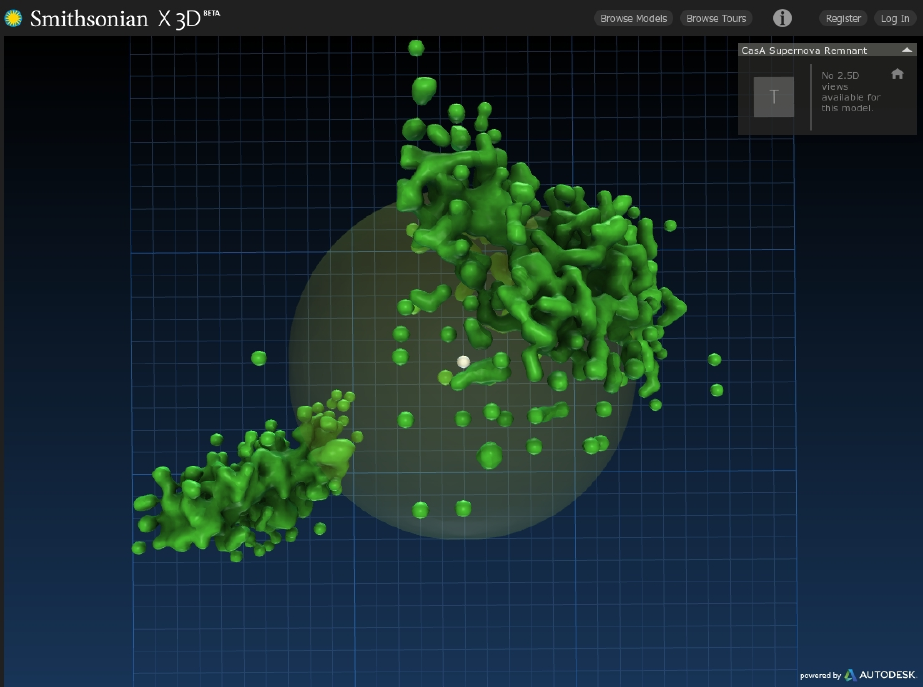}}\\
\vspace{5pt}
\resizebox{0.40\hsize}{!}{\includegraphics{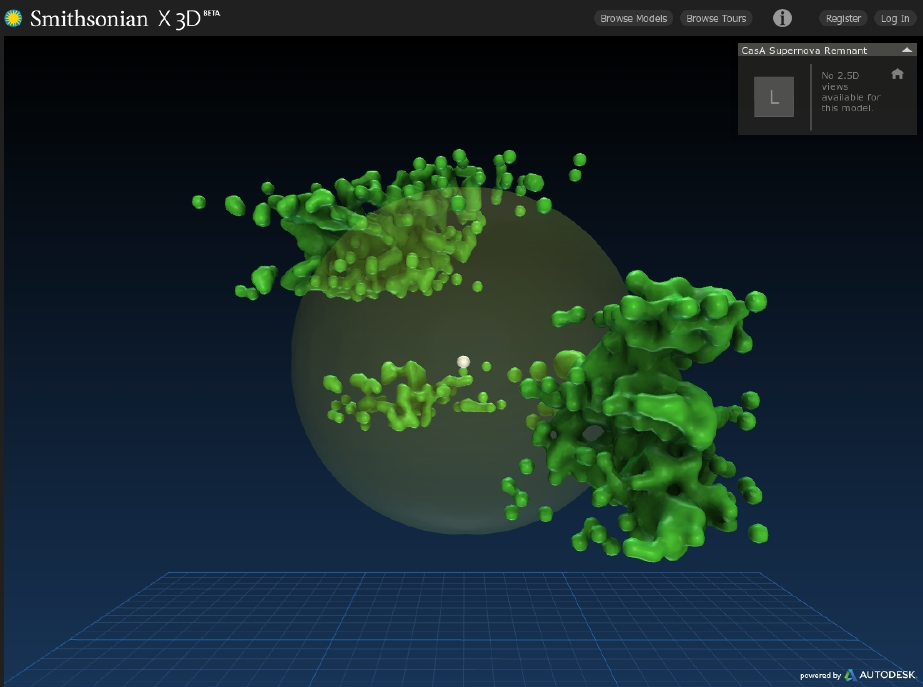}}
\hspace{20pt}
\resizebox{0.40\hsize}{!}{\includegraphics{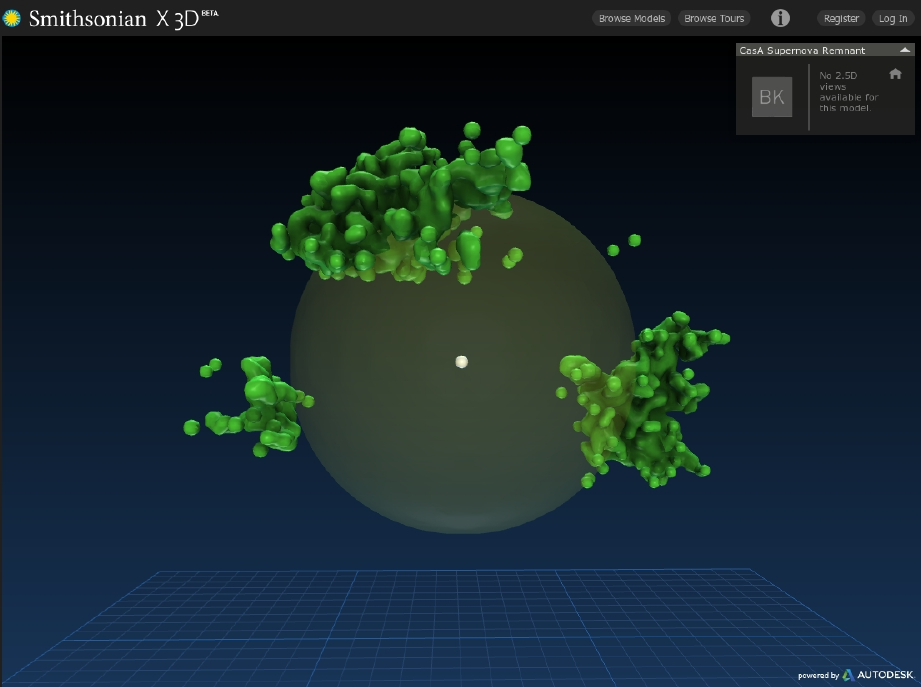}}
\caption{
Iron K-shell emission around the reverse-shock sphere of the
Cas~A SN remnant as measured by the Chandra X-ray observatory
\citep{Delaneyetal10,HwangLaming2003}
according to a 3D visualization available at
{\tt http://3d.si.edu/explorer?modelid=45}. Together with other
material (Si, Ar, Ne, S, O) the iron is assembled in a thick, 
ring-like belt girding the reverse-shock sphere (see the 
images for the combined composition information on the 
mentioned website). The NS is shown as
white ball at the center. The plots (from {\em top left} to
{\em bottom right}) give the ``front view'', ``top view'',
``left view'', and ``back view''. The three regions of high
iron concentration have remarkable similarities to the three
extended ``fingers'' of iron-rich and $^{44}$Ti-rich matter 
visible in model W15-2-cw-IIb (Figs.~\ref{fig:niti3D}, 
\ref{fig:viewingangles}),
although the neutrino-driven explosion of the latter model 
was not tuned to reproduce the properties of Cas~A.
}
\label{fig:smithonian}
\end{figure*}

\subsection{Comparison with 3D observational data}
\label{sec:comparisontoobservations}

The ``three-finger'' geometry of the radioactive ejecta
displayed in Figs.~\ref{fig:niti3D} and \ref{fig:viewingangles}
reminds one of the three
extended iron-rich structures \citep[called ``shrapnels'' by][]
{Orlandoetal16} that are seen outside of the 
reverse-shock radius in Cas~A \citep[e.g.,][]{Hwangetal04}.
In order to explore this morphological resemblence of our
3D model W15-2-cw-IIb with Cas~A in more detail, we compare
the properties of the innermost SN ejecta with the 3D
visualization of the observational data of Cas~A available
at the Smithsonian X3D website
{\tt http://3d.si.edu/explorer?modelid=45} 
(see Fig.~\ref{fig:smithonian}) and at
{\tt https://www.cfa.harvard.edu/$\sim$dmilisav/\-casa-web\-app/model.html}.
Our main focus in this Section is on the basic aspects of
the $^{56}$Ni (or, after decay, Fe) distribution, 
which carries direct information
on the geometry of the explosion in its earliest phase.

In Fig.~\ref{fig:viewingangles} we present images of the 
iron and $^{44}$Ti distributions from four different 
observational perspectives corresponding to the ``front view''
(left panels), ``top view'' (second column), ``left view''
(third column), and ``back view'' (right panels) defined by
the Smithsonian visualization (Fig.~\ref{fig:smithonian}). 
For optimal resemblance we
chose the front view similar to the observer direction of
Figs.~\ref{fig:combmap} and \ref{fig:niti3D}, however with a 
slightly different inclination of the NS kick vector relative
to the line-of-sight. Instead of being perpendicular to the
line-of-sight as in the previous Figures, the kick vector 
points toward the observer and out of the plane of the page 
with an angle of about 37$^\circ$. We note in passing that
the corresponding velocity component of the NS motion 
perpendicular to the line-of-sight is around 460\,km\,s$^{-1}$.
This is in the ballpark of the transverse velocity of the NS
in Cas~A, which was estimated to be about 350\,km\,s$^{-1}$
\citep{Fesenetal06} and thus typical of young pulsars
\citep[e.g.,][]{LyneLorimer94,Arzoumanianetal02}. 

The large angle of 53$^\circ$ between the NS kick and the 
observer direction for the front view of our model
implies that the hemispheric asymmetry of the 
ejecta is mapped well from this perspective, where most of the
$^{44}$Ti and $^{56}$Ni mass are seen in the northern part while
the NS velocity vector points southward. The distribution of
the radionuclei along the line-of-sight is far less one-sided.
This can be concluded from the ``top view'' panels in the 
second column and the ``left view'' panels in the 
third column of Fig.~\ref{fig:viewingangles}, where one of
the lower images permits the view on the NS kick vector.
There are roughly similar amounts of matter moving toward the
observers (who are located in the $(-y)$-direction) and
away from them. This seems to be compatible with the
observed redshift of the $^{44}$Ti line in Cas~A. 
The corresponding bulk line-of-sight Doppler
velocity was determined to be 1100--3000\,km\,s$^{-1}$ 
\citep{Grefenstetteetal14}. \citet{Boggsetal15} estimated
a ``look-back'' redshift velocity (associated with the
finite lifetime of $^{44}$Ti and a longer light-travel time
from the far side of Cas~A) of $\sim$1400\,km\,s$^{-1}$
for the age and expansion speed of the SNR. They reasoned
that the look-back effect is consistent with the measured
redshift of the $^{44}$Ti line and that the spatially
integrated $^{44}$Ti spectrum alone does not support any
statistically significant asymmetry. Rather than having a
large intrinsic mass-distribution asymmetry along the 
line-of-sight, Cas~A reveals major asymmetry by the 
spatial brightness distribution of the $^{44}$Ti emission
\citep{Boggsetal15}. This look-back argument has been 
revised by the recent 3D analysis of \citet{Grefenstetteetal17}.
Nevertheless, the overall conclusions of these authors on the 
ejecta asymmetry in Cas~A are in line with those of
\citet{Boggsetal15} (see below).

Besides the similarities between our model and 
Cas~A discussed already, the different views in
Figs.~\ref{fig:viewingangles} and \ref{fig:smithonian} 
reveal yet another
resemblance. The three extended, iron and titanium containing
fingers lie, by chance, in the same plane. For this reason,
the ejecta fill a much narrower volume when looked at from
the top and left views (panels in the two columns in the 
middle of Fig.~\ref{fig:viewingangles}).
The configuration is reminiscent of Cas~A, where 
reverse-shock-heated metals (Si, Ar, Ne, S, O) are arranged 
in a thick, ring-like belt girding the reverse-shock sphere,
and the unshocked interior ejecta \citep[seen in infrared
via Si~II;][]{Delaneyetal10} are
apparently assembled in a ``tilted thick disk'' 
\citep{Grefenstetteetal17}. This outer belt 
also encompasses the iron-rich regions as well
as the fiducial, Si-and Mg-rich ``jets'' 
(see the 3D visualizations
of the Cas~A observations on the websites mentioned above). 
Since this peculiar structure of the Cas~A ejecta
is not perpendicular to the ``jets'' but, quite the
contrary, its central plane cuts the northeast and southwest
``jets'' of Cas~A, rapid rotation seems highly unlikely as an 
explanation of the Cas~A morphology. Adding to this argument,
the NS kick in Cas~A is perpendicular to the NE-SW direction,
which contradicts expectations for a jet-driven explosion
\citep{Hwangetal04}.

Our long-time simulation demonstrates that the origin 
of the three iron
fingers in our (nonrotating) 3D model W15-2-cw-IIb
goes back to three dominant high-entropy plumes arising from
postshock convective instability at the time when the 
explosion was launched \citep[see left panel in the top row
of figure~7 of][]{Wongwathanaratetal15}. The morphological
resemblance to the iron ``shrapnels'' of Cas~A supports the
interpretation that this remnant carries clear fingerprints 
of the initial explosion asymmetries, and neither a
dynamically relevant amount of rotation nor a jet-driven
explosion need to be invoked to explain the overall structure
of the iron distribution in the Cas~A remnant.

After the submission of our paper, a 3D analysis of the
$^{44}$Ti distribution in Cas~A was posted on arXiv by
\citet{Grefenstetteetal17}.
In basic aspects their findings and conclusions are, 
again, in remarkably good agreement with 
our 3D results for model W15-2-cw-IIb as discussed 
above: 
\begin{itemize}
\item[(i)] The estimated initial mass of $^{44}$Ti, 
$(1.54\pm 0.21)\times 10^{-4}\,M_\odot$, with an Fe/Ti
ratio around 500 is in the 
ballpark of our model yields, although our nucleosynthesis
calculations suggest that the iron content of Cas~A
could be somewhat higher than inferred by 
\citet{Grefenstetteetal17}.
\item[(ii)] In excellent correspondence to our
simulated remnant geometry visualized in 
Figs.~\ref{fig:niti3D} and \ref{fig:viewingangles}, 
the average momentum of the $^{44}$Ti in Cas~A
(estimated on grounds of the flux-weighted 
average of the ejecta velocities) is oriented in the plane
of the sky almost precisely opposite to the direction
of motion of the central compact object, and the NS
is determined to have a significant line-of-sight
velocity component toward the observer, even with
an angle similar to the one shown in 
Fig.~\ref{fig:viewingangles}.
\item[(iii)] $^{44}$Ti knots near and exterior to the reverse
shock are associated with emission from shock-heated
iron (which should mostly be the product of radioactive
$^{56}$Ni synthesized along with $^{44}$Ti); while the
existence of titanium-rich regions without iron emission
exterior to the reverse shock is not finally assured,
the presence of shock-heated iron without observed 
titanium requires that the $^{44}$Ti yield is suppressed
relative to the iron yield by at least a factor of two.
This is easily compatible with predictions from our model,
for which a detailed analysis reveals that the color 
variations visible in Fig.~\ref{fig:combmap} correspond
to large-scale variations of the titanium-to-iron ratio
up to factors of 2--4 (and small-scale variations with
even larger amplitudes).
\item[(iv)] Significant amounts of 
$^{44}$Ti and $^{56}$Ni are diagnosed to lie within
the reverse shock of Cas~A, as suggested by our 
simulations. While our preferred model implies that $\sim$43\%
of the titanium could be in the unshocked inner sphere
(Fig.~\ref{fig:cumul}), \citet{Grefenstetteetal17} found
roughly 40\% to be located clearly interior to the reverse 
shock. But current observational constraints (in
particular of the part of the iron that is still hidden,
i.e.\ unshocked, 
at the age of the Cas~A remnant) are very uncertain,
and better information from explosion models requires
late-time simulations that follow the effects of 
radioactive decay heating by $^{56}$Ni and $^{44}$Ti
as well as the evolution of the reverse shock from the
ejecta interaction with the circumstellar medium.
\item[(v)] \citet{Grefenstetteetal17} did not see any 
evidence for $^{44}$Ti ejecta associated with the 
blueshifted half of the thick disk of interior, unshocked
ejecta, though redshifted titanium
regions were found to lie in the redshifted half
of the disk. However, the identity and location
of the blueshifted knot 20b are disputable as discussed
by \citet{Grefenstetteetal17} and mentioned in a 
footnote in Sect.~\ref{sec:veldistribution}. 
Maybe some part of this matter belongs to the 
blueshifted half of the thick disk.
Different from \citet{Boggsetal15},
\citet{Grefenstetteetal17} did not diagnose any
significant ``look-back'' effect (associated with
the difference in light-travel time between blueshifted
and redshifted ejecta) affecting the results, but 
concluded a moderate asymmetry of the $^{44}$Ti
distribution with some excess of redshifted material (mean
line-of-sight velocity of $920\pm 510$\,km\,s$^{-1}$).
This seems to be basically compatible with the asymmetric
distribution of titanium in our model, as reflected by
the line-of-sight component of the NS velocity toward the 
observer in Fig.~\ref{fig:viewingangles}.
\end{itemize}

In future work we plan to perform a closer comparison of 
our model predictions and the Cas~A data with respect
to a wider set of nucleosynthesis products and their spatially
inhomogeneous distribution, e.g.\ in the fiducial Si/Mg-rich,
high-velocity, wide-angle ``jets'' \citep[e.g.][]{Fesenetal06},
in the form of large cavities and ``bubbles'' in the unshocked
inner sphere \citep{MilisavljevicFesen15}, and as crown-like
and ring-like features (containing Ar)
that also surround the iron-rich 
``shrapnels'' in the Cas~A SNR's optical main shell 
\citep[e.g.][]{Fesenetal2001,Delaneyetal10,MilisavljevicFesen2013}.
For this goal we will continue
the present 3D explosion simulations toward the remnant stage.
Such calculations will have to include the effects of 
radioactive decay heating, which may inflate the regions
containing the radionuclei against their surroundings.
They will also have to take into account
the reverse shock from the ejecta-environment interaction,
which will have an impact on the radial velocity profile
and is likely to modify the detailed shape of the shock-compressed
ejecta regions.

\section{Conclusions and discussion}
\label{sec:conclusions}

In this work we presented 3D simulations of a neutrino-driven
SN explosion with remarkable resemblance to basic
properties of Cas~A. The discussed model (W15-2/W15-2-cw)
was one of the cases
studied by \citet{Wongwathanaratetal13,Wongwathanaratetal15}.
Neither the progenitor star, a nonrotating 15\,$M_\odot$ 
model (whose hydrogen envelope we removed by hand down to a 
small rest of $\sim$0.3\,$M_\odot$ to account for the SN~IIb
case of Cas~A), nor the explosion 
parameters were iterated for optimal agreement with the
observations. The morphological similarities between the
simulation results and Cas~A are consequences of stochastic
processes (convective overturn and SASI activity) that take 
place in the SN core at the onset of the explosion. Since 
the nonradial hydrodynamic instabilities grow from small,
random, initial seed perturbations in a chaotic way, the 
final explosion geometry is not under control by input 
parameters of the simulation, and larger sets of 3D calculations
may be needed to obtain results of very close similarity
for other progenitor stars or different explosion energies.

Our 3D simulations were started shortly after core bounce, and
the accretion phase and beginning of the SN explosion until
1.3\,s were computed with a simplified, gray treatment of the
neutrino transport and with prescribed, time-dependent neutrino
luminosities at the inner grid boundary. Due to its free 
parameters this neutrino engine allowed us to tune
the energy of the neutrino-driven explosion to a chosen value.
Although approximative, the description of the neutrino effects
was sufficiently realistic to adequately capture the development
of hydrodynamic instabilities in the neutrino-heating layer
and the subsequent evolution of
asymmetries and mixing instabilities during the SN explosion.
The discussed 3D simulations followed the evolution until
about one day, at which time the expansion of the ejecta
already closely approached a homologous state in the 
investigated SN~IIb-like explosion.

Based on tracer-particle trajectories, we
post-processed our 3D model with respect to the 
nucleosynthesis of radioactive nuclei, focusing 
particularly on $^{44}$Ti and $^{56}$Ni, which decay to stable
$^{44}$Ca and $^{56}$Fe, respectively. 
These radionuclides are especially 
interesting for a comparison with Cas~A, because they are
assembled near the very center of the explosion and therefore
carry imprints of the asymmetries that play a role for the
physical mechanism triggering the blast wave of the SN.

Our main findings can be summarized as follows:

\smallskip\noindent
{\em 1. The production of radionuclei.}\\
The explosion with an energy of $\sim$1.5\,B and an ejecta 
mass of about 3.3\,$M_\odot$ produces
between 0.043\,$M_\odot$ and 0.96\,$M_\odot$ of $^{56}$Ni and
between $1.6\times 10^{-5}$\,$M_\odot$ and 
$1.6\times 10^{-4}$\,$M_\odot$ of $^{44}$Ti. The $^{44}$Ti
yield is 50\% higher than these values (up to
$\sim$$2.4\times 10^{-4}$\,$M_\odot$) when our standard rate of
the $^{44}$Ti$(\alpha,p)^{47}$V reaction from \citet{Cyburtetal10}
is reduced by a factor of two as suggested by recent experimental 
results of \citet{Margerinetal14}.
Major contributions to the $^{56}$Ni and $^{44}$Ti nucleosynthesis
come from $\alpha$-particle-rich freeze-out in neutrino-processed,
high-entropy ejecta. 
The uncertainties of the nickel and titanium yields are
connected to uncertainties of the electron fraction $Y_e$ in
this matter. The lowest
yields are based on $Y_e$-values slightly less than 0.5, as
computed with our approximative neutrino handling; the upper
bounds for the yields were obtained when we adopted
$Y_e = 0.5$ as present in the progenitor star. For values
slightly higher than 0.5, one can still expect efficient
creation of $^{56}$Ni but noticeably less $^{44}$Ti,
whereas the production of both $^{56}$Ni and $^{44}$Ti
drops steeply below $Y_e = 0.5$
\citep{Magkotsiosetal10,Wanajoetal13}.
A large production of $^{44}$Ti therefore hinges sensitively
on the $Y_e$ of the expelled matter, and is possible in 
principle because neutrino-driven explosions genericly eject
considerable amounts of matter with a favorable range of 
entropies (10\,$k_\mathrm{B}$$\lesssim s\lesssim$30\,$k_\mathrm{B}$
per nucleon). 

It is important to note that the neutron-to-proton ratio in
neutrino-processed ejecta can only be determined with accurate,
energy-dependent neutrino transport. However, even with the 
best numerical treatment, considerable uncertainties
prevent reliable theoretical predictions at the present 
time. Since $Y_e$ in neutrino-heated SN outflows
is set by a delicate competition of $\nu_e$ and 
$\bar\nu_e$ absorptions (and initially also the inverse 
emission processes), a variety of factors can have important
influence, easily leading to shifts of $Y_e$ by a few percent 
above or below 0.5. Examples of such sensitive effects include,
for example, the detailed and progenitor-dependent
flow dynamics, which determines
the expansion time scale and thus decides how long the ejecta 
can stay in reactive equilibrium with respect to the mentioned
processes; remaining uncertainties of the neutrino 
opacities at high densities interior to the neutrinospheres,
which could modify the radiated luminosities and spectra of
$\nu_e$ and $\bar\nu_e$ relative to each other and thus
the neutrino absorption at larger radii;
incompletely understood physics phenomena such as 
neutrino oscillations, in particular collective flavor
transformations \citep[for a review, see, e.g.,][]{Mirizzietal16};
or the new lepton-emission self-sustained asymmetry 
\citep[LESA;][]{Tamborraetal14}, which could lead to considerable
changes of the neutrino emission from SN cores.

For a somewhat higher
explosion energy and a moderately larger ejecta mass than
considered in our simulation, both of
which seem likely for Cas~A \citep[$E_\mathrm{exp}\approx 2.3$\,B,
$M_\mathrm{ej}\approx 4\,M_\odot$;][]{Vink2004,Orlandoetal16},
neutrino-driven explosions produce higher nickel yields
\citep[see the correlation in 
figure 17 of][]{Sukhboldetal16}. Since a higher
explosion energy goes hand in hand with a larger mass of 
high-entropy ejecta, also more titanium can be
expected, provided $Y_e$ in the neutrino-heated matter
stays in a favorable, narrow interval around 0.5. We stress again
that we did not seek for a perfect quantitative match of Cas~A
observations and estimates here, but our discussion was 
focussed on a proof-of-principle.

We conclude that neutrino-driven explosions, without the need
to invoke rapid rotation or (jet-induced) bipolar deformation,
can well account for the initial
$^{44}$Ti masses and titanium-to-nickel mass-ratios (around
$10^{-3}$) deduced from observations of Cas~A and SN~1987A.
For the former, current measurements give a $^{44}$Ti yield of
$\sim$(1.0\,...\,1.7)$\times 10^{-4}$\,$M_\odot$ 
\citep{Grefenstetteetal14,Grefenstetteetal17,Siegertetal15,Tsygankovetal16,WangLi16}, 
whereas in the case of SN~1987A the recent determinations of the
expelled amount of $^{44}$Ti exhibit a wider spread, namely between
$(0.55\pm 0.17)\times 10^{-4}$\,$M_\odot$ \citep{Seitenzahletal14}
and roughly (2\,...\,$4)\times 10^{-4}$\,$M_\odot$ 
\citep{Grebenevetal12}, although the value was narrowed down to 
$(1.5\pm 0.3)\times 10^{-4}\,M_\odot$ by \citet{Boggsetal15}
\citep[for a result consistent with that, see][]{Jerkstrandetal11}. 

The rarity of pointlike sources in the Galaxy
that produce of order $10^{-4}$\,$M_\odot$ of $^{44}$Ti,
as inferred from an analysis of the observed gamma-ray
sky compared to theoretical expectations \citep{Theetal06,
Renaudetal06,Tsygankovetal16},
implies that only a small fraction of stellar explosions
eject sufficient amounts of matter with $Y_e$ close to 0.5 and 
temperatures and entropies high enough for efficient $\alpha$-rich 
freeze-out nucleosynthesis of $^{44}$Ti. 
In the context of neutrino-driven explosions, considerable
masses of high-entropy ejecta are a generic property,
but smaller differences of the explosion dynamics or
neutrino-emission characteristics, which could vary with the 
progenitor or the physics of the nascent NS, 
may be responsible to provide or prevent 
conditions where atypically large amounts of $^{44}$Ti
can be synthesized. Cas~A and SN~1987A, which both originated
from stars in the 15--20\,$M_\odot$ range, seem to be
representative of such atypical collapse events, in which 
special and yet undetermined conditions allowed $Y_e$ in the
neutrino-processed, high-entropy matter to stay in the 
favorable vicinity of 0.5. Independent of the physics 
needed to achieve this, large amounts of $^{44}$Ti 
as observed in Cas~A and SN~1987A indicate sizable
ejecta masses with entropies between
$\sim$10\,$k_\mathrm{B}$ and $\sim$30\,$k_\mathrm{B}$ per 
nucleon. This may be interpreted as a characteristic 
fingerprint that is expected for SN explosions powered 
by energy stored in neutrino-heated matter. In this context
it is notable that SN~1987A as well as Cas~A were fairly 
energetic explosions, both having energies toward the high
side of the range that may be explained by the neutrino-driven
mechanism.

\smallskip\noindent
{\em 2. The velocities of radioactive ejecta.}\\
The remnant of Cas~A resulted from an asymmetric Type~IIb SN
\citep{Krauseetal08,Restetal11}. For this 
reason we repeated explosion calculations for our model
W15-2/W15-2-cw after removing the hydrogen envelope of the
original red supergiant progenitor except for a minor rest of 
$\sim$0.3\,$M_\odot$. The new simulation was named W15-2-cw-IIb.

The absence of a massive hydrogen envelope had two important 
consequences. First, for the same explosion energy, the bulk
of the ejecta (dominated by the helium shell) 
was expelled with significantly higher velocities
compared to the red supergiant. This could be expected because
of the scaling of the average ejecta velocity with 
$\sqrt{E_\mathrm{exp}/M_\mathrm{ej}}$. Without the deceleration
by a reverse shock from the He/H interface, also the innermost
SN ejecta as traced by the radionuclei of $^{44}$Ti and $^{56}$Ni,
could retain much higher mean bulk velocities up to roughly 
7000\,km\,s$^{-1}$ (and smaller amounts of material being
even faster), close to the velocity of the fastest 
$^{44}$Ti material reported 
\citep{Grefenstetteetal14,Grefenstetteetal17,Siegertetal15}.

The second consequence of the stripped nature of the
SNIIb progenitor was the preservation of large,
coherent iron and titanium ``clumps'' or plumes, which 
in the case of SN-shock deceleration by a massive H-envelope
would have fragmented because of a RT unstable
density inversion near the He/H interface; see 
\citet{Kifonidisetal03} for a detailed discussion and
3D results for the red supergiant simulation of model W15-2-cw
in \citet{Wongwathanaratetal15}.

Because the progenitor of Cas~A did not possess a massive
hydrogen envelope, the asymmetric Ni and Ti-rich structures 
imposed by the explosion
mechanism at the very beginning of the SN blast could 
survive intact, with much less perturbative influence of secondary
hydrodynamic instabilities that develop at the composition 
interfaces of the progenitor after the passage of the outgoing
SN shock. As shown by \citet{Wongwathanaratetal15} and in
our simulation of model W15-2-cw-IIb, the 
RT unstable layer at the C-O/He boundary leads
to considerably less fragmentation of the initial structures
than the instability at the He/H transition. Therefore our
results support speculations by \citet{Hwangetal04} and
\citet{Vink2004} that
Cas~A is a very young remnant whose morphology still carries 
the imprints of the asymmetries at the onset of the explosion.
This property makes Cas~A an extremely precious test case for
studying the explosion mechanism of core-collapse SNe.

\smallskip\noindent
{\em 3. 3D spatial distribution of the radionuclei.}\\ 
Basic morphological features of the iron and $^{44}$Ti ejecta
in our model W15-2-cw-IIb exhibit amazing similarity to
overall properties of the spatial distribution of these
elements in Cas~A. 

The production (by (in)complete Si-burning and the 
$\alpha$-particle-rich freeze-out process)
and distribution of $^{44}$Ti are closely linked
to those of $^{56}$Ni. Most of the mass of these radionuclei
is condensed in widely distributed clumps and knots of different
sizes, which contain high mass fractions [$X$($^{56}$Ni),
$X$($^{44}$Ti)] of more than $\sim$40--50\%
for iron and more than roughly 0.1\% for titanium.
Considerable spatial variations of the relative mix of iron
and titanium can occur, and clumpy
structures exist with high concentrations of $^{44}$Ti
but comparatively small amounts of admixed iron, and vice versa.
On large spatial scales, $X(^{44}\mathrm{Ti})/X(^{56}\mathrm{Ni})$
can vary between $\sim$0.001 and $\sim$0.004, on small scales
even with larger amplitudes.
The central regions of our simulated explosion are more
homogeneously filled with lower-velocity material of the
two nuclear species.

Most of the synthesized $^{44}$Ti and $^{56}$Ni mass in
model W15-2-cw-IIb are expelled in the hemisphere opposite
to the NS kick direction, although a bigger mass concentration
shares the same hemisphere with the kick vector. This model
feature, once more, is similar to the situation observed
in Cas~A, where the NS moves in a direction pointing away
from the region where most of the $^{44}$Ti clumps are 
located (compare figure~2 of \citealt{Grefenstetteetal14}
with Fig.~\ref{fig:combmap} for our model and the 3D 
$^{44}$Ti distribution analyzed by 
\citealt{Grefenstetteetal17} with our Figs.~\ref{fig:niti3D} 
and \ref{fig:viewingangles}). This observation is
perfectly compatible with expectations on grounds of the 
gravitational tug-boat mechanism, which predicts the NS
to be recoiled by explosion asymmetries and to be 
accelerated opposite to the direction of the stronger 
SN blast wave, i.e., away from the hemisphere where 
elements from silicon to the iron group, including
$^{44}$Ti, are produced more efficiently by explosive
nucleosynthesis \citep{Wongwathanaratetal13}.

Three big, iron-rich fingers with mushroom-head-like
tips (see Figs.~\ref{fig:niti3D} and \ref{fig:viewingangles})
could have their correspondence in the three,
extended regions seen in the Fe~K-shell emission outside of
the reverse shock in Cas~A (Fig.~\ref{fig:smithonian}). 
The fact that these fingers
in our model, by chance, lie essentially in the same plane,
leads to an ejecta structure that is considerably less
extended perpendicular to this plane than in the plane
itself. This geometry strongly resembles the spatial
arrangement of the iron ``shrapnels'' \citep{Orlandoetal16}
and of intermediate-mass elements (Si, Ar, Ne, S, O) within
a thick, ring-like belt around the reverse-shock sphere of
Cas~A \citep{Delaneyetal10,MilisavljevicFesen2013}. 
Closest resemblance between model and observations is
found when the plane of the three iron fingers in our
simulation is oriented roughly perpendicular to the 
line-of-sight. This implies that a considerable component 
of the NS space velocity could lie along the line-of-sight
toward the observer (Fig.~\ref{fig:viewingangles}).

Based on their recent, spatially
resolved spectroscopic analysis of the 3D distribution of
$^{44}$Ti in Cas~A, \citet{Grefenstetteetal17} came to
the same conclusion and described an ejecta and NS kick 
geometry with very close similarity to our model results.
Also other main findings of \citet{Grefenstetteetal17}
are in very nice agreement with our simulations. 
In particular, this
recent work eliminates tension between some observational
aspects reported by \citet{Grefenstetteetal14} and the
theoretical expectations. $^{44}$Ti is now
observed clearly interior to, near, and exterior to the 
reverse shock of Cas~A, and shock-heated iron is found 
where titanium is observed in the shocked shell. Reversely,
the fact that some iron-rich regions are not associated
with a detection of $^{44}$Ti does not pose a major 
problem. Since the model-predicted mass-fraction ratio, 
$X(^{44}\mathrm{Ti})/X(^{56}\mathrm{Ni})$, varies 
between $\sim$0.001 and $\sim$0.004 on large spatial 
scales,\footnote{Iron-group species different from
$^{56}$Ni are synthesized abundantly especially in
shock-heated ejecta (i.e., via incomplete silicon 
burning) and can also be formed in the neutrino-heated
ejecta under $Y_e$ conditions that do not allow for
abundant $^{44}$Ti synthesis. These non-$^{56}$Ni nuclei 
contribute several 
$10^{-2}$\,$M_\odot$ or globally about 25\% (locally
up to $\sim$35\%) of the total iron yield 
of our model W15-2-cw-IIb. They may enhance the
spatial variations of the Ti/Fe ratio 
\citep{Grefenstetteetal17}, although
efficient mixing tends to work in the opposite direction
and tries to erase pronounced differences of
the $X(^{44}\mathrm{Ti})/X(^{56}\mathrm{Ni})$ and
$X(^{44}\mathrm{Ti})/X(\mathrm{iron})$ distributions
in extended volumes.}
it is well able to account for the suppression of
the production of $^{44}$Ti by at least a factor of two in
some iron-rich regions as concluded by \citet{Grefenstetteetal17}.

While the NuSTAR data of \citet{Grefenstetteetal14} were
interpreted to imply that at least 80\% of the observed
$^{44}$Ti emission are contained within the reverse-shock
radius of Cas~A as projected on the plane of the sky,
\citet{Grefenstetteetal17} were now able to refine this 
picture. Based on the present-day flux they estimated that
40\% of the $^{44}$Ti mass are clearly interior to the reverse
shock, 40\% are at or near the reverse-shock radius, and
roughly 20\% are clearly exterior to it. This is nicely
compatible with expectations from our model W15-cw-IIb,
which predicts $\sim$43\% of the $^{44}$Ti to be associated
with the slower half of the iron (below about
4000\,km\,s$^{-1}$), if we assume that up to as much as
50\% (or up to $\sim$0.1\,$M_\odot$)
of the total Fe yield of the Cas~A SN could reside in
the remnant's unshocked central volume 
\citep[e.g.,][]{MilisavljevicFesen15,Orlandoetal16}.
\citet{Grefenstetteetal17} estimated that only about 
0.02\,$M_\odot$ of the iron are ``hidden'' in the unshocked
interior of Cas~A. This value appears to be on the low
side compared to the previous estimates and also compared
to our model results. However, one should keep in mind 
that it is based on the assumption of a model-dependent
and uncertain (constant) Fe/Ti ratio of 500.

\smallskip
Although our current 3D SN explosion model is not fully
self-consistent, but invokes a parametrized neutrino engine,
the presented results are assuring and may, cautiously,
be considered 
as support of the possibility that Cas~A is the remnant of
a neutrino-driven explosion with its generic hydrodynamic
instabilities. Rotation does not seem to be needed to
explain basic morphological properties of Cas~A, and some
of the discussed observations even disfavor its
relevance. The fiducial ``jet'' and ``counter-jet'' are
spatially connected to the thick metal belt that girds
the reverse-shock sphere, and the direction of the NS kick
is perpendicular to the NE-SW direction defined by the jets.
Moreover, the jets are Si-rich but do not contain much
iron, are very wide \citep{Fesenetal06,MilisavljevicFesen2013},
and their kinetic energy \citep[of order $10^{50}$\,B;][]
{Lamingetal2006,FesenMilisavljevic16} is too low to
attribute to them a crucial
role in the explosion mechanism. All of these facts are
not compatible with theoretical expectations for a classical
``jet-driven'' explosion with the axis in the NE-SW direction
\citep[e.g.,][]{Khokhlovetal1999,Lamingetal2006,Wheeleretal2008,Hwangetal04,Isenseeetal2010,Grefenstetteetal14}.

Despite the reported first successes in understanding the
origin of some basic properties and morphological
peculiarities of Cas~A, a large
variety of features of Cas~A remain to be explained and
require more, longer, and better (in particular, more
self-consistent) simulations.
Important questions, for example, are connected to
the following problems.

The fairly high explosion energy estimated for Cas~A (around
2.3\,B) may pose a challenge for the neutrino-driven mechanism.
Presently no modern self-consistent multi-dimensional 
simulation has achieved to obtain such energetic explosions
(see the recent reviews by \citealt{Jankaetal16} and 
\citealt{Mueller16}).\footnote{This is different from
older two-dimensional smoothed-particle-hydrodynamics simulations
with gray, flux-limited neutrino diffusion, where neutrino-driven
explosions up to $\sim$3\,B could be obtained 
\citep{Fryer99,FryerKalogera01}, and it also disagrees 
with parametric neutrino-wind-driven explosions in spherical
symmetry, which have produced explosion energies up to more
than 7\,B \citep{Youngetal06,PejchaThompson15}.} 
However, 1D explosion models with a 
parametric treatment of the neutrino-heating mechanism
(where the parameters were calibrated by employing
constraints set by observed properties of SN~1987A)
could reach these energies for the most extreme cases
\citep{Uglianoetal12,Ertletal16,Muelleretal16b}. 
Interestingly, the corresponding progenitors are in the
18--22\,$M_\odot$ range and thus around the suspected
mass of the Cas~A progenitor.
This justifies hope that the energy problem could ultimately
be solved with a better quantitative understanding of the
explosion physics of the neutrino-driven mechanism in 3D.

Of course, the
origin of the fiducial jets in Cas~A, which are actually
rather wide, high-velocity structures \citep{Fesenetal06},
deserves special attention. Could they be
connected to large-scale, low-mode asymmetries in the
convective Si- and/or O-burning shells of the collapsing
star such as the quadrupolar convection mode recently found
by \citet{Muelleretal16} for an 18\,$M_\odot$ star
at the onset of core collapse? Or do the high-velocity flows
point to a phase of bipolar (magnetic) spin-down activity
of the newly formed NS, possibly shortly after the
explosion was launched by the neutrino mechanism?
Or are they ``simply'' caused by direction and 
element-dependent differences of the interaction of
asymmetrically expelled inner ejecta with the reverse
shock from the SN expansion into the environment?
Si-rich ejecta with the highest velocities (in our
moderately energetic explosion model in excess of 
$\sim$10000\,km\,s$^{-1}$) are far outside
of the fastest iron and are likely to be affected by
their collision with the reverse shock in a different
(possibly weaker) way than the lower-velocity 
bulk of the iron deeper inside the expanding SN debris,
which encounters a fully developed reverse shock moving
inward with higher speed.

Finally, it is intriguing that both Cas~A and SN~1987A 
exhibit atypically high $^{44}$Ti masses and are attributed
to progenitors in the 15--20\,$M_\odot$ range \citep[or, possibly,
even in the 18--20\,$M_\odot$ interval, see the discussion
for SN~1987A in][]{Sukhboldetal16}. Do both cases 
share a common reason for the exceptionally high production
of $^{44}$Ti? Could there also be morphological similarities
of both explosions and remnants? 
The Doppler redshifts of the $^{44}$Ti lines in SN~1987A
were interpreted by a single-lobe explosion oriented at
an angle pointing away from us \citep{Boggsetal15}, whereas
\citet{Larssonetal16} reported a resemblance to a broken dipole
structure, analysing in detail spectra and images of SN~1987A
taken by HST/STIS and VLT/SINFONI at optical and near-infrared
wavelengths. We speculate that our 3D explosion model discussed
in the present paper, although no single-lobe explosion, might
exhibit sufficient asymmetry to explain the redshifting of the
NuSTAR spectrum with its $^{44}$Ti lines \citep{Boggsetal15} 
and of the gamma-ray and infrared emission associated with nickel
and iron \citep{Haasetal90,Spyromilioetal90}.
In order to account for this observed asymmetry of SN~1987A,
our model would predict a NS kick pointing toward
us with a rather small angle relative to the line-of-sight (in 
contrast to the case of Cas~A, where we inferred this
angle to be fairly large). In such a case the far majority
of the synthesized $^{44}$Ti and $^{56}$Ni in our model
would move away from the observer, consistent with the 
redshifted lines of $^{44}$Ti in SN~1987A. Further studies 
of this SN on the basis of the set of 3D simulations of
\citet{Wongwathanaratetal13,Wongwathanaratetal15}
will be a topic of future research.

\acknowledgments
HTJ thanks B.~Grefenstette, 
R.~Fesen, and D.~Milisavljevic for numerous discussions
and education about Cas~A observations, 
S.~Couch, M.~Modjaz, and J.~Murphy for an inspiring dinner
conversation about Cas~A, R.~Diehl for useful comments on the
manuscript, and C.~Fransson, A.~Jerkstrand, J.~Larsson, and
B.~Leibundgut for information on SN~1987A. The authors are also
grateful to an anonymous referee for knowledgable comments.
The project was supported by the Deutsche Forschungsgemeinschaft through
the Excellence Cluster ``Universe'' EXC 153, by the 
European Research Council through grant ERC-AdG No.\ 341157-COCO2CASA, 
by the JSPS Grants-in-Aid for Scientific Research (KAKENHI Grant Numbers 
26400232,26400237), and by the RIKEN iTHES project.
The computations were performed on Hydra of the Max Planck 
Computing and Data Facility.\\

\bibliographystyle{aasjournal}
\bibliography{ti44}

\end{document}